\documentclass[12pt]{amsart}

\newcommand {\supplus}{\mathop{{\supset}\llap{\raise 
0.5pt\hbox{\normalfont\small+}\hskip 0.5pt}}} 

\newcommand {\subplus}{\mathop{{\subset}\llap{\raise 
0.5pt\hbox{\normalfont\small+}\hskip 0.5pt}}}  

\newcommand {\Cee}    {{\mathbb  C}}

\newcommand {\Zee}    {{\mathbb  Z}}

\newcommand {\fa}     {{\mathfrak{a}}}

\newcommand {\fab}    {{\mathfrak{ab}}} 

\newcommand {\fas}    {{\mathfrak{as}}}
\newcommand {\faut}   {{\mathfrak{aut}}} 
\newcommand {\fb}     {{\mathfrak{b}}}
\newcommand {\fc}    {{\mathfrak{c}}}
\newcommand {\fcg}    {{\mathfrak{cg}}}

\newcommand {\fcvect}   {{\mathfrak{cvect}}}
\newcommand {\fder}   {{\mathfrak{der}}}   %

\newcommand {\fd}     {{\mathfrak{d}}}
\newcommand {\fe}     {{\mathfrak{e}}}

\newcommand {\fg}     {{\mathfrak{g}}}    %
\newcommand {\fgl}    {{\mathfrak{gl}}}  %
\newcommand {\fG}    {{\mathfrak{G}}}  %
\newcommand {\fh}     {{\mathfrak{h}}}
\newcommand {\fhei}   {{\mathfrak{hei}}}
\newcommand {\fk}     {{\mathfrak{k}}}

\newcommand {\fle}    {{\mathfrak{le}}}
\newcommand {\fm}     {{\mathfrak{m}}}
\newcommand {\fn}     {{\mathfrak{n}}}

\newcommand {\fo}     {{\mathfrak{o}}}
\newcommand {\fosp}   {{\mathfrak{osp}}}
\newcommand {\fpe}    {{\mathfrak{pe}}}   %
\newcommand {\fpg}    {{\mathfrak{pg}}}

\newcommand {\fpgl}   {{\mathfrak{pgl}}}
\newcommand {\fpo}    {{\mathfrak{po}}}

\newcommand {\fpsl}   {{\mathfrak{psl}}}
\newcommand {\fpq}    {{\mathfrak{pq}}}

\newcommand {\fpsq}   {{\mathfrak{psq}}}
\newcommand {\fq}     {{\mathfrak{q}}}

\newcommand {\fs}     {{\mathfrak{s}}}

\newcommand {\fsh}    {{\mathfrak{sh}}}

\newcommand {\fsl}    {{\mathfrak{sl}}}
\newcommand {\fsle}   {{\mathfrak{sle}}}
\newcommand {\fsm}    {{\mathfrak{sm}}}
\newcommand {\fsp}    {{\mathfrak{sp}}}
\newcommand {\fspe}   {{\mathfrak{spe}}}
\newcommand {\fspo}   {{\mathfrak{spo}}}
\newcommand {\fsq}    {{\mathfrak{sq}}}

\newcommand {\fsvect} {{\mathfrak{svect}}}

\newcommand {\fv}     {{\mathfrak{v}}}     %
\newcommand {\fvect}  {{\mathfrak{vect}}}   %
\newcommand {\fvir}   {{\mathfrak{vir}}}

\newcommand {\fwitt}  {{\mathfrak{witt}}}

\newcommand {\cal} {\mathcal}

\newcommand {\cC}     {{\cal C}}

\newcommand {\cF}     {{\cal F}}

\newcommand {\cL}     {{\cal L}}
\newcommand {\cM}     {{\cal M}}

\newcommand {\cO}     {{\cal O}}

\def \opname#1#2%
  {\expandafter\newcommand \csname #1\endcsname {{\mathop{#2}\nolimits}}}

\newcommand{\rmname}[1]
  {\expandafter\newcommand \csname #1\endcsname {{\operatorname{#1}}}}

\newcommand{\rmnameii}[2]
  {\expandafter\newcommand \csname #1\endcsname {{\operatorname{#2}}}}

\rmname{act}
\rmname{Ad}
\rmname{Add}
\rmname{ad}
\rmname{Alt}
\rmname{alt}
\rmname{Ann}
\rmname{antidiag}
\rmname{Ber}
\rmname{ber}
\rmname{Br}
\rmname{card}
\rmname{ch}
\rmname{Char}
\rmname{cem}
\rmname{cj}
\rmname{Cliff}
\rmname{cntr}
\rmname{codim}
\rmname{coind}
\rmname{const}
\rmname{col}
\rmname{cork}
\rmname{cpr}
\rmname{diag}
\rmnameii{Div}{div}
\rmname{Def}
\rmname{Der}
\rmname{Dim}
\rmname{End}
\rmname{Even}
\rmname{Ext}
\rmname{gr}
\rmname{Hom}
\rmname{HT}
\rmnameii{Ht}{ht}
\rmname{hwt}
\rmname{Id}
\rmname{id}
\rmname{ind}
\rmname{Ind}
\rmname{Inf}
\rmname{irr}
\rmname{Le}
\rmname{Lie}
\rmname{lwt} 
\rmname{mult}
\rmname{Mor}
\rmname{nm}
\rmname{Ob}
\rmname{Odd}
\rmname{Osc}
\rmname{per}
\rmname{Pic}
\rmname{pr}
\rmname{pro}
\rmname{Prime}
\rmname{Proj}
\rmname{prt}
\rmname{pt}
\rmname{Q}
\rmname{qet}
\rmname{qtr}
\rmname{rd}
\rmname{rk}
\rmname{row}
\rmname{Res}
\rmname{salt}
\rmname{Sch}
\rmname{SBr}
\rmname{scalar}
\rmname{Ser}
\rmname{sign}
\rmname{Smbl}
\rmname{spin}
\rmname{ssym}
\rmname{str}
\rmname{st}
\rmname{sgn}
\rmname{sq}
\rmname{symm}
\rmname{supp}
\rmname{Supp}
\rmname{St}
\rmname{Spec}
\rmname{Spm}
\rmname{tr}
\rmname{vpt}
\rmname{weyl}
\rmname{Weyl}
\rmname{Witt}

\opname{vvol}  {{v\hspace{-0.1ex}o\hspace{-0.02ex}l\/}}
\opname{pnt}  {\text{\normalfont pt}}
\opname{Span} {{Span}}
\opname{slim} {\overline{\lim}}
\opname{Vol}  {{V\hspace{-0.55ex}o\hspace{-0.02ex}l\/}}
\opname{QVol} {{Q\hspace{-0.3ex}V\hspace{-0.55ex}o\hspace{-0.02ex}l\/}}
\opname{PoVol}{{P\hspace{-0.35ex}o\hspace{-0.25ex}V\hspace{-0.55ex}o\hspace{-0.02ex}l\/}}
\opname{BVol} {{B\hspace{-0.2ex}V\hspace{-0.55ex}o\hspace{-0.02ex}l\/}}
\opname{Par}  {{P\hspace{-0.3ex}a\hspace{-0.05ex}r\/}}

\rmname{Mat}
\rmname{Bil}
\rmname{Diff}
\rmname{Ker}
\rmname{Herm}
\rmname{Coker}
\rmname{Conn}
\rmname{Covect}
\rmname{Vect}
\rmname{Int}

\rmnameii {IM} {Im}
\rmnameii {RE} {Re}

\opname{Aut} {{A\hspace{-0.2ex}u\hspace{-0.1ex}t\/}}
\opname{GL} {{G\hspace{-0.3ex}L}}
\opname{SL} {{S\hspace{-0.3ex}L}}
\opname{Exp} {{E\hspace{-0.2ex}x\hspace{-0.1ex}p\/}}
\opname{GQ} {{G\hspace{-0.2ex}Q}}
\opname{OSp} {{O\hspace{-0.25ex}S\hspace{-0.15ex}p\/}}
\opname{Out} {{O\hspace{-0.25ex}u\hspace{-0.15ex}t\/}}
\opname{Spp} {{S\hspace{-0.2ex}p\/}}
\opname{SpO} {{S\hspace{-0.2ex}p\hspace{-0.02ex}O\/}}
\opname{Pe} {{P\hspace{-0.25ex}e\/}}
\opname{SPe} {{S\hspace{-0.25ex}P\hspace{-0.25ex}e\/}}
\opname{Spin} {{S\hspace{-0.25ex}p\hspace{-0.05ex}i\hspace{-0.1ex}n\/}}
\opname{Iso} {{I\hspace{-0.25ex}s\hspace{-0.1ex}o\/}}
\opname{SSPe} {{S\hspace{-0.25ex}S\hspace{-0.15ex}P\hspace{-0.25ex}e\/}}
\opname{PeU} {{P\hspace{-0.25ex}e\hspace{-0.1ex}U\/}}
\opname{QU} {{Q\hspace{-0.15ex}U\/}}
\opname{U} {{U\/}}

\opname{cGQ} {{\cal G \hspace{-0.2em} Q \/}}
\opname{cSL} {{\cal S \hspace{-0.2em} L \/}}
\opname{cGL} {{\cal G \hspace{-0.2em} L \/}}
\opname{cOSp} {{\cal O \hspace{-0.2em} S \hspace{-0.3em} \it p\/}}
\opname{cPe} {{\cal P \hspace{-1.5pt} \it e\/}}
\opname{cVect} {{\cal V \hspace{-1.5pt} \it e\hspace{-0.1ex}c\hspace{-0.1ex}t\/}}
\opname{cVol} {{\cal V \hspace{-1.5pt} \it o\hspace{-0.1ex}l\/}}
\opname{cAut} {{\cal A \hspace{-0.2em} \it u\hspace{-0.1em}t\/}}
\opname{cCovect} {{\cal C \hspace{-1.5pt}
     \it o\hspace{-0.1ex}v\hspace{-0.1ex}e\hspace{-0.1ex}c\hspace{-0.1ex}t\/}}
\opname{CW} {{C\hspace{-0.15ex}W}}

\opname {Ab}   {{\sf Ab}}
\opname {Alg}   {{\sf Alg}}
\opname {ASch}  {{\sf Aff\;Sch}}
\opname {Funct}   {{\sf Funct}}
\opname {Gr}   {{\sf Gr}}
\opname {Grf}  {{{\sf Gr}_f}}
\opname {Mods}   {{\sf mods}}
\opname {Rings}   {{\sf Rings}}
\opname {Salg}   {{\sf Salg}}
\opname {Sets} {{\sf Sets}}
\opname {SSMan} {{\sf SMan}}
\opname {Top}  {{\sf Top}}
\opname {Vebun}   {{\sf Vebun}}

\newcommand {\ev} {{\bar0}}
\newcommand {\od} {{\bar1}}
\newcommand {\eps} {\varepsilon}
\newcommand {\degree}  {{}^\circ}
\newcommand {\tto} {\longrightarrow}
\newcommand {\pder}[1] {{\frac{\partial}{\partial {#1}}}}
\newcommand {\pderf}[2] {{\frac{\partial {#1}}{\partial {#2}}}}

\newcommand {\bcdot}   {\mathbin{\hbox{\raise.4ex\hbox{\bf.}}}} 

\newcommand {\secno} {}

\newtheorem{Theorem}{\secno Theorem}

\newenvironment {th*}[1]
    {\gdef\thname{#1} \begin{thn}}%
    {\end{thn}}
\newtheorem{thn}[Theorem] {\thname}

\theoremstyle{definition}

\newenvironment {ex*}[1]
    {\gdef\thname{#1} \begin{exn}}%
    {\end{exn}}
\newtheorem{exn}[Theorem]{\thname}

\theoremstyle{remark}

\newtheorem{Remark}[Theorem]{\secno Remark}

\newenvironment {rem*}[1]
    {\gdef\thname{#1} \begin{remn}}%
    {\end{remn}}
\newtheorem{remn}[Theorem]{\thname}

\newcommand {\ssec}{\subsection*}

\begin{document}

\title[Relations in Lie superalgebras]{Defining relations for the
exceptional Lie superalgebras of vector fields pertaining to The
Standard Model}

\author{Pavel Grozman${}^1$, Dimitry Leites${}^1$, Irina 
Shchepochkina${}^2$} 

\address{${}^1$Deptartment of Mathematics, University of Stockholm,
Kr\"aftriket hus 6, SE-106 91, Stockholm, Sweden;
mleites@matematik.su.se; ${}^2$The Independent University of Moscow,
Bolshoj Vlasievsky per, dom 11, RU-121 002 Moscow, Russia;
Ira@Pa\-ra\-mo\-no\-va.mccme.ru}

\thanks{Financial support of NFR and RFBR grant 99-01-00245 is
thankfully acnowledged by D.~L. and I.~Shch., respectively.  We are
also thankful to ESI for hospitality in 2000 and 2001, in particular, to
A.~Kirillov and V.~Kac, organizers of a miniprogram in 2000.}



\begin{abstract} We list defining relations for the four of the five
exceptional simple Lie superalgebras some of which, as David
Broadhurst conjectured and Kac demonstrated, may pertain to The
Standard Model of elementary particles.  For the fifth superalgebra
the result is not final: there might be infinitely many relations. 
Contrarywise, for the same Lie superalgebra with Laurent polynomials
as coefficients there are only finitely many relations.
\end{abstract} 

\maketitle

The ground field is $\Cee$.  

David Broadhurst conjectured \cite{Ka2} that some of the exceptional
infinite dimensional simple vectorial Lie superalgebras might be
related to a Standard Model, and Kac laboriously elucidated and
popularized this conjecture \cite{Ka3}.  Here we describe the
exceptional algebras in terms of generators and relations.  To make
the paper of interest to a wider audience, we give an elucidation of
the list of simple vectorial superalgebras and append with some not
very well known definitions.

\footnotesize
Our choice of contribution to Kirillov's Fest was prompted in 1977 by
A.~A.~Kirillov who suggested to one of us a joint project: calculation
of certain Lie superalgebra cohomology.  For various reasons the
project was put aside.  Still, we remember even passing wishes of our
teacher especially if we can not answer immediately; another reason
was to thank one of the editors by reminding of the problems from the
happy childhood, see \cite{Ki}.
\normalsize

Observe that even the conventional simple finite dimensional Lie
algebras admit several distinct presentations.  For example, there are
(1) the Serre relations between {\it Chevalley generators} (convenient
and simple looking but numerous: for the non-exceptional algebras
($\fg=\fsl(n)$, $\fsp(2n)$, $\fo(n)$) the number of generators and
relations grows with $\rk \fg\sim n$); (2) relations between just a
pair of {\it Jacobson generators}, see \cite{GL3} (the number of the
relations between the Jacobson generators does not depend on $n$ but,
though indispensable in some questions, they are less convenient than
Serre relations).  There are other natural choices of generators even
for $\fsl(n)$.

Among various presentations, Chevalley generators (satisfying Serre
relations) are the first choice because in terms of them we split the
algebra into the sum of its ``positive'' and ``negative'' parts
$\fg_{\pm}\simeq \fn$ which are nilpotent.  For any nilpotent Lie
algebra the very notions ``generators'' and ``relations'' are most
transparently defined.  Moreover, they admit a homological
interpretation.  Suppose, as will be the case in our examples, there
is a set of outer derivations acting on $\fn$ so that $\fn$ splits
into the direct sum of 1-dimensional eigenspaces.  Then the choice of
a basis is unique up to scalar factors; moreover, usually,
$\fn\simeq\fn /[\fn , \fn ]\oplus [\fn , \fn ]$ so any basis of the
space $\fn /[\fn, \fn] = H_{1}(\fn)$ is a set of {\it generators} of
$\fn$.

To describe relations between the generators, consider the standard
homology complex for $\fn$ with trivial coefficients (\cite{Fu}):
$$
0  \longleftarrow  \fn \stackrel{d_{1}}{\longleftarrow}   \fn \wedge \fn
\stackrel{d_{2}}{\longleftarrow}   \fn \wedge \fn
\wedge \fn  \stackrel{d_{3}}{\longleftarrow}  \dots \, .
$$
By definition
$$
d_{1}(x\wedge y)=[x, y],\quad d_{2}(x\wedge y\wedge z)=[x, y]\wedge z +
[y, z]\wedge x + [z, x]\wedge y.
$$
The condition $d_{1}d_{2}=0$ expresses the Jacobi identity. 
Obviously, the elements of $\Ker d_{1}$ are relations.  The
consequences of the Jacobi identity are considered as trivial
relations, they constitute $\IM\; d_{2}$.  Thus, for a nilpotent Lie
algebra $\fn$, we identify a convenient basis of $H_{1}(\fn)$ with the
generators and that of $H_{2}(\fn)$ with {\it defining relations}. 
Observe that the same arguments apply as well to {\it locally
nilpotent} Lie (super)algebras (as considered below).

There is, actually, one more, ``zeroth'' piece, $\fg_{0}$, consisting
of Cartan subalgebra $\fh$, but the relations expressing the elements
of $\fh$ via positive and negative generators, as well as the action
of $\fh$ on these generators, though vital and introduce Cartan matrix
(or its analog), are obvious or, at least, so easy to compute, that we
skip them.  The most difficult to describe are relations between
generators ``of the same sign'' (positive or negative).  Such
presentations for Lie algebras became very fashionable in connection
with $q$-quantizations.  For Lie superalgebras, even with Cartan
matrix, there are some subtleties: it turned out that there are
non-Serre relations even between analogs of Chevalley generators, see
\cite{FLV}, \cite{Y}, \cite{GL2}.

For simple finite dimensional Lie superalgebras, even with Cartan
matrix, there are {\it several} inequivalent sets of Chevalley
generators and, respectively, analogs of Serre relations, see
\cite{GL2}.  For the Lie superalgebras without Cartan matrix the
analogs of such sets are not described very explicitely at the moment,
besides, there are infinitely many (non-equivalent) of them for
infinite dimensional Lie (super)algebras.  So we only confine
ourselves to ``a most simple'' (or most convenient) one; for the
series it corresponds to the so-called standard grading, for the
exceptional vectorial Lie superalgebras the grading consistent with
parity (if exists) is the most convenient one.

In \S 1, we describe simple vectorial Lie superalgebras, \cite{LSh0,
LSh1}, \cite{Sh5}, \cite{Sh14}; \cite{CK2}.  Five of them, if
considered as abstract algebras, are exceptional; these 5 algebras
have exactly 15 ``incarnations'' as particularly graded or filtered
Lie superalgebras of vector fields on some supermanifolds.  There are,
of course, lots of gradings, but only finitely many (just one for
simple Lie algebras) are determined by a maximal subalgebra of finite
codimension.  Such particular (Weisfeiler) grading is described below;
it is of special interest.  Among these 15 realizations some are
distinguished by the fact that the corresponding $\Zee$-grading is
consistent (with parity), these realizations are easier to deal with
in our problem --- description of defining relations.

Actually, our shining goal is the description of defining relations
for any realization; unlike the finite dimensional case this might be
impossible: there does not seem to be a way to pass from relations in
terms of one set of simple roots to that in terms of another (even 
for $\fsl(m|n)$).

In \S 2 we state our main result.  To formulate it, we represent the Lie
superalgebras we consider as $\fg=\mathop{\oplus}\limits_{i\geq
-d}\fg_{i}$; we further set $\fg_{-}=\mathop{\oplus}\limits_{i<0}\fg_{i}$
and $\fg_{+}=\mathop{\oplus}\limits_{i>0}\fg_{i}$.

Observe that $\fg_{0}$ has its own $\Zee$-grading, so it can be
represented as $\fg_{0}=\mathop{\oplus}\limits_{i}\fg_{0, i}$, where
$\fg_{0, 0}$ is the Cartan subalgebra in $\fg_{0}$ and in $\fg$.  Set
$\fg_{0-}=\mathop{\oplus}\limits_{i<0}\fg_{0, i}$ and
$\fg_{0+}=\mathop{\oplus}\limits_{i>0}\fg_{0, i}$.

Finally, set $\fG_{+}=\fg_{0+}\oplus \fg_{+}$ and
$\fG_{-}=\fg_{0-}\oplus \fg_{-}$.

{\bf Our main result} is a description of defining relations for the
locally nilpotent Lie superalgebras $\fG_{+}$ and $\fG_{-}$.  

Observe a remarkable difference in presentations for $\fk\fas$
generated by polynomials (in this paper) and $\fk\fas^L$ generated by
Laurent polynomials, see \cite{GLS1}.  Our results (description of
defining relations for $\fG_{+}$ and $\fG_{-}$) are final in all the
cases, except $\fk\fas_{+}$; we even suspect that $\fk\fas_{+}$ might
be not finitely presented.  This never happens with Lie algebras of
class SZGPG (simple $\Zee$-graded of polynomial growth) we are
studying, but does happen with certain most innocent-looking loop
superalgebras, cf.  \cite{GL2}.

The results are verified by means of Grozman's {\it Mathematica}-based
package SuperLie, see \cite{GL1}.  We could not apply the general
arguments of \cite{FF1} to the unfinished case of $\fk\fas_{+}$;
observe that these arguments should be applied very carefully: the
original idea \cite{FF2} that all relations for $\fg_{+}$ for any
simple vectorial Lie algebras are of degree 2 is wrong, as pointed by
Kochetkov, for the Hamiltonian series which always has relations of
degree 3, in particular, for Lie superalgebras of Hamiltonian fields,
cf.  \cite{GLP}.

{\bf Related results}.  For description of defining relations in other
cases see \cite{LP} (simple and close to simple Lie algebras of vector
fields with polynomial coefficients), \cite{GLP} (nonexceptional Lie
superalgebras; for series $\fq$ see \cite{LSe}.

\section*{\S 1. Description of simple vectorial Lie 
superalgebras} 

\ssec {0.0.  On setting of the problem} Selection of Lie algebras with
reasonably nice properties is a matter of taste and is under the
influence of the problem considered.  One of the usual choices is the
class of {\it simple} algebras: they are easier to study and
illuminate important symmetries.

Of simple Lie algebras, finite dimensional algebras are the first to
study.  They can be neatly encoded by very simple graphs --- Dynkin
diagrams --- or Cartan matrices.

Next on the agenda are $\Zee$-{\it graded Lie algebras of polynomial
growth} (let us call them SZGLAPGs for short).  They resemble finite
dimensional simple Lie algebras very much and their theory is very
similar, see \cite{Ka1}; they proved very useful in various branches
of mathematics and theoretical physics.  Some of them have no Cartan
matrix, but are no less useful; e.g., such are vectorial Lie
superalgebras.

\ssec {0.0.1.  Types of the classical Lie algebras} Let us
qualitatively describe the simple Lie algebras of polynomial growth to
better visualize them.  They break into the disjoint union of the
following types:

1)  {\it finite dimensional} (growth 0).

2) {\it loop algebras}, perhaps, twisted (growth 1); more important in
applications are their ``relatives'' called affine {\it Kac--Moody}
algebras, cf.  \cite{Ka1}.

3) {\it vectorial algebras}, i.e., Lie algebras of vector
fields\footnote{These algebras are sometimes known under the recently
introduced clumsy name ``algebras of Cartan type": just imagine a
\lq\lq Cartan subalgebra in a Lie algebra of Cartan type".} with
polynomial coefficients (growth is equal to the number of
indeterminates) or their completions with formal power series as
coefficients.

4) {\it the stringy algebra}\footnote{The term induced by the lingo of
imaginative physicists who now play with (either cherish, as we do,
\cite{GSW} or take a calmer attitude \cite{WL}) the idea that an
elementary particle is not a point but rather a slinky springy string;
the term ``stringy algebra'' means ``pertaining to string theory" but
also mirrors their structure as a collection of several strings ---
the modules over the Witt algebra.  In our setting this means
``vectorial (super)algebra on a supermanifold whose base is a circle
or a ``ralative'' of such an algebra (its central extension, or an
algebra of differentiatioins, etc.".} $\fvect^{L}(1) =
\fder~\Cee[t^{-1}, t]$ (the superscript stands for Laurent).  This
algebra is often called {\it Witt algebra} \index{Witt
algebra}\index{algebra Witt} $\fwitt$ and by physicists {\it the
centerless Virasoro algebra} because its nontrivial central extension,
$\fvir$, is called the {\it Virasoro algebra}.

Strictly speaking, stringy algebras are vectorial, but we retain the
generic term {\it vectorial} for algebras with {\it polynomial} or
formal coefficients.

The algebras of types 1)--4) are $\Zee$-graded. Several of the 
algebras of types 2) and 3) (in supercase all four types) have 
deformations and some of the deformed algebras are not $\Zee$-graded 
of polynomial growth. These deformations (studied insufficiently so 
far) naturally indicate one more type: filtered Lie (super)algebras of 
polynomial growth:

5)  {\it the Lie algebra of matrices of complex size and its
generalizations}, cf.  \cite{GL3, LS}.  These algebras are simple
{\it filtered} Lie algebras of polynomial growth whose associated graded are
not simple.

\ssec {0.0.2.  Superization} After Wess and Zumino made importance of
supersymmetries manifest, cf.  \cite{D, WZ}, it was natural to list
simple ``classical'' Lie superalgebras. 

--- {\it Finite dimensional superalgebras}.  All except vectorial ones
were classified by Scheunert, Nahm, Rittenberg \cite{SNR} and
I.~Kaplansky.  The unpublished preprint-1975 by Kaplansky appeared
with some extensions as \cite{FK}, \cite{K}; it helped Kac to repare
gaps in his independent proof, cf.  \cite{K1C}.  For details of the
proof (further elucidated in \cite{Sch}) in all cases see \cite{Ka4}
where Kac also considered representation theory, infinite dimensional
case and real forms.

--- {\it (Twisted) loop superalgebras}.  For an intrinsic
characterization of loop superalgebras without apeal to Cartan matrix
(rewritten from O.~Mathieu) together with same for stringy
superalgebras see \cite{GLS1}: both types of algebras
$\fg=\mathop{\oplus}\limits_{ i=-d}^{\infty}\fg_{i}$ are of infinite
depth $d$ but
$$
\renewcommand{\arraystretch}{1.4}
\begin{array}{l}
\text{{\sl for the loop algebras every root vector corresponding to 
the real root}}\\
\text{{\sl  acts
locally nilpotently in the adjoint representation,}}\\
\text{{\sl for the stringy algebras this is not so.}}\end{array}
$$

Leites conjectured that, as for Lie algebras, simple twisted loop
superalgebras correspond to outer automorphisms of $\fg$.  Serganova
listed these automorphisms and amended the conjecture, see \cite{S1}. 
J.~van de Leur classified twisted loop superalgebras with
symmetrizable Cartan \cite{vdL}; his list supports Leites-Serganova's
conjecture \cite{S1}.

--- {\it Stringy superalgebras}.  For their conjectural list and
partial proof of the completeness of this list see \cite{GLS1, KvdL}. 
Observe that some of the stringy superalgebras possess Cartan matrix,
though nonsymmetrizable ones \cite{GLS1}.

In this paper we consider the remaining type of simple graded Lie 
superalgebras of polynomial growth:

--- {\it Vectorial Lie superalgebras}.  Main examples to keep before
mind's eye are the filtered Lie algebra $\cL=\fder~\Cee[[x]]$ of
formal vector fields and $L=\fder~\Cee[x]$ of polynomial vector fields
with grading and filtration given by setting $\deg x_{i}=1$ for all
$i$. 

More exactly, we present the exceptional simple vectorial 
superalgebras in terms of generators and defining relations.

\ssec{1.0.  Linear algebra in superspaces.  Generalities} A {\it 
superspace} is a $\Zee /2$-graded space; for any superspace 
$V=V_{\ev}\oplus V_{\od }$ denote by $\Pi (V)$ another copy of the 
same superspace: with the shifted parity, i.e., $(\Pi(V))_{\bar i}= 
V_{\bar i+\od }$.  The {\it superdimension} of $V$ is $\dim 
V=p+q\eps $, where $\eps ^2=1$ and $p=\dim V_{\ev}$, 
$q=\dim V_{\od }$.  (Usually, $\dim V$ is expressed as a pair $(p,q)$ 
or $p|q$; this obscures the fact that $\dim V\otimes W=\dim V\cdot 
\dim W$; this fact is clear with the use of $\eps $.)

A superspace structure in $V$ induces the superspace structure in the 
space $\End (V)$.  A {\it basis of a superspace} is always a basis 
consisting of {\it homogeneous} vectors; let $\Par=(p_1, \dots, 
p_{\dim V})$ be an ordered collection of their parities.  We call 
$\Par$ the {\it format} of (the basis of) $V$.  A square {\it 
supermatrix} of format (size) $\Par$ is a $\dim V\times \dim V$ matrix 
whose $i$th row and $i$th column are of the same parity $p_i$.  

\footnotesize One usually considers one of the simplest formats 
$\Par$, e.g., $\Par$ of the form $(\ev , \dots, \ev ; \od , \dots, \od 
)$ is called {\it standard}.  In this paper we can do without 
nonstandard formats.  But they are vital in the classification of 
systems of simple roots that the reader might be interested in in 
connection with applications to $q$-quantization or integrable 
systems.  Besides, systems of simple roots corresponding to distinct 
nonstandard formats are related by odd reflections --- analogs of our 
nonstandard regradings. (For an approach to superroots see \cite{S2}.)

\normalsize

The matrix unit $E_{ij}$ is supposed to be of parity $p_i+p_j$ and the 
bracket of supermatrices (of the same format) is defined via {\bf Sign 
Rule}:

{\it if something of parity $p$ moves past something of parity $q$ the 
sign $(-1)^{pq}$ accrues; the formulas defined on homogeneous elements 
are extended to arbitrary ones via linearity}.

Examples of how to apply the Sign Rule: setting $[X, 
Y]=XY-(-1)^{p(X)p(Y)}YX$ we get the notion of the supercommutator and 
the ensuing notions of the supercommutative superalgebra and the Lie 
superalgebra (which in addition to superskew-commutativity satisfies 
the super Jacobi identity, i.e., the Jacobi identity amended with the 
Sign Rule).  The derivation (better say, superderivation) of a 
superalgebra $A$ is a linear map $D: A\tto A$ that satisfies the 
Leibniz rule (and Sign rule)
$$
D(ab)=D(a)b+(-1)^{p(D)p(a)}aD(b). 
$$
In particular, let $A=\Cee[x]$ be the free supercommutative polynomial 
superalgebra in $x=(x_{1}, \dots , x_{n})$, where the superstructure 
is determined by the parities of the indeterminates: $p(x_{i})=p_{i}$.  
Partial derivatives are defined (with the help of super Leibniz Rule) 
by the formulas $\pderf{x_{i}}{x_{j}}=\delta_{i,j}$.  Clearly, the 
collection $\fder A$ of all superdifferentiations of $A$ is a Lie 
superalgebra whose elements are of the form $\sum f_i(x)\pder{x_{i}}$.

Given the supercommutative superalgebra $\cF$ of ``functions'' in 
indeterminates $x$, define the supercommutative superalgebra $\Omega$ 
of differential forms as polynomial algebra over $\cF$ in $dx$, where 
$p(d)=\od$.  Since $dx$ is even for an odd $x$, we can consider not 
only polynomials in $dx$.  Smooth or analytic functions in the 
differentials of the $x$ are called {\it pseudodifferential forms} on 
the supermanifold with coordinates $x$, see \cite{BL1}.  We will need 
them to interpret $\fh_{\lambda}(2|2)$. The exterior differential is 
defined on (pseudo) differential forms by the formulas (mind Leibniz 
and Sign Rules):
$$
d(x_{i})=dx_{i}\text{ and } d^2=0.
$$
The Lie derivative is defined (minding same Rules) by the formula
$$
L_{D}(df)=(-1)^{p(D)}d(D(f)).
$$
In particular,
$$
L_{D}\left 
((df)^{\lambda}\right)=\lambda(-1)^{p(D)}d(D(f))(df)^{\lambda-1}\text{ 
for any }\lambda\in\Cee.
$$

The {\it general linear} Lie superalgebra of all supermatrices of size 
$\Par$ is denoted by $\fgl(\Par)$; usually, $\fgl(\ev, \dots, \ev, 
\od, \dots, \od)$ is abbreviated to $\fgl(\dim V_{\bar 0}|\dim V_{\bar 
1})$.  Any matrix from $\fgl(\Par)$ can be expressed as the sum of its 
even and odd parts; in the standard format this is the following block 
expression:
$$
\begin{pmatrix}A&B\\ C&D\end{pmatrix}=\begin{pmatrix}A&0\\
0&D\end{pmatrix}+\begin{pmatrix}0&B\\ C&0\end{pmatrix},\quad 
p\left(\begin{pmatrix}A&0\\
0&D\end{pmatrix}\right)=\ev, \; p\left(\begin{pmatrix}0&B\\
C&0\end{pmatrix}\right)=\od.
$$

The {\it supertrace} is the map $\fgl (\Par)\tto \Cee$, 
$(A_{ij})\mapsto \sum (-1)^{p_{i}}A_{ii}$.  Since $\str [x, y]=0$, the 
subsuperspace of supertraceless matrices constitutes the {\it special 
linear} Lie subsuperalgebra $\fsl(\Par)$.

There are, however, two super versions of $\fgl(n)$, not one.  The 
other version is called the {\it queer} Lie superalgebra and is defined 
as the one that preserves the complex structure given by an {\it odd} 
operator $J$, i.e., is the centralizer $C(J)$ of $J$:
$$
\fq(n)=C(J)=\{X\in\fgl(n|n)\mid [X, J]=0 \}, \text{ where } J^2=-\id.
$$
It is clear that by a change of basis we can reduce $J$ to the form 
$J_{2n}=\begin{pmatrix}0&1_n\\ -1&0\end{pmatrix}$.  In the standard 
format we have
$$
\fq(n)=\left \{\begin{pmatrix}A&B\\ B&A\end{pmatrix}\right\}.
$$
On $\fq(n)$, the {\it queertrace} is defined: $\qtr: 
\begin{pmatrix}A&B\\
B&A\end{pmatrix}\mapsto
\tr B$. Denote by $\fsq(n)$ the Lie superalgebra of {\it queertraceless}
matrices.

Observe that the identity representations of $\fq$ and $\fsq$ in $V$, 
though irreducible in supersetting, are not irreducible in the 
nongraded sense: take homogeneous (with respect to parity) and 
linearly independent vectors $v_1$, \dots , $v_n$ from $V$; then 
$\Span (v_1+J(v_1), \dots , v_n+J(v_n))$ is an invariant subspace of 
$V$ which is not a subsuperspace.

A representation is {\it irreducible} \index{representation of Lie 
superalgebra irreducible} \index{$G$-type irreducible representation 
of Lie superalgebra} of {\it general type} or just of {\it $G$-type} 
if there is no invariant subspace, otherwise it is called {\it 
irreducible of $Q$-type} \index{$Q$-type irreducible representation of 
Lie superalgebra} ($Q$ is after the general queer Lie superalgebra --- 
a specifically ``superish'' analog of $\fgl$); an irreducible 
representation of $Q$-type has no invariant sub{\it super}space but 
{\it has} an invariant subspace.

{\bf Lie superalgebras that preserve bilinear forms: two types}.  To 
the linear map $F: V\tto W$ of superspaces there corresponds the dual 
map $F^*: W^*\tto V^*$ between the dual superspaces.  In a basis 
consisting of the vectors $v_{i}$ of format $\Par$, the formula 
$F(v_{j})=\mathop{\sum}\limits_{i}v_{i}A_{ij}$ assigns to $F$ the 
supermatrix $A$. In the dual bases, to $F^*$ the {\it supertransposed} matrix $A^{st}$ 
corresponds:
$$
(A^{st})_{ij}=(-1)^{(p_{i}+p_{j})(p_{i}+p(A))}A_{ji}.
$$

The supermatrices $X\in\fgl(\Par)$ such that 
$$
X^{st}B+(-1)^{p(X)p(B)}BX=0\quad \text{for an homogeneous matrix 
$B\in\fgl(\Par)$}
$$
constitute the Lie superalgebra $\faut (B)$ that preserves the 
bilinear form $B^f$ on $V$ whose matrix $B$ is given by the formula
$B_{ij}=(-1)^{p(B^{f})p(v_{i})}B^f(v_{i}, v_{j})$ for the basis vectors 
$v_{i}$.

Recall that the {\it supersymmetry} of the homogeneous form $B^f$ 
means that its matrix $B$ satisfies the condition $B^{u}=B$, where 
$B^{u}=
\begin{pmatrix} 
R^{t} & (-1)^{p(B)}T^{t} \\ (-1)^{p(B)}S^{t} & -U^{t}\end{pmatrix}$ 
for the matrix $B=\begin{pmatrix} R& S \\ T & U\end{pmatrix}$. 
Similarly, {\it skew-supersymmetry} of $B$ means that $B^{u}=-B$. 
Thus, we see that the {\it upsetting} of bilinear forms $u: \Bil (V, 
W)\tto\Bil(W, V)$, which for the {\it spaces} and when $V=W$ is expressed on 
matrices in terms of the transposition, is a new operation.

Most popular canonical forms of the nondegenerate supersymmetric form 
are the ones whose supermatrices in the standard format are the 
following canonical ones, $B_{ev}$ or $B'_{ev}$:
$$
B'_{ev}(m|2n)= \begin{pmatrix} 
1_m&0\\
0&J_{2n}
\end{pmatrix},\quad \text{where 
$J_{2n}=\begin{pmatrix}0&1_n\\-1_n&0\end{pmatrix}$},
$$
or
$$
B_{ev}(m|2n)= \begin{pmatrix} 
\antidiag (1, \dots , 1)&0\\
0&J_{2n}
\end{pmatrix}. 
$$
The usual notation for $\faut (B_{ev}(m|2n))$ is $\fosp(m|2n)$ or, 
more precisely, $\fosp^{sy}(m|2n)$.  Observe that the passage from $V$ 
to $\Pi (V)$ sends the supersymmetric forms to superskew-symmetric 
ones, preserved by the \lq\lq symplectico-orthogonal" Lie 
superalgebra, $\fsp'\fo (2n|m)$ or, better say, $\fosp^{sk}(m|2n)$, 
which is isomorphic to $\fosp^{sy}(m|2n)$ but has a different matrix 
realization.  We never use notation $\fsp'\fo (2n|m)$ in order not to 
confuse with the special Poisson superalgebra.

In the standard format the matrix realizations of these algebras
are: 
$$
\begin{matrix} 
\fosp (m|2n)=\left\{\left (\begin{matrix} E&Y&X^t\\
X&A&B\\
-Y^t&C&-A^t\end{matrix} \right)\right\};\quad \fosp^{sk}(m|2n)=
\left\{\left(\begin{matrix} A&B&X\\
C&-A^t&Y^t\\
Y&-X^t&E\end{matrix} \right)\right\}, \\
\text{where}\; 
\left(\begin{matrix} A&B\\
C&-A^t\end{matrix} \right)\in \fsp(2n),\quad E\in\fo(m).\end{matrix} 
$$

A nondegenerate supersymmetric odd bilinear form $B_{odd}(n|n)$ can be 
reduced to a canonical form whose matrix in the standard format is 
$J_{2n}$.  A canonical form of the superskew odd nondegenerate form in 
the standard format is $\Pi_{2n}=\begin{pmatrix} 
0&1_n\\1_n&0\end{pmatrix}$.  The usual notation for $\faut 
(B_{odd}(\Par))$ is $\fpe(\Par)$.  The passage from $V$ to $\Pi (V)$
establishes an isomorphism $\fpe^{sy}(\Par)\cong\fpe^{sk}(\Par)$.  
This Lie superalgebra is called, as A.~Weil suggested, {\it 
periplectic}.  The matrix realizations in the 
standard format of these superalgebras is shorthanded to:
$$
\begin{matrix}
\fpe ^{sy}\ (n)=\left\{\begin{pmatrix} A&B\\
C&-A^t\end{pmatrix}, \; \text{where}\; B=-B^t,\; 
C=C^t\right\};\\
\fpe^{sk}(n)=\left\{\begin{pmatrix}A&B\\ C&-A^t\end{pmatrix}, \;
\text{where}\; B=B^t,\;  C=-C^t\right\}.
\end{matrix}
$$
Observe that though the Lie superalgebras $\fosp^{sy} (m|2n)$ and 
$\fpe ^{sk} (2n|m)$, as well as $\fpe ^{sy} (n)$ and $\fpe ^{sk} (n)$, 
are isomorphic, the difference between them is sometimes crucial, see
\cite{Sh5}.

The {\it special periplectic} superalgebra is
$\fspe(n)=\{X\in\fpe(n)\mid \str X=0\}$.  Of particular interest to us
will be also $\fspe(n)_{a, b}=\fspe(n)\subplus\Cee(az+bd)$, where
$z=1_{2n}$, $d=\diag(1_{n}, -1_{n})$.  Hereafter $\fa\subplus \fb$ or
$\fb\subplus \fa$ denote the semidirect product in which $\fa$ is the
ideal.

\ssec{1.0.1.  What a Lie superalgebra is} Dealing with superalgebras 
it sometimes becomes useful to know their definition.  Lie superalgebras were
distinguished in topology in 1930's or earlier.  So when somebody
offers a ``better than usual'' definition of a notion which seemed to
have been established about 70 year ago this might look strange, to
say the least.  Nevertheless, the answer to the question ``what is a
Lie superalgebra?''  is still not a common knowledge.  Indeed, the
naive definition (``apply the Sign Rule to the definition of the Lie
algebra'') is manifestly inadequate for considering the (singular)
supervarieties of deformations and applying representation theory to
mathematical physics, for example, in the study of the coadjoint
representation of the Lie supergroup which can act on a supermanifold
but never on a superspace (an object from another category).  So, to
deform Lie superalgebras and apply group-theoretical methods in
``super'' setting, we must be able to recover a supermanifold from a
superspace, and vice versa.

A proper definition of Lie superalgebras is as follows, cf. 
\cite{L3}--\cite{L5}.  The {\it Lie superalgebra} in the category of
supermanifolds corresponding to the ``naive'' Lie superalgebra $L=
L_{\ev} \oplus L_{\od}$ is a linear supermanifold $\cL=(L_{\ev},
\cO)$, where the sheaf of functions $\cO$ consists of functions on
$L_{\ev}$ with values in the Grassmann superalgebra on $L_{\od}^*$;
this supermanifold should be such that for \lq\lq any" (say, finitely
generated, or from some other appropriate category) supercommutative
superalgebra $C$, the space $\cL(C)=\Hom (\Spec C, \cL)$, called {\it
the space of $C$-points of} $\cL$, is a Lie algebra and the
correspondence $C\tto \cL(C)$ is a functor in $C$.  (A.~Weil
introduced this approach in algebraic geometry in 1954; in super
setting it is called {\it the language of points} or {\it families},
see \cite{D}, \cite{L4}.)  This definition might look terribly complicated, but
fortunately one can show that the correspondence
$\cL\longleftrightarrow L$ is one-to-one and the Lie algebra $\cL(C)$,
also denoted $L(C)$, admits a very simple description: $L(C)=(L\otimes
C)_{\ev}$.

A {\it Lie superalgebra homomorphism} $\rho: L_1 \tto L_2$ in these 
terms is a functor morphism, i.e., a collection of Lie algebra 
homomorphisms $\rho_C: L_1 (C)\tto L_2(C)$ compatible with morphisms 
of supercommutative superalgebras $C\tto C'$.  In particular, a {\it 
representation} of a Lie superalgebra $L$ in a superspace $V$ is a 
homomorphism $\rho: L\tto \fgl (V)$, i.e., a collection of Lie algebra 
homomorphisms $\rho_C: L(C) \tto ( \fgl (V )\otimes C)_{\ev}$.

\begin{rem*}{Example} Consider a representation $\rho:\fg\tto\fgl(V)$. 
The tangent space of the moduli superspace of deformations of $\rho$
is isomorphic to $H^1(\fg; V\otimes V^*)$.  For example, if $\fg$ is
the $0|n$-dimensional (i.e., purely odd) Lie superalgebra (with the
only bracket possible: identically equal to zero), its only
irreducible representations are the trivial one, {\bf 1}, and
$\Pi({\bf 1})$.  Clearly, ${\bf 1}\otimes {\bf 1}^*\simeq \Pi({\bf
1})\otimes \Pi({\bf 1})^*\simeq {\bf 1}$, and because the superalgebra
is commutative, the differential in the cochain complex is trivial. 
Therefore, $H^1(\fg; {\bf 1})=E^1(\fg^*)\simeq\fg^*$, so there are
$\dim\,\fg$ odd parameters of deformations of the trivial
representation.  If we consider $\fg$ ``naively'' all of the odd
parameters will be lost.

Which of these infinitesimal deformations can be
extended to a global one is a separate much tougher question, usually
solved {\it ad hoc}. \end{rem*}

Examples that lucidly illustrate why one should always remember that a 
Lie superalgebra is not a mere linear superspace but a linear 
supermanifold are, e.g., deforms $\widetilde{\fsvect}(0|2n+1)$ and 
$\widetilde{\fs\fb}_{\mu}(2^{2n}-1|2^{2n})$ with odd parameters 
considered below and viewed as Lie algebras. In the category of 
supermanifolds these are simple Lie superalgebras.

\ssec{1.0.2.  Projectivization} If $\fs$ is a Lie algebra of scalar 
matrices, and $\fg\subset \fgl (n|n)$ is a Lie subsuperalgebra 
containing $\fs$, then the {\it projective} Lie superalgebra of type 
$\fg$ is $\fpg= \fg/\fs$.  Lie superalgebras $\fg_1\bigodot \fg_2$ 
described in sect. 3.1 are also projective.

Projectivization sometimes leads to new Lie superalgebras, for 
example: $\fpgl (n|n)$, $\fpsl (n|n)$, $\fpq (n)$, $\fpsq (n)$; 
whereas $\fpgl (p|q)\cong \fsl (p|q)$ if $p\neq q$.

\ssec{1.0.3.  What is a semisimple Lie superalgebra} These algebras
are needed in description of primitive Lie superalgebras of vector
fields --- a geometrically natural problem though wild for Lie
superalgebras, see \cite{ALSh}.  Recall that the Lie superalgebra $\fg$ without proper
ideals and of dimension $>1$ is called {\it simple}.  Examples:
$\fsl(m|n)$ for $m> n\geq 1$, $\fpsl(n|n)$ for $n>1$, $\fpsq(n)$ for
$n>2$, $\fosp(m|2n)$ for $mn\neq 0$ and $\fspe(n)$ for $n>2$.

\footnotesize

We will not need the remaining simple finite dimensional Lie
superalgebras of non-vectorial type.  These superalgebras, discovered
by I.~Kaplansky (a 1975-preprint, see \cite{K}) are
$\fosp_{\alpha}(4|2)$, the deforms of $\fosp(4|2)$, and the two
exceptions that we denote by $\fa\fg_{2}$ and $\fa\fb_{3}$.  For their
description we refer to \cite{K}, \cite{FK}, \cite{Sud}, see also \cite{GL2} for the
description of the system of simple roots see \cite{Ka4} completed in
\cite{vdL, S1, S2}.

\normalsize

We say that $\fh$ is {\it almost simple} if it can be sandwiched 
(non-strictly) between a simple Lie superalgebra $\fs$ and the Lie 
superalgebra $\fder~\fs$ of derivations of $\fs$: 
$\fs\subset\fh\subset\fder~\fs$.

By definition, $\fg$ is {\it semisimple} if its radical is zero. 
Literally following the description of semisimple Lie algebras over
the fields of prime characteristic, V.~Kac \cite{Ka4} gave the
following description of semisimple Lie superalgebras.  Let $\fs_1$,
\dots , $\fs_k$ be simple Lie superalgebras, let $n_1$, \dots , $n_k$
be {\it pairs} of non-negative integers $n_j=(n_j^\ev, n_j^\od)$, let
$\cF(n_j)$ be the supercommutative superalgebra of polynomials in
$n_j^\ev$ even and $n_j^\od$ odd indeterminates, and
$\fs=\mathop{\oplus}\limits_{j}\left (\fs_j\otimes\cF(n_j)\right)$. 
Then
$$
\fder~\fs=\mathop{\oplus}\limits_{j}\left((\fder~\fs_j)\otimes\cF(n_j)
\subplus \id_{\fs_j}\otimes
\fvect(n_j)\right). \eqno{(0.1.3)}
$$
{\sl Let $\fg$ be a subalgebra of $\fder\fs$ containing $\fs$.  If the 
projection of $\fg$ on $1\otimes\fvect(n_j)_{-1}$ is onto for each 
$j$, then $\fg$ is semisimple and all semisimple Lie superalgebras 
arise in the manner indicated}, i.e., as sums of {\it almost simple 
superalgebras} corresponding to the summands of $(0.1.3)$.

\ssec{1.0.4.  A.~Sergeev's central extension} In 70's A.~Sergeev
proved that there is just one nontrivial central extension of
$\fspe(n)$.  It exists only for $n=4$ and we denote it by $\fas$.  Let
us represent an arbitrary element $A\in\fas$ as a pair $A=x+d\cdot z$,
where $x\in\fspe(4)$, $d\in{\Cee}$ and $z$ is the central element. 
The bracket in $\fas$ is
$$
\left[\begin{pmatrix} a & b \cr 
c & -a^t \end{pmatrix}+d\cdot z, \begin{pmatrix}
a' & b' \cr 
c' & -a'{}^t \end{pmatrix} +d'\cdot z\right]=
\left[\begin{pmatrix} a & b \cr 
c & -a^t \end{pmatrix}, \begin{pmatrix}
a' & b' \cr 
c' & -a'{}^t \end{pmatrix}\right]+\tr~c\tilde c'\cdot z,\eqno{(0.1.4.1)}
$$
where $\, \tilde{}\, $ is extended via linearity from matrices
$c_{ij}=E_{ij}-E_{ji}$ on which $\tilde c_{ij}=c_{kl}$
for any even permutation $(1234)\mapsto(ijkl)$.

The Lie superalgebra $\fas$ can also be described by means of the
spinor representation.  For this we need several vectorial
superalgebras defined in sect.  0.3.  Consider $\fpo(0|6)$, the Lie
superalgebra whose superspace is the Grassmann superalgebra
$\Lambda(\xi, \eta)$ generated by $\xi_1, \xi_2, \xi_3, \eta _1, \eta
_2, \eta_3$ and the bracket is the Poisson bracket (0.3.6).  Recall
that $\fh(0|6)=\Span (H_f\mid f\in\Lambda (\xi, \eta))$.

Now, observe that $\fspe(4)$ can be embedded into $\fh(0|6)$.  Indeed, 
setting $\deg \xi_i=\deg \eta _i=1$ for all $i$ we introduce a 
$\Zee$-grading on $\Lambda(\xi, \eta)$ which, in turn, induces a 
$\Zee$-grading on $\fh(0|6)$ of the form 
$\fh(0|6)=\mathop{\oplus}\limits_{i\geq -1}\fh(0|6)_i$.  Since 
$\fsl(4)\cong\fo(6)$, we can identify $\fspe(4)_0$ with $\fh(0|6)_0$.

It is not difficult to see that the elements of degree $-1$ in the 
standard gradings of $\fspe(4)$ and $\fh(0|6)$ constitute isomorphic 
$\fsl(4)\cong\fo(6)$-modules.  It is subject to a direct verification 
that it is possible to embed $\fspe(4)_1$ into $\fh(0|6)_1$.

Sergeev's extension $\fas$ is the result of the restriction to 
$\fspe(4)\subset\fh(0|6)$ of the cocycle that turns $\fh(0|6)$ into 
$\fpo(0|6)$.  The quantization deforms $\fpo(0|6)$ into 
$\fgl(\Lambda(\xi))$; the through maps $T_\lambda: 
\fas\tto\fpo(0|6)\tto\fgl(\Lambda (\xi))$ are representations of 
$\fas$ in the $4|4$-dimensional modules $\spin_\lambda$ isomorphic to 
each other for all $\lambda\neq 0$.  The explicit form of $T_\lambda$ 
is as follows:
$$
T_\lambda: \begin{pmatrix} a & b \cr 
c & -a^t \end{pmatrix}+d\cdot z\mapsto \begin{pmatrix}
a & b-\lambda \tilde c \cr 
c & -a^t \end{pmatrix}+\lambda d\cdot 1_{4|4}, \eqno{(0.1.4.2)}
$$
where $1_{4|4}$ is the unit matrix and $\tilde c$ is defined in 
formula (0.1.4.1).  Clearly, $T_\lambda$ is an irreducible 
representation for any $\lambda$.

\ssec{1.0.5.  Vectorial Lie superalgebras.  The standard realization} 
The elements of the Lie algebra $\cL=\fder\; \Cee [[u]]$ are 
considered as vector fields.  The Lie algebra $\cL$ has only one 
maximal subalgebra $\cL_0$ of finite codimension (consisting of the 
fields that vanish at the origin).  The subalgebra $\cL_0$ determines 
a filtration of $\cL$: set
$$
\cL_{-1}=\cL\; \text{ and }\; \cL_i =\{D\in \cL_{i-1}\mid [D, 
\cL]\subset\cL_{i-1}\}\; \text{for }i\geq 1.\eqno{(0.2.1)}
$$
The associated graded Lie algebra $L=\mathop{\oplus}\limits_{i\geq 
-1}L_i$, where $L_i=\cL_{i}/\cL_{i+1}$, consists of the vector fields 
with {\it polynomial} coefficients.

{\bf Superization}.  For a simple Lie superalgebra $\cL$ (for example, 
take $\cL=\fder\, \Cee [u, \xi]$), suppose $\cL_0\subset\cL$ is a 
maximal subalgebra of finite codimension.  Let $\cL_{-1}$ be a minimal 
subspace of $\cL$ containing $\cL_0$, different from $\cL_0$ and 
$\cL_0$-invariant.  A {\it Weisfeiler filtration} of $\cL$ is 
determined by setting for $i\geq 1$:
$$
\cL_{-i-1}=[\cL_{-1}, \cL_{-i}]+\cL_{-i}\; \text{ and }\; \cL_i 
=\{D\in \cL_{i-1}\mid [D, \cL_{-1}]\subset\cL_{i-1}\}.\eqno{(0.2.2)}
$$
Since the codimension of $\cL_0$ is finite, the filtration takes the form
$$
\cL=\cL_{-d}\supset\dots\supset\cL_{0}\supset\dots \eqno{(0.2.3)}
$$
for some $d$.  This $d$ is the {\it depth} of $\cL$ and of the 
associated graded Lie superalgebra $L$.  

Considering the subspaces $(0.2.1)$ as the basis of a topology, we can 
complete the graded or filtered Lie superalgebras $L$ or $\cL$; the 
elements of the completion are the vector fields with formal power 
series as coefficients.  Though the structure of the graded algebras 
is easier to describe, in applications the completed Lie superalgebras 
are usually needed. 

Observe that not all filtered or graded Lie superalgebras of finite 
depth are {\it vectorial}, i.e., realizable with vector fields on a 
supermanifold of the same dimension as that of $\cL/\cL_{0}$; only 
those with faithful $L_{0}$-action on $L_{-}$ are.

Unlike Lie algebras, simple vectorial {\it super}algebras possess {\it 
several} nonisomorphic maximal subalgebras of finite codimension, see 
1.2.2.  

{\bf 1) General algebras}.  Let $x=(u_1, \dots , u_n, \theta_1, \dots 
, \theta_m)$, where the $u_i$ are even indeterminates and the 
$\theta_j$ are odd ones.  Set $\fvect (n|m)=\fder\; \Cee[x]$; it is 
called {\it the general vectorial Lie superalgebra}.  \index{$\fvect$ 
general vectorial Lie superalgebra}\index{ Lie superalgebra general 
vectorial}

{\bf 2) Special algebras}.  The {\it divergence}\index{divergence} of 
the field $D=\sum\limits_if_i\pder{u_{i}} + \sum\limits_j 
g_j\pder{\theta_{j}}$ is the function (in our case: a polynomial, or a 
series)
$$
\Div D=\sum\limits_i\pderf{f_{i}}{u_{i}}+
\sum\limits_j (-1)^{p(g_{j})}
\pderf{g_{i}}{\theta_{j}}.\eqno{(0.2.2)}
$$

$\bullet$ The Lie superalgebra $\fsvect (n|m)=\{D \in \fvect (n|m)\mid 
\Div D=0\}$ is called the {\it special} (or {\it divergence-free}) 
{\it vectorial superalgebra}.  \index{$\fsvect$ general vectorial Lie 
superalgebra}\index{ Lie superalgebra special vectorial}\index{ Lie 
superalgebra divergence-free}

It is clear that it is also possible to describe $\fsvect(n|m)$ as $\{ 
D\in \fvect (n|m)\mid L_D\vvol _x=0\}$, where $\vvol_x$ is the volume 
form with constant coefficients in coordinates $x$ (see 0.6) and $L_D$ 
the Lie derivative with respect to $D$.

$\bullet$ The Lie superalgebra $\fsvect_{\lambda}(0|m)=\{D \in \fvect 
(0|m)\mid \Div (1+\lambda\theta_1\cdot \dots \cdot \theta_m)D=0\}$, where 
$p(\lambda)\equiv m\pmod 2$, --- the deform of $\fsvect(0|m)$ --- is 
called the {\it deformed special} (or {\it divergence-free}) {\it vectorial 
superalgebra}.  Clearly, $\fsvect_{\lambda}(0|m)\cong 
\fsvect_{\mu}(0|m)$ for $\lambda\mu\neq 0$.  So we briefly denote 
these deforms by $\widetilde{\fsvect}(0|m)$.

Observe that for odd $m$ the parameter of deformation, $\lambda$, is 
odd.

\begin{rem*}{Remark} As is customary in differential geometry, where 
meaningful notations prevail, we sometimes write $\fvect (x)$ or 
$\fvect (V)$ if $V=\Span(x)$ and use similar notations for the 
subalgebras of $\fvect$ introduced below.  Some algebraists sometimes 
abbreviate $\fvect (n)$ and $\fs\fvect (n)$ to $W_n$ (in honor of 
Witt) and $S_n$, respectively.
\end{rem*}

{\bf 3) The algebras that preserve Pfaff equations and differential 
2-forms}.

$\bullet$ Set $u=(t, p_1, \dots , p_n, q_1, \dots , q_n)$; let
$$
\tilde \alpha_1 = dt +\sum\limits_{1\leq i\leq n}(p_idq_i - q_idp_i)\ 
+ \sum\limits_{1\leq j\leq m}\theta_jd\theta_j\quad\text{and}\quad 
\tilde \omega_0=d\tilde \alpha_1\ .
$$
The form $\tilde \alpha_1$ is called {\it contact}, the form $\tilde 
\omega_0$ is called {\it symplectic}.\index{form differential 
contact}\index{form differential symplectic} Sometimes it is more 
convenient to redenote the $\theta$'s and set
$$
\xi_j=\frac{1}{\sqrt{2}}(\theta_{j}-i\theta_{r+j});\quad \eta_j=\frac{1}{
\sqrt{2}}(\theta_{j}+i\theta_{r+j})\; \text{ for}\; j\leq r= [m/2]\; (\text{here}\;
i^2=-1),
\quad
\theta =\theta_{2r+1} 
$$ 
and in place of $\tilde \omega_0$ or $\tilde \alpha_1$ take $\alpha_1$
and $\omega_0=d\alpha_1$, respectively, where 
$$
\alpha_1=dt+\sum\limits_{1\leq i\leq n}(p_idq_i-q_idp_i)+
\sum\limits_{1\leq j\leq r}(\xi_jd\eta_j+\eta_jd\xi_j)
\left\{\begin{matrix}&
\text{ if }\ m=2r\\
+\theta d\theta&\text{ if }\ m=2r+1.\end{matrix}\right. 
$$

The Lie superalgebra that preserves the {\it Pfaff equation}
\index{Pfaff equation} $\alpha_1=0$, i.e., the superalgebra
$$
\fk (2n+1|m)=\{ D\in \fvect (2n+1|m)\mid L_D\alpha_1=f_D\alpha_1\text{ 
for some }f_D\in \Cee [t, p, q, \theta]\}, 
$$
is called the {\it contact superalgebra}.\index{$\fk$ contact 
superalgebra} \index{Lie superalgebra contact} The Lie superalgebra
$$
\begin{array}{c}
\fpo (2n|m)=\{ D\in \fk (2n+1|m)\mid L_D\alpha_1=0\}\end{array}
$$
is called the {\it Poisson} superalgebra.\index{$\fpo$ Poisson 
superalgebra} (A geometric interpretation of the Poisson superalgebra: 
it is the Lie superalgebra that preserves the connection with form 
$\alpha$ in the line bundle over a symplectic supermanifold with the 
symplectic form $d\alpha$.)

$\bullet$ Similarly, set $u=q=(q_1, \dots , q_n)$,
let $\theta=(\xi_1, \dots , \xi_n; \tau)$ be odd. Set
$$
\begin{array}{c}
\alpha_0=d\tau+\sum\limits_i(\xi_idq_i+q_id\xi_i), \qquad\qquad
\omega_1=d\alpha_0
\end{array}
$$
and call these forms the {\it odd-contact} and {\it periplectic}, 
respectively.\index{form differential contact odd}\index{form 
differential periplectic}

The Lie superalgebra that preserves
the Pfaff equation $\alpha_0=0$, i.e., the superalgebra 
$$
\fm (n)=\{ D\in \fvect (n|n+1)\mid L_D\alpha_0=f_D\cdot \alpha_0\text{ for 
some }\; f_D\in \Cee [q, \xi, \tau]\}
$$
is called the {\it odd-contact superalgebra}.\index{$\fm$ odd-contact
superalgebra}

The Lie superalgebra \index{$\fb$ Buttin superalgebra}
$$
\begin{array}{c}
\fb (n)=\{ D\in \fm (n)\mid L_D\alpha_0=0\}\end{array}
$$
is called the {\it Buttin} superalgebra (\cite{L4}).  (A geometric 
interpretation of the Buttin superalgebra: it is the Lie superalgebra 
that preserves the connection with form $\alpha_1$ in the line bundle 
of rank $\eps=(0|1)$ over a {\it periplectic} supermanifold, i.e., over a 
supermanifold with the periplectic form $d\alpha_0$.)

The Lie superalgebras
$$
\begin{array}{c}
\fsm (n)=\{ D\in \fm (n)\mid \Div\ D=0\}\ , \; \fs\fb (n)=\{ D\in
\fb (n)\mid\Div\ D=0\} 
\end{array}
$$
are called the {\it divergence-free} (or {\it special}) {\it odd-contact} and
{\it special Buttin} superalgebras, respectively.

\begin{rem*}{Remark} A relation with finite dimensional geometry is as 
follows.  Clearly, $\ker \alpha_1= \ker \tilde\alpha_1$.  The 
restriction of $\tilde \omega_0$ to $\ker \alpha_1$ is the 
orthosymplectic form $B_{ev}(m|2n)$; the restriction of $\omega_0$ to 
$\ker \tilde \alpha_1$ is $B'_{ev}(m|2n)$.  Similarly, the restriction of 
$\omega _1$ to $\ker \alpha_0$ is $B_{odd}(n|n)$.
\end{rem*}

\ssec{1.0.5.1. Generating functions} A laconic way to describe
$\fk$, $\fm$ and their subalgebras is via generating functions.

$\bullet$ Odd form $\alpha_1$.  For $f\in\Cee [t, p, q, \theta]$ 
set\index{$K_f$ contact vector field} \index{$H_f$ Hamiltonian vector 
field}:
$$
K_f=(2-E)(f)\pder{t}-H_f +
\pderf{f}{t} E, \eqno{(0.3.1)}
$$
where $E=\sum\limits_i y_i \pder{y_{i}}$ (here the $y_{i}$ are all the 
coordinates except $t$) is the {\it Euler operator} (which counts the 
degree with respect to the $y_{i}$), and $H_f$ is the hamiltonian field 
with Hamiltonian $f$ that preserves $d\tilde \alpha_1$:
$$
H_f=\sum\limits_{i\leq n}(\pderf{f}{p_i}
\pder{q_i}-\pderf{f}{q_i}
\pder{p_i}) -(-1)^{p(f)}\sum\limits_{j\leq m}\pderf{
f}{\theta_j} \pder{\theta_j}. \eqno{(0.3.2)}
$$

The choice of the form $\alpha_1$ instead of $\tilde\alpha_1$ only 
affects the shape of $H_f$ that we give for $m=2k+1$:
$$
H_f=\sum\limits_{i\leq n} (\pderf{f}{p_i}
\pder{q_i}-\pderf{f}{q_i}
\pder{p_i}) -(-1)^{p(f)}\left(\sum\limits_{j\leq
k}(\pderf{f}{\xi_j} \pder{\eta_j}+
\pderf{f}{\eta_j} \pder{\xi_j})+
\pderf{f}{\theta} \pder{\theta}\right ). 
$$

$\bullet$ Even form $\alpha_0$. For $f\in\Cee [q, \xi, \tau]$ set:
$$
M_f=(2-E)(f)\pder{\tau}- Le_f
-(-1)^{p(f)} \pderf{f}{\tau} E, \eqno{(0.3.3)}
$$
where $E=\sum\limits_iy_i 
\pder{y_i}$ (here the $y_i$ are all the coordinates except
$\tau$), and
$$
Le_f=\sum\limits_{i\leq n}( \pderf{f}{q_i}\ 
\pder{\xi_i}+(-1)^{p(f)} \pderf{f}{\xi_i}\ 
\pder{q_i}).\eqno{(0.3.4)}
$$
\index{$M_f$ contact vector field} \index{$Le_f$ periplectic vector field} 

Since 
$$
\renewcommand{\arraystretch}{1.4}
\begin{array}{l}
	L_{K_f}(\alpha_1)=2 \pderf{f}{t}\alpha_1=K_1(f)\alpha_1, \\
L_{M_f}(\alpha_0)=-(-1)^{p(f)}2 \pderf{
f}{\tau}\alpha_0=-(-1)^{p(f)}M_1(f)\alpha_0,
\end{array}\eqno{(0.3.5)}
$$
it follows that $K_f\in \fk (2n+1|m)$ and $M_f\in \fm (n)$. Observe that
$$
p(Le_f)=p(M_f)=p(f)+\od.
$$

$\bullet$ To the (super)commutators $[K_f, K_g]$ or $[M_f, M_g]$ there 
correspond {\it contact brackets}\index{Poisson bracket}\index{contact 
bracket} of the generating functions:
$$
[K_f, K_g]=K_{\{f, \; g\}_{k.b.}};\quad\quad [M_f, M_g]=M_{\{f, \; g\}_{m.b.}}
$$
The explicit formulas for the contact brackets are as follows.  Let us 
first define the brackets on functions that do not depend on $t$ 
(resp.  $\tau$).

The {\it Poisson bracket} $\{\cdot , \cdot\}_{P.b.}$ (in the realization with the form
$\omega_0$) is given by the formula 
$$
\{f, g\}_{P.b.}=\sum\limits_{i\leq n}\ \bigg(\pderf{f}{p_i}\ 
\pderf{g}{q_i}-\ \pderf{f}{q_i}\ 
\pderf{g}{p_i}\bigg)-(-1)^{p(f)}\sum\limits_{j\leq m}\ 
\pderf{f}{\theta_j}\ \pderf{g}{\theta_j}. \eqno{(0.3.6)}
$$
and in the realization with the form
$\omega_0$ for $m=2k+1$ it is given by the formula 
$$
\renewcommand{\arraystretch}{1.4}
\begin{array}{l}
	\{f, g\}_{P.b.}=\sum\limits_{i\leq n}\ (\pderf{f}{p_i}\ 
\pderf{g}{q_i}-\ \pderf{f}{q_i}\ 
\pderf{g}{p_i})-\\
(-1)^{p(f)}\left(\sum\limits_{j\leq m}( \pderf{f}{\xi_j}\ \pderf{ 
g}{\eta_j}+\pderf{f}{\eta_j}\ \pderf{ g}{\xi_j})+\pderf{f}{\theta}\ 
\pderf{ g}{\theta}\right ).
\end{array}
$$

The {\it Buttin bracket} $\{\cdot , \cdot\}_{B.b.}$ \index{Buttin 
bracket $=$ Schouten bracket} is given by the formula
$$
\{ f, g\}_{B.b.}=\sum\limits_{i\leq n}\ \left(\pderf{f}{q_i}\ 
\pderf{g}{\xi_i}+(-1)^{p(f)}\ \pderf{f}{\xi_i}\ 
\pderf{g}{q_i}\right).\eqno{(0.3.7)}
$$

\footnotesize \begin{rem*}{Remark} The what we call here ``Buttin 
bracket'' was discovered in pre-super era by Schouten; Buttin was the 
first to prove that this bracket establishes a Lie superalgebra 
structure.  The interpretations of the Buttin superalgebra similar to 
that of the Poisson algebra and of the elements of $\fle$ as analogs 
of Hamiltonian vector fields was given in \cite{L1}.  The Buttin 
bracket and ``odd mechanics'' introduced in \cite{L1} was rediscovered 
by Batalin with Vilkovisky; it gained a great deal of currency under 
the name {\it antibracket}, cf.  \cite{GPS}.  The {\it Schouten 
bracket} was originally defined on the superspace of polyvector 
fields on a manifold, i.e., on the superspace of sections of the 
exterior algebra (over the algebra $\cF$ of functions) of the tangent 
bundle, 
$\Gamma(\Lambda^{\bcdot}(T(M)))\cong\Lambda^{\bcdot}_\cF(Vect(M))$.  
The explicit formula of the Schouten bracket (in which the hatted slot 
should be ignored, as usual) is
$$
\renewcommand{\arraystretch}{1.4}
\begin{array}{c}
[X_1\wedge\dots \wedge\dots \wedge X_k, Y_1\wedge\dots \wedge Y_l]=\\ 
\sum_{i, j}(-1)^{i+j}[X_i, Y_j]\wedge X_1\wedge\dots\wedge \hat 
X_i\wedge \dots\wedge X_k\wedge Y_1\wedge\dots\wedge \hat Y_j\wedge 
\dots \wedge Y_l.
\end{array}\eqno{(*)}
$$
With the help of Sign Rule we easily superize formula $(*)$ for the 
case when $M$ is replaced with a supermanifold $\cM$.  Let $x$ and 
$\xi$ be the even and odd coordinates on $\cM$.  Setting 
$\theta_i=\Pi(\pder{x_{i}})=\check x_{i}$, 
$q_j=\Pi(\pder{\xi_{j}})=\check \xi_{j}$ we get an identification of 
the Schouten bracket of polyvector fields on $\cM$ with the Buttin 
bracket of functions on the supermanifold $\check\cM$ whose 
coordinates are $x, \xi$ and $\check x$, $\check \xi$.
\end{rem*}
\normalsize

In terms of the Poisson and Buttin brackets, respectively, the contact
brackets are
$$
\{ f, g\}_{k.b.}=(2-E) (f)\pderf{g}{t}-\pderf{f}
{t}(2-E) (g)-\{ f, g\}_{P.b.}\eqno{(0.3.8)}
$$
and
$$
\{ f, g\}_{m.b.}=(2-E) (f)\pderf{g}{\tau}+(-1)^{p(f)}
\pderf{f}{\tau}(2-E) (g)-\{ f, g\}_{B.b.}.\eqno{(0.3.9)}
$$

The Lie superalgebras of {\it Hamiltonian fields}\index{Hamiltonian
vector fields} (or {\it Hamiltonian 
superalgebra}) and its special subalgebra (defined only if $n=0$) are
$$
\fh (2n|m)=\{ D\in \fvect (2n|m)\mid\ L_D\omega_0=0\}\; \text{ and} \;
\fsh (m)=\{H_f\in \fh (0|m)\mid \int f\vvol_{\theta}=0\}.
$$
The ``odd'' analogues of the Lie superalgebra of Hamiltonian fields 
are the Lie superalgebra of vector fields $\Le_{f}$ introduced in 
\cite{L1} and its special subalgebra:
$$
\fle (n)=\{ D\in \fvect (n|n)\mid L_D\omega_1=0\} \; \text{ and} \;
\fsle (n)=\{ D\in \fle (n)\mid \Div D=0\}.
$$

It is not difficult to prove the following isomorphisms (as superspaces): 
$$
\renewcommand{\arraystretch}{1.4}
\begin{array}{rclrcl}
\fk (2n+1|m)&\cong&\Span(K_f\mid f\in \Cee[t, p, q, \xi]);&\fle
(n)&\cong&\Span(Le_f\mid f\in \Cee [q, \xi]);\\
\fm (n)&\cong&\Span(M_f\mid f\in \Cee [\tau, q, \xi]);&
\fh (2n|m)&\cong&\Span(H_f\mid f\in
\Cee [p, q, \xi]).
\end{array}
$$
Set $\fspo (m)=\{ K_f\in \fpo (0|m)\mid\int fvol_\xi=0\}$ and $\fsh 
(m)=\fspo (m)/\Cee\cdot K_1$.

\ssec{1.0.5.2.  Divergence-free subalgebras} Since, as is easy to
calculate,
$$
\Div K_f =(2n+2-m)K_1(f), 
$$
it follows that the divergence-free subalgebra of the contact Lie 
superalgebra either coincides with it (for $m=2n+2$) or is the Poisson 
superalgebra.  For the ``odd'' contact series the situation is more 
interesting: the divergence free subalgebra is simple and new (as 
compared with the above list).

Since
$$
\Div M_f =(-1)^{p(f)}2\left ((1-E)\pderf{f}{\tau} - \sum\limits_{i\leq
n}\frac{\partial^2 f}{\partial q_i \partial\xi_i}\right ), 
\eqno{(0.4.1)}
$$
it follows that
$$
\fsm (n) = \Span\left (M_f \in \fm (n)\mid (1-E)\pderf{f}{\tau}
=\sum\limits_{i\leq n}\frac{\partial^2 f}{\partial q_i
\partial\xi_i}\right ).
$$

In particular, 
$$
\Div Le_f = (-1)^{p(f)}2\sum\limits_{i\leq n}\frac{\partial^2 f}{\partial 
q_i \partial\xi_i}.\eqno{(0.4.2)}
$$
The odd analog of the Laplacian, namely, the operator
$$
\Delta=\sum\limits_{i\leq n}\frac{\partial^2 }{\partial 
q_i \partial\xi_i}\eqno{(0.4.3)}
$$
on a periplectic supermanifold appeared in physics under the name of 
{\it BRST operator}, cf.  \cite{GPS}.  The divergence-free vector fields 
from $\fsle (n)$ are generated by {\it harmonic} functions, i.e., such 
that $\Delta(f)=0$.

Lie superalgebras $\fsle (n)$, $\fs\fb (n)$ and $\fsvect (1|n)$
have ideals $\fsle \degree(n)$, $\fs\fb \degree(n)$ and $\fsvect
\degree(n)$ of codimension 1 defined from the exact sequences 
$$ 
\renewcommand{\arraystretch}{1.4}
\begin{array}{c}
0\tto \fsle \degree(n)\tto \fsle 
(n)\tto \Cee\cdot Le_{\xi_1\dots\xi_n} \tto 0, 
\\
0\tto \fs\fb \degree(n)\tto \fs\fb 
(n)\tto \Cee\cdot M_{\xi_1\dots\xi_n} \tto 0, \\
\displaystyle 0\tto \fsvect \degree(n)\tto 
\fsvect (1|n)\tto \Cee 
\cdot\xi_1\dots\xi_n\pder{t}\tto 0.\end{array}
$$

\ssec{1.0.5.3.  The Cartan prolongs} We will repeatedly use the Cartan 
prolong.  So let us recall the definition and generalize it somewhat.  
Let $\fg$ be a Lie algebra, $V$ a $\fg$-module, $S^i$ the operator of 
the $i$-th symmetric power.  Set $\fg_{-1} = V$, $\fg_0 = \fg$ and for 
$i > 0$ define the $i$-th {\it Cartan prolong} (the result of Cartan's 
{\it prolongation}) of the pair $(\fg_{-1}, \fg_0)$ as
$$
\fg_k= \{X\in \Hom(\fg_{-1}, \fg_{k-1})\mid X(v_0)(v_1, v_2,\dots v_k) = 
X(v_1)(v_0, v_2,
\dots, v_k)\;
\text{ for any }\; v_i\in \fg_{-1}\}.
$$
Equivalently, let 
$$
i: S^{k+1}(\fg_{-1})^*\otimes \fg_{-1}\tto 
S^{k}(\fg_{-1})^*\otimes \fg_{-1}^*\otimes\fg_{-1}\eqno{(0.5.1)}
$$ 
be the natural embedding and
$$
j: S^{k}(\fg_{-1})^*\otimes \fg_{0}\tto 
S^{k}(\fg_{-1})^*\otimes \fg_{-1}^*\otimes\fg_{-1}\eqno{(0.5.2)}
$$ 
the natural map.  Then $\fg_k=i(S^{k+1}(\fg_{-1})^*\otimes 
\fg_{-1})\cap j(S^{k}(\fg_{-1})^*\otimes \fg_{0})$.

The {\it Cartan prolong} of the pair $(V, \fg)$ is $(\fg_{-1}, 
\fg_{0})_* = \mathop{\oplus}\limits_{k\geq -1} \fg_k$.  

\footnotesize

(In what follows ${\bcdot}$ in superscript denotes, as is now 
customary, the collection of all degrees, while $*$ is reserved for 
dualization; in the subscripts we retain the old-fashioned $*$ instead 
of ${\bcdot}$ to avoid too close a contact with the punctuation 
marks.)

\normalsize

Suppose that the $\fg_0$-module $\fg_{-1}$ is {\it faithful}. Then, clearly, 
\begin{multline*}
(\fg_{-1}, \fg_{0})_*\subset \fvect (n) = \fder~
\Cee[x_1, ... , x_n],\; \text{ where }\; n = dim~ \fg_{-1}\; \text{ and }\\
\fg_i = \{D\in \fvect(n)\mid \deg D=i, [D, X]\in\fg_{i-1}\text{ for any }
X\in\fg_{-1}\}. 
\end{multline*}
It is subject to an easy verification that the Lie algebra structure on
$\fvect (n)$ induces same on $(\fg_{-1}, \fg_{0})_*$. 

Of the four simple vectorial Lie algebras, three are Cartan prolongs:
$\fvect(n)=(\id, \fgl(n))_*$, $\fsvect(n)=(\id, \fsl(n))_*$ and
$\fh(2n)=(\id, \fsp(n))_*$. The fourth one --- $\fk(2n+1)$ --- is the
result of a trifle more general construction described as follows.

\ssec{1.0.5.4. A generalization of the Cartan prolong} Let 
$\fg_-=\mathop{\oplus}\limits_{-d\leq i\leq -1}\fg_i$ be a nilpotent 
$\Zee$-graded Lie algebra and $\fg_0\subset \fder_0\fg$ a Lie 
subalgebra of the $\Zee$-grading-preserving derivations.  
Let 
$$
i: S^{k+1}(\fg_{-})^*\otimes \fg_{-}\tto 
S^{k}(\fg_{-})^*\otimes \fg_{-}^*\otimes\fg_{-}\eqno{(0.5.1')}
$$ 
and
$$
j: S^{k}(\fg_{-})^*\otimes \fg_{0}\tto 
S^{k}(\fg_{-})^*\otimes \fg_{-}^*\otimes\fg_{-}\eqno{(0.5.2')}
$$ 
be the natural embeddings similar to (0.5.1) and (0.5.2), 
respectively.  For $k>0$ define the $k$-th prolong of the pair 
$(\fg_-, \fg_0)$ to be:
$$ 
\fg_k = (j(S^{\bcdot}(\fg_-)^*\otimes \fg_0)\cap 
i(S^{\bcdot}(\fg_-)^*\otimes \fg_-))_k,
$$ 
where the subscript $k$ in the right hand side singles out the 
component of degree $k$.

Set $(\fg_-, \fg_0)_*=\mathop{\oplus}\limits_{i\geq -d} \fg_i$; then, 
as is easy to verify, $(\fg_-, \fg_0)_*$ is a Lie algebra.

What is the Lie algebra of contact vector fields in these terms?  
Denote by $\fhei(2n)$ the Heisenberg Lie algebra: its space is 
$W\oplus {\Cee}\cdot z$, where $W$ is a $2n$-dimensional space endowed 
with a nondegenerate skew-symmetric bilinear form $B$ and the bracket 
in $\fhei(2n)$ is given by the following relations:
$$
\text{$z$ is in the center and $[v, w]=B(v, w)\cdot z$ for any $v, w\in
W$.}
$$
Clearly, $ \fk(2n+1)\cong (\fhei(2n), \fc\fsp(2n))_*$.

\ssec{1.0.5.5.  Lie superalgebras of vector fields as Cartan's prolongs} 
The superization of the constructions from sec.  0.5 are 
straightforward: via Sign Rule.  We thus get infinite dimensional Lie 
superalgebras
$$
\begin{matrix}
\fvect(m|n)=(\id, \fgl(m|n))_*; \; \fsvect(m|n)=(\id, \fsl(m|n))_*; \\
	\fh(2m|n)=(\id, \fosp^{sk}(m|2n))_*; \\
\fle(n)=(\id, \fpe^{sk}(n))_*; \; 
\fs\fle(n)=(\id, \fspe^{sk}(n))_*. 
\end{matrix}
$$
{\it Remark}. Observe that the Cartan prolongs $(\id, \fosp^{sy} (m|2n))_*$
and $(\id, \fpe ^{sy} (n))_*$ are finite dimensional. 

The generalization of Cartan's prolongations described in sec.  0.5.1 
was first defined in \cite{ALSh} (and repeatedly used later, e.g., in 
\cite{CK1}).  Observe that after superization it has {\bf two} analogs 
associated with the contact series $\fk$ and $\fm$, respectively.

$\bullet$ Define the Lie superalgebra $\fhei(2n|m)$ on the direct sum 
of a $(2n, m)$-dimensional superspace $W$ endowed with a nondegenerate 
skew-symmetric bilinear form and a $(1, 0)$-dimensional space spanned 
by $z$.

Clearly, we have $\fk(2n+1|m)=(\fhei(2n|m), \fc\fosp^{sk}(m|2n))_*$ 
and, given $\fhei(2n|m)$ and a subalgebra $\fg$ of 
$\fc\fosp^{sk}(m|2n)$, we call $(\fhei(2n|m), \fg)_*$ the {\it 
$k$-prolong} of $(W, \fg)$, where $W$ is the identity 
$\fosp^{sk}(m|2n)$-module.

$\bullet$ The ``odd'' analog of $\fk$ is associated with the following 
``odd'' analog of $\fhei(2n|m)$.  Denote by $\fab(n)$ the {\it 
antibracket} Lie superalgebra: its space is $W\oplus \Cee\cdot z$, 
where $W$ is an $n|n$-dimensional superspace endowed with a 
nondegenerate skew-symmetric odd bilinear form $B$; the bracket in 
$\fab(n)$ is given by the following relations:
$$
\text{$z$ is odd and lies in 
the center; $[v, w]=B(v, w)\cdot z$ for $v, w\in W$.}
$$

Set $\fm(n)=(\fab(n), \fc\fpe^{sk}(n))_*$ and, given $\fab(n)$ and
a subalgebra $\fg$ of $\fc\fpe^{sk}(n)$, we call $(\fab(n),
\fg)_*$ the {\it $m$-prolong} of $(W, \fg)$, where $W$ is the identity
$\fpe^{sk}(n)$-module.

Generally, given a nondegenerate form $B$ on a superspace $W$ and a
superalgebra $\fg$ that preserves $B$, we refer to the above
generalized prolongations as to {\it $mk$-prolongation} of the pair
$(W, \fg)$.

\ssec{1.0.5.6.  A partial Cartan prolong of
$(\mathop{\oplus}\limits_{i\leq 0}\fg_i, \fh_1)$} Consider the
generalized Cartan prolong $(\fg_-\ , \fg_0)_*$.  Take a
$\fg_0$-submodule $\fh_1$ in $\fg_1$ such that $[\fg_{-1},
\fh_1]=\fg_0$.  If such $\fh_1$ exists (usually, the inclusion
$[\fg_{-1}, \fh_1]\subset\fg_0$ is strict), define the $i$th
prolongation of $(\mathop{\oplus}\limits_{i\leq 0}\fg_i, \fh_1)$ for
$i\geq 2$ to be $\fh_{i}=\{D\in\fg_{i}\mid [D, \fg_{-1}]\in \fh_{i-1}\}$. 
Set $\fh_i=\fg_i$ for $i<0$ and $\fh_*=\sum\fh_i$.

{\it Examples}: $\fvect(1|n)$ is a subalgebra of $\fk(1|2n)$.  
The former is obtained as Cartan's prolong of the same nonpositive 
part as $\fk(1|2n)$ and a submodule of $\fk(1|2n)_1$, cf.  Table 
1.2.1.  The simple exceptional superalgebra $\fk\fas$ introduced in 1.2.3 
is another example.

\ssec{1.0.6.  The modules of tensor fields} To advance further, we have 
to recall the definition of the modules of tensor fields over 
$\fvect(m|n)$ and its subalgebras, see \cite{BL1}, \cite{L2}.  
For any other $\Zee$-graded vectorial Lie superalgebra the 
construction is identical.

Let $\fg=\fvect(m|n)$ and $\fg_{\geq}=\mathop{\oplus}\limits_{i\geq 
0}\fg_{i}$.  Clearly, $\fvect_0(m|n)\cong \fgl(m|n)$.  Let $V$ be the 
$\fgl(m|n)$-module with the {\it lowest} weight $\lambda=\lwt(V)$.  
Make $V$ into a $\fg_{\geq}$-module setting $\fg_{+}\cdot V=0$ for 
$\fg_{+}=\mathop{\oplus}\limits_{i> 0}\fg_{i}$.  Let us realize $\fg$ 
by vector fields on the $m|n$-dimensional linear supermanifold 
$\cC^{m|n}$ with coordinates $x=(u, \xi)$.  The superspace 
$T(V)=\Hom_{U(\fg_{\geq})}(U(\fg), V)$ is isomorphic, due to the 
Poincar\'e--Birkhoff--Witt theorem, to ${\Cee}[[x]]\otimes V$.  Its 
elements have a natural interpretation as formal {\it tensor fields of 
type} $V$.  When $\lambda=(a, \dots , a)$ we will simply write $T(\vec 
a)$ instead of $T(\lambda)$.  We will usually consider $\fg$-modules 
induced from irreducible $\fg_0$-modules.

{\it Examples}: $\fvect(m|n)$ as $\fvect(m|n)$- and 
$\fsvect(m|n)$-modules is $T(\id)$.  More examples: $T(\vec 0)$ is the 
superspace of functions; $\Vol(m|n)=T(1, \dots , 1; -1, \dots , -1)$ 
(the semicolon separates the first $m$ (``even'') coordinates of the 
weight with respect to the matrix units $E_{ii}$ of $\fgl(m|n)$) is 
the superspace of {\it densities} or {\it volume forms}.  We denote 
the generator of $\Vol(m|n)$ corresponding to the ordered set of 
coordinates $x$ by $\vvol(x)$.  The space of $\lambda$-densities is 
$\Vol^{\lambda}(m|n)=T(\lambda, \dots , \lambda; -\lambda, \dots , 
-\lambda)$.  In particular, $\Vol^{\lambda}(m|0)=T(\vec \lambda)$ but 
$\Vol^{\lambda}(0|n)=T(\overrightarrow{-\lambda})$.  We set: 
$\Vol_0(0|m)=\{v\in \Vol\mid\int v=0\}$ and $T_0(\vec 
0)=\Lambda(m)/\Cee\cdot 1$.

If the generator $\vvol$ of $\Vol$ is fixed, then $\Vol\cong T(\vec 
0)$, as $\fsvect(m|n)$-modules.  We set $T^0_{0}(\vec 0)$ to denote 
the $\fsvect(m|n)$-module $\Vol_0(0|m)/\Cee~\vvol(\xi)$.

\footnotesize

{\it Remark}.  To view the volume element as \lq\lq $d^mud^n\xi$" is 
totally wrong: the superdeterminant can never appear as a factor under 
the changes of variables.  We can try to use the usual notations of 
differentials provided {\it all} the differentials anticommute.  Then 
the linear transformations that do not intermix the even $u$'s with 
the odd $\xi$'s multiply the volume element $\vvol(x)$, viewed as the 
fraction $\frac{du_1\cdot ...\cdot du_m}{d\xi_1\cdot ...\cdot 
d\xi_n}$, by the Berezinian of the transformation.  But how could we 
justify this?  Let $X=(x, \xi)$.  If we consider the usual, exterior, 
differential forms, then the $dX_{i}$'s super anti-commute, hence, the 
$d\xi_i$ commute; whereas if we consider the {\it symmetric} product 
of the differentials, as in the metrics, then the $dX_{i}$'s 
supercommute, hence, the $dx_i$ commute.  However, the $\pder{\xi_i}$ 
anticommute and, from transformations' point of view, 
$\pder{\xi_i}=\frac{1}{d\xi_{i}}$.  The notation, $du_1\cdot ...\cdot 
du_m\cdot\pder{\xi_1}\cdot \ldots \cdot \pder{\xi_n}$, is, 
nevertheless, still wrong: almost any transformation $A: (u, 
\xi)\mapsto (v, \eta)$ sends $du_1\cdot ...\cdot 
du_m\cdot\pder{\xi_1}\cdot ...\cdot \pder{\xi_n}$ to the correct 
element, $\ber (A)(du^m\cdot\pder{\xi_1}\cdot ...\cdot \pder{\xi_n})$, 
plus extra terms.  Indeed, the fraction $du_1\cdot ...\cdot 
du_m\cdot\pder{\xi_1}\cdot ...\cdot \pder{\xi_n}$ is the highest 
weight vector of an {\it indecomposable} $\fgl(m|n)$-module and 
$\vvol(x)$ is the notation of the image of this vector in the 
1-dimensional quotient module modulo the invariant submodule that 
consists precisely of all the extra terms.

\normalsize

\ssec{1.0.7.  Deformations of the Buttin superalgebra} (After
\cite{ALSh}.)  As is clear from the definition of the Buttin bracket, there 
is a regrading (namely, $\fb (n; n)$ given by $\deg\xi_i=0, \deg 
q_i=1$ for all $i$) under which $\fb(n)$, initially of depth 2, takes 
the form $\fg=\mathop{\oplus}\limits_{i\geq -1}\fg_i$ with 
$\fg_0=\fvect(0|n)$ and $\fg_{-1}\cong \Pi(\Cee[\xi])$.  Replace now 
the $\fvect(0|n)$-module $\fg_{-1}$ of functions (with inverted 
parity) with the module of $\lambda$-densities, i.e., set 
$\fg_{-1}\cong \Pi(\Vol(0|n)^\lambda)$, where
$$
L_D(vol_\xi)^\lambda =\lambda \Div D\cdot vol_\xi^\lambda\; \text{ and 
}\; p(vol_\xi)^\lambda=\od.
$$

\ssec{0.7.1}  Define $\fb_{\lambda}(n; n)$ as the Cartan's prolong
$(\fg_{-1}, \fg_0)_*=(\Pi(\Vol(0|n)^\lambda), \fvect(0|n))_*$. 
Clearly, it is a deform of $\fb(n; n)$.  The collection of these
$\fb_{\lambda}(n; n)$ for all $\lambda$'s is called the {\it main
deformation}.\index{deformation of the Buttin bracket, main} (Though
main, this deformation is not the quantization of the Buttin bracket,
cf.  \cite{Ko1}, \cite{L4}, \cite{LSh3}.)

The deform $\fb_{\lambda}(n)$ of $\fb(n)$ is a regrading of 
$\fb_{\lambda}(n; n)$ described as follows.  Set
$$
\fb_{\lambda}(n) =\{M_f\in \fm (n)\mid\ a\; 
\Div M_f=(-1)^{p(f)}2(a-bn)\pderf{f}{\tau}\}.\eqno{(0.7)}
$$
Taking into account the explicit form (0.4.1) of the divergence of 
$M_{f}$ we get
$$
\fb_{\lambda}(n) =\{M_f\in \fm (n)\mid (bn-aE)\pderf{f}{\tau} =a\Delta 
f\}=\{D\in\fvect(n|n+1) \mid L_{D}(\vvol_{q, \xi, 
\tau}^a\alpha_{0}^{a-bn})=0\}.
$$
It is subject to a direct check that $\fb_{a, b}(n)$ is isomorphic to  
$\fb_\lambda(n)$ for $\lambda =\frac{2a}{n(a-b)}$.  This isomorphism 
shows that $\lambda$ actually runs over $\Cee P^1$, not $\Cee$.  

Observe that $\fb_{nb, b}(n)\cong \fsm(n)$.  Observe also that 
$\fb_{a, -a}(2; 2)\cong \fh_{1/2}(2|2)$, see (OI).

As follows from the description of $\fvect(m|n)$-modules (\cite{BL1}) 
and the criteria for simplicity of $\Zee$-graded Lie superalgebras 
(\cite{Ka4}), the Lie superalgebras $\fb_\lambda(n)$ are simple for 
$n>1$ and $\lambda\neq 0$, 1, $\infty$.  It is also clear that the 
$\fb_{\lambda}(n)$ are nonisomorphic for distinct $\lambda$'s for 
$n>2$.

The Lie superalgebra $\fb(n)=\fb_{0}(n)$ is not simple: it has an 
$\eps$-dimensional, i.e., $(0|1)$-dimensional, center.  At $\lambda=1$ 
and $\infty$ the Lie superalgebras $\fb_{\lambda}(n)$ are not simple 
either: they has an ideal of codimension $\eps^n$ and $\eps^{n+1}$, 
respectively.  The corresponding exact sequences are
$$ 
\renewcommand{\arraystretch}{1.4}
\begin{array}{c}
0\tto \Cee M_{1} \tto \fb(n)\tto \fle(n)\tto 0,\\
0\tto \fb_{1} \degree(n)\tto \fb_{1}(n)\tto \Cee\cdot 
M_{\xi_1\dots\xi_n} \tto 0,\\
0\tto \fb_{\infty} \degree(n)\tto \fb_{\infty}(n)\tto \Cee\cdot 
M_{\tau\xi_1\dots\xi_n} \tto 0.\\
\end{array}
$$

\ssec{1.0.7.2. A correction} G.~Shmelev \cite{Sm} interpreted
$\fh_{\lambda}(2|2)$ of the Lie superalgebra $\fh(2|2)$ of Hamiltonian 
vector fields as preserving either the pseudodifferential form
$$
d\eta^{\frac{2\lambda-1}{1-\lambda}}\left(\lambda dqdp+
(1-\lambda)d\xi d\eta\right)
$$
or, equivalently, as preserving the pseudodifferential form
$$
d\eta^{\frac{1}{\lambda}-2}\left(dqdp+d\xi d\eta\right).
$$
Careful calculations reveal that this interpretations (that we 
rewrote, e.g. in \cite{ALSh}, are incorrect.

\ssec{1.0.8.  The exceptional simple vectorial Lie superalgebras as
Cartan's prolongs} For a more detailed description of the
``standard'' realizations of the exceptions see \cite{Sh14},
\cite{ShP}.

\ssec{1.0.9.  The structures preserved} It is always desirable to find
the structure preserved by the Lie superalgebra under the study.  To
see what do the vectorial superalgebras in nonstandard realizations
preserve, we have to say, first of all, what is the structure that
$\fg_{0}=\fvect(0|n)$ preserves on $\fg_{-1}=\Lambda(n)$.

Let $\fg=\fvect(0|n)$, set further
$$
W=\Lambda(n),\; V=\Lambda(n)/\Cee\cdot 1,\; 
V_{0}=\{\varphi\in \Lambda(n)\mid \varphi(0)=0\}.
$$
The projection $p: W\tto V$ establishes a natural isomorphism 
between $V$ and $V_{0}$. Let $i: V_{0}\tto W$ be the ``inverse'' 
embedding.

Denote by $\mult: W\otimes W\tto W$ the tensor of valency $(2, 1)$ on 
$W$ which determines the multiplication on $W$.  Since $V_{0}$ is an 
ideal in the associative supercommutative superalgebra $W$, the image 
$\mult|_{V_{0}\otimes V_{0}}$ is contained in $V_{0}$.  Denote by 
$\mult\degree$ the tensor which coincides with $\mult$ on 
$V_{0}\otimes V_{0}$ and vanishes on $\Cee\cdot 1\otimes W\oplus 
W\otimes \Cee\cdot 1$.  By means of the projection $p$ and the 
embedding $i$ we can $\fg$-invariantly transport $\mult\degree$ to 
$V$. The tensor obtained will be also denoted by $\mult\degree$.

For any monomial $\varphi\in W$ denote by $\varphi^{*}$ the dual 
functional (in the monomial basis $B(W)$). Then
$$
\mult=\mult\degree+\sum_{\varphi\in B(W)}1^*\otimes \varphi^{*}\otimes 
\varphi= \mult\degree+1^*\otimes \sum_{\varphi\in 
B(W)}\varphi^{*}\otimes \varphi.
$$
By definition of $\fg$, it preserves $\mult$, i.e., $L_{D}(\mult)=0$ 
for any $D\in\fg$. Hence,
$$
L_{D}(\mult\degree)=-L_{D}(1^{*})\otimes \sum_{\varphi\in 
B(W)}\varphi^{*}\otimes \varphi-(1^{*})\otimes 
L_{D}\left(\sum_{\varphi\in B(W)}\varphi^{*}\otimes \varphi\right).
$$
Under the restriction onto $V_{0}\otimes V_{0}$ the second summand 
vanishes. Observe that $\sum_{\varphi\in B(W)}\varphi^{*}\otimes 
\varphi$ is the identity operator on $W$. Thus,
$$
L_{D}(\mult\degree|_{V_{0}\otimes V_{0}})=-L_{D}(1^{*})\otimes 
\id|_{V_{0}}.
$$
The lift of this identity operator to $W$ reads as follows:
$$
L_{D}(\mult)=\alpha(D)\otimes \id|_{W}\text{ for a 1-form $\alpha$ on 
$W$}.
$$

Thus, all the structures preserved by $\fg_{0}$ on $\fg_{-1}$ are 
clear, except for those preserved by several of the exceptional 
algebras.  Namely, these structures are: (1) the tensor products 
$B\otimes\mult$ of a bilinear or a volume form $B$ preserved (perhaps, 
conformally, up to multiplication by a scalar) in the fiber of a 
vector bundle over a $0|r$-dimensional supermanifold on which the 
structure governed by $\mult$ is preserved, (2) $\mult\degree$, or 
$\mult$ twisted by divergence with factor $\lambda$.  Observe that the 
volume element $B$ may be not just $\vvol(\xi)$ but 
$(b+\alpha\xi_{1}\ldots\xi_{n})\vvol(\xi)$ as well.

The structures of another type, namely certain pseudodifferential 
forms, preserved by $\fh_{\lambda}(2|2)$, are already described.

\ssec{1.1.  Description of algebras} Consider infinite dimensional complex
filtered Lie superalgebras $\tilde \cL$ with decreasing filtration of the
form
$$
\tilde \cL= \tilde \cL_{-\tilde d}\supset \tilde \cL_{-\tilde d+1}\supset \dots \supset \tilde \cL_{0}\supset 
\tilde \cL_{1}\supset \dots\eqno{(PF)}
$$ 
where  {\it depth} $\tilde d$ is finite and where 

1) $\tilde \cL_{0}$ is a maximal subalgebra of finite codimension; 

2) for any non-zero $x\in \tilde L_{k}$ for $k\geq 0$, where $\tilde
L_{k}=\tilde \cL_{k}/\tilde \cL_{k+1}$, there exists $y\in \tilde
L_{-1}$ such that $[x, y]\neq 0$;

3) $\tilde \cL_{0}$ does not contain ideals.

The pair $(\tilde \cL, \tilde \cL_{0})$ is called a {\it primitive Lie
algebra}. 

It turns out that there are virtually as many primitive Lie algebras as
there are simple ones of polynomial growth and finite depth. 
Contrarywise, the classification problem of primitive Lie
superalgebras is wild, see \cite{ALSh}.

We assume that these Lie superalgebras $\tilde \cL$ are complete with
respect to a natural topology whose basis of neighborhoods of zero is
formed by the spaces of finite codimension, e.g., the $\tilde
\cL_{i}$.  (In the absence of odd indeterminates this topology is the
most natural one: we consider two vector fields {\it $k$-close} if
their coefficients coincide up to terms of degree $>k$.)  This
topology is naturally (see.  \S 1) called {\it projective limit
topology} but a more clumsy term ``linearly compact topology'' is also
used.  

Observe that the very term ``filtered algebra" implies that $[\tilde
\cL_{i}, \tilde \cL_{j}]\subset \tilde \cL_{i+j}$ whereas conditions
1) and 2) manifestly imply that $\dim \tilde L_{i}<\infty$ for all $k$
and the $\Zee$-graded Lie superalgebra $\tilde
L=\mathop{\oplus}\limits_{k\geq -\tilde d} \tilde L_{k}$ associated with
$\tilde \cL$ grows {\it polynomially}, i.e., $\dim
\mathop{\oplus}\limits_{k\leq n} \tilde L_{k}$ grows as a polynomial
in $n$.

Weisfeiler endowed every such filtered Lie algebra $\tilde \cL$ with
another, refined, filtration $\tilde \cL=\cL$:
$$
\cL= \cL_{-d}\supset \cL_{-d+1}\supset \dots \supset \cL_{0}\supset 
\cL_{1}\supset \dots\eqno{(WF1)}
$$ 
where $\cL_{-1}$ is a minimal $\cL_{0}$-invariant subspace and the 
other terms are defined by the formula (for $i\geq 1$):
$$
\cL_{-i-1}=[\cL_{-1}, \cL_{-i}]+\cL_{-i}\; \text{ and }\; \cL_i 
=\{D\in \cL_{i-1}\mid [D, \cL_{-1}]\subset\cL_{i-1}\}.\eqno{(WF2)}
$$
This $d$ is the {\it depth} of $\cL$ and of the 
associated graded Lie superalgebra $L$.  

An advantage of the {\it Weisfeiler filtrations} is that for the
corresponding regraded Lie superalgebra $\cL$ the $L_{0}$-action on
$L_{-1}$ is irreducible (see sec.  1.0.2).  These refined filtrations
are called {\it Weisfeiler filtrations} even for Lie
superalgebras, where Weisfeiler's construction is literally applied;
we will shortly write {\it W-filtrations} and call the gradings
associated with W-filtrations {\it W-gradings}.

When the $L_{0}$-module $L_{-1}$ is faithful, as is always the case
for simple Lie superalgebras $\cL$, such filtered Lie superalgebras
$\cL$ (and the associated with them graded ones, $L$) can be realized
by {\it vector fields} on $(\cL/\cL_{0})^*$ with formal (resp. 
polynomial) coefficients.  So, being primarily interested in simple Lie
superalgebras, we will refer to W-filtered and W-graded Lie
superalgebras as {\it vectorial} ones for short.

\vskip 0.2 cm 

In \cite{LSh0} we have anounced and in \cite{LSh1} we have classified
simple {\bf W-graded} vectorial Lie superalgebras.  Observe that the
classification results of Cheng and Kac concerning simple W-filtered
{\it complete} vectorial Lie superalgebras (\cite{Ka7}, \cite{CK1})
should be completed: they have only considered filtered deformations
corresponding to {\it one} filtration; other filtrations might produce
new filtered deformations....

{\bf On notations}.  To simplify grasping the general picture 
from the displayed formulas of the following theorem, let us 
immediately inform that the prime example, $\fvect(m|n; r)$, is the 
Lie algebra of vector fields whose coefficients are formal power 
series (or polynomials, depending on the problem) in $m$ commuting and 
$n$ anticommuting indeterminates with the filtration (and grading) 
determined by equating the degrees of $r$ ($0\leq r\leq m$) odd 
indeterminates to 0, the degrees of the remaining indeterminates being 
equal to 1.  The regradings of other series are determined similarly, 
see 1.2.2.  We omit indicating parameter $r$ if $r=0$.

In the sequel for any $\fg$ we write 
$\fcg=\fg \oplus {\Cee}\cdot z$ or $\fc(\fg)$ to denote the trivial 
central extension with the 1-dimensional even center generated by $z$.  
Any nontrivial central extension is shorthanded as $\fe(\fg)$.

If $d$ is the operator that determines the $\Zee$ grading of $\fg$ and
does not belong to $\fg$, then the Lie algebra $\fg\subplus\Cee\cdot
d$ is denoted by $\fd(\fg)$.  If the operators $d$ and $z$ described
above may be considered as having the same degree with respect to some
grading, we write $\fg_{a, b}$ to shorthand $\fg\subplus\Cee(az+bd)$.

In proofs we use notations of roots in the simple finite dimensional Lie 
algebras from \cite{OV}, Table 5. The standard (identity) representation of 
a matrix Lie superalgebra $\fg$, i.e., a subalgebra of $\fgl(V)$, in $V$ is 
denoted by $\id$ or, for clarity, $\id_{\fg}$.

{\bf Further elucidations}.  1) In formulas $(S)$ and $(E)$ below:
parentheses contain the superdimension of the superspace of
indeterminates.  The semicolon separates the superdimension from a
shorthand description of the regrading.

Passing from one regrading to another one we take a ``minimal''
realization (i.e., with a minimal $\dim\cL/\cL_{0}$) as the point of
reference.  For the exceptional Lie superalgebras another point of
reference is often more convenient, the consistent regrading $K$.

The regradings of the series are governed by a parameter $r$ described
in sec.  1.2.  All regradings are given in suggestive notations, e.g.,
$\fk\fas^\xi (; 3\eta)$ means that having taken $\fk\fas^\xi$ as the
point of reference we set $\deg\eta=0$ for each of the three $\eta$'s
(certainly, this imposes some conditions on the degrees of the other
indeterminates).  The exceptional grading $E_{H}$ of
$\fh_{\lambda}(2|2)$ is described in passing in the list of occasional
isomorphisms (OI); it is described in detail for another incarnation
of this algebra (1.2.1.1).  By $K$ we initially denoted {\it
consistent} gradings spelled with Russian accent, later we decided to
retain it in order to acknowledge Kac's skill in using it, $CK$ is an
exceptional grading found by Cheng and Kac.

2) Several algebras are ``drop-outs'' from the series.  For example:
$\fsvect\degree(1|n; r)$ are ``drop-outs'' from the series
$\fsvect(m|n; r)$ since the latter are not simple for $m=1$ but
contain the simple ideal $\fsvect\degree(1|n; r)$.  Similarly,
$\fle(n|n; r)$, $\fb_{1}\degree(n|n+1; r)$ and
$\fb_{\infty}\degree(n|n+1; r)$ are ``drop-outs'' from the series
$\fb_{\lambda}(n|n+1; r)$ corresponding to $\lambda= 0$, 1 and
$\infty$, respectively.  Though $\fsm$ is not a dropout due to the
above reason, it is singled out by its divergence free property,
hence, deserves a separate line.

The finite dimensional Lie superalgebra $\fh(0|n)$ of hamiltonian
vector fields is not simple either and contains a simple ideal
$\fsh(0|n)$.

\begin{Theorem} The simple W-graded
vectorial Lie superalgebras $L$ constitute the following series $(S)$
and  five exceptional families of fifteen individual algebras
$(E)$.  They are pairwise non-isomorphic, as graded and filtered
superalgebras, bar for occasional isomorphisms $(OI)$.  Parameters at
which $\dim L$ becomes finite are marked by ``FD''.

All these algebras are either Cartan prolongs or results of the
generalized Cartan prolongations (described in sec 0.5) and,
therefore, are determined by the terms $L_{i}$ with $i\leq 0$ (or
$i\leq 1$ in some cases).  These terms are listed in sec.  1.2 {\em
(and their form might drastically vary with $n$ and $r$)}.
\end{Theorem}
$$
\renewcommand{\arraystretch}{1.4}
\begin{tabular}{|l|}
\hline
$\fvect(m|n; r)$ for $(m|n)\neq (0|1)$ and $0\leq r\leq n$; FD for 
$m=0$, $n>1$\cr
\hline
$\fsvect(m|n; r)$ for $m>1$, $0\leq r\leq n$; FD for 
$m=0$ and $n>2$ \cr
$\fsvect\degree(1|n; r)$ for $n>1$, $0\leq r\leq n$\cr
$\widetilde{\fsvect}(0|n)$ for $n>2$ (FD)\cr
\hline
\end{tabular}\eqno{(S_{vect})}
$$
$$
\renewcommand{\arraystretch}{1.4}
\begin{tabular}{|l|}
\hline
$\fk(2m+1|n; r)$ for $0\leq r\leq [\frac{n}{2}]$ unless $(m|n)=(0|2k)$\cr
$\fk(1|2k; r)$ for $0\leq r\leq k$ except $r=k-1$\cr
\hline
$\fh(2m|n; r)$ for $m>0$ and $0\leq r\leq [\frac{n}{2}]$\cr
$\fh_{\lambda}(2|2; r)$ for $\lambda\neq 0, 1, \infty$, and $r=0$, $1$ 
and $E_{H}$ (see (OI) and sec. 1.2.1.1)\cr
$\fsh(0|n)$ for $n>3$ (FD)\cr
\hline
\end{tabular}\eqno{(S_{k})}
$$
$$
\renewcommand{\arraystretch}{1.4}
\begin{tabular}{|l|}
\hline
$\fm(n|n+1; r)$ for $0\leq r\leq n$ except $r= n-1$\cr
\hline
$\fsm(n|n+1; r)$ for $0\leq r\leq n$ except $r= n-1$ and except $n=3$\cr
\hline
$\fb_{\lambda}(n|n+1; r)$ for $n>1$, where $\lambda\neq 0, 1, \infty$ 
and $0\leq r\leq n$ except $r= n-1$\cr
$\fb_{\lambda}(2|3; r)$, where $\lambda \neq 0, 1, \infty$ 
for $r=0$, $2$ and $E$ (see sec. 1.2.1.1)\cr
\hline
$\fb_{1}\degree(n|n+1; r)$ for $n>1$ and $0\leq r\leq n$ except $r= 
n-1$\cr 
\hline
$\fb_{\infty}\degree(n|n+1; r)$ for $n>1$ and $0\leq r\leq n$ except 
$r= n-1$\cr 
\hline
$\fle(n|n; r)$ for $n>1$ and $0\leq r\leq n$ except $r= n-1$\cr
\hline
$\fsle\degree(n|n; r)$ for $n>1$ and $0\leq r\leq n$ except $r= 
n-1$\cr \hline
$\widetilde{\fs\fb}_{\mu}(2^{n-1}-1|2^{n-1})$ for $\mu\neq 0$ and $n>2$\cr
\hline
\end{tabular}\eqno{(S_{m})}
$$

Our original notations of exceptional simple vectorial superalgebras,
though reflect the way they are constructed and the geometry
preserved, are rather long.  But to write just $\fe(\dim)$ is to
create confusion: the superdimensions of the superspaces
$(\cL/\cL)^{*}$ on which the algebra $\cL$ is realized coincide
sometimes for different regradings.  So we simplify notations only for
$\fk\fs\fle$ and, in accordance with description of nonstandard
gradings in sec.  1.2, we set:
$$
\renewcommand{\arraystretch}{1.4}
\begin{tabular}{|l|l|}
\hline
Lie superalgebra&its regradings (shorthand)\cr
\hline
$\fv\fle(4|3; r)$, $r=0, 1, K$&$\fv\fle(4|3)$,  $\fv\fle(5|4)$,   and $\fv\fle(3|6)$\cr
\hline
	$\fv\fa\fs(4|4)$&$\fv\fa\fs(4|4)$\cr
\hline
$\fk\fa\fs(1|6; r)$, $r=0, 1, 3\xi, 3\eta$&$\fk\fa\fs(1|6)$, $\fk\fa\fs(5|5)$, $\fk\fa\fs(4|4)$,  
and  
$\fk\fa\fs(4|3)$\cr \hline
$\fm\fb(4|5; r)$, $r=0, 1, K$&$\fm\fb(4|5)$, $\fm\fb(5|6)$,  and $\fm\fb(3|8)$\cr
\hline
$\fk\fs\fle(9|6; r)$, $r=0, 2, K, CK$&$\fk\fs\fle(9|6)$, $\fk\fs\fle(11|9)$,
$\fk\fs\fle(5|10)$, and $\fc\fk(9|11)$\cr \hline
\end{tabular}\eqno{(E)}
$$

Hereafter we inconsistently abbreviate $\fm(n|n+1; r)$,
$\fb_{\lambda}(n|n+1; r)$, $\fle(n|n; r)$, etc., i.e., algebras from
$(S_{m})$, to $\fm(n; r)$, $\fb_{\lambda}(n; r)$, etc., respectively.

Sometimes instead of $\fb_{\lambda}(n; r)$, where
$\lambda=\frac{2a}{n(a-b)}\in\Cee\cup\{\infty\}$, we write $\fb_{a,
b}(n; r)$ for clarity.  Parameters $a, b$, as natural as $\lambda$, are
introduced in sec.  0.7.  Observe that the exceptional regrading $E$
of $\fb_{\lambda}(2)$ and isomorphisms $\fh_{\lambda}(2|2)\cong
\fb_{\lambda}(2; 2)$ determine the exceptional grading $E_{H}$ of
$\fh_{\lambda}(2|2)$.  

{\bf (OI): Occasional isomorphisms}:
$$
\renewcommand{\arraystretch}{1.4}
\begin{array}{l}
\fvect(1|1)\cong \fvect(1|1; 1);\\
\fsvect(2|1)\cong \fle(2; 2); \; \fsvect(2|1; 1)\cong \fle(2)\\
\fsm(n)\cong\fb_{2/(n-1)}(n); 
\text{ in particular, 
$\fsm(2)\cong\fb_{2}(2)$, and} \\
\text{$\fsm(3)\cong\fb_{1}(3)$, hence, 
$\fsm(3)$ is not 
simple};\\
\fb_{1/2}(2; 2)\cong 
\fh_{1/2}(2|2)=\fh(2|2);\; \; \fh_{\lambda}(2|2)\cong \fb_{\lambda}(2; 
2); \; \; \fh_{\lambda}(2|2; 1)\cong \fb_{\lambda}(2);\\
\fb_{a, b}(2; E)\cong\fb_{-b, -a}(2)\cong \fb_{b, a}(2);\; 
\text{for $a\neq b$ and $b\neq 0$};\\
\text{$\fb_{a, 0}(2; E)\cong\fle(2)$ and
$\fb_{a, a}(2; E) \cong \fb_{\infty}^\circ(2)$};\\
\fh_{\lambda}(2|2)\cong \fh_{-1-\lambda}(2|2);\\
\widetilde{\fs\fb}_{\mu}(2^{n-1}-1|2^{n-1})\cong 
\widetilde{\fs\fb}_{\nu}(2^{n-1}-1|2^{n-1}) \text{ for $\mu\nu\neq 0$}.
\end{array}\eqno{(OI)}
$$
Though $\fb_{\lambda}(2)$ and $\fh_{\lambda}(2|2)$ are isomorphic, we 
consider them separately because they preserve very distinct 
structures.

{\bf Warning}.  Isomorphic abstract Lie superalgebras might be 
quite distinct as filtered: e.g., regradings provide us with 
isomorphisms (\cite{ALSh})
$$
\fk(1|2)\cong\fvect(1|1)\cong\fm(1)\text{ as abstract algebras}.
$$
Observe that of the above three nonisomorphic filtered 
algebras only one is W-filtered.

Even if the regraded algebra can be realized by vector fields on the
superspace of the same dimension, the structures preserved are totally
different: e.g., $\fk(1|2)\cong\fm(1)$ and $\fk\fs\fle(9|6;
2)=\fk\fs\fle(11|9)\cong\fc\fk(9|11)$.

\begin{Remark} 1) The excluded regradings $\fk(1|2k; k-1)$ as well as
$\fm(n|n+1; n-1)$ and the ones they induce on $\fb_{\lambda}$, $\fle$
and $\fsle$ are often considered in applications (at least for small
values of $k$ and $n$) though these gradings/filtrations are not
Weisfeiler ones: for them the $\fg_0$-module $\fg_{-1}$ is reducible.

2) $\widetilde{\fsvect}(0|n)$, as well as 
$\widetilde{\fs\fb}_{\mu}(2^{n-1}-1|2^{n-1})$, depend on an odd 
parameter if $n$ is odd.

3) The Lie superalgebras $\widetilde{\fsvect}$ and $\fsh$, as well as 
$\fvect$, $\fsvect$ for $m=0$, are finite dimensional.
\end{Remark}

The above Lie superalgebras sometimes admit deformations that do not 
possess Weisfeiler filtrations.  These deformations are considered in 
detail in \cite{L4}, \cite{Ko1}, \cite{Ko2}, \cite{LSh3}, \cite{LSh4}.

The five types of exceptional Lie superalgebras are given below in
their minimal realizations as Cartan's prolongs $(\fg_{-1},
\fg_{0})_{*}$ or generalized (see sect.  0.5) Cartan's prolongs
$(\fg_{-}, \fg_{0})_{*}^{mk}$ for
$\fg_{-}=\mathop{\oplus}\limits_{-d\leq i\leq -1}\fg_{i}$ expressed 
for $d=2$ as $(\fg_{-2}, \fg_{-1}, \fg_{0})_{*}^{mk}$ together with
one of the Lie superalgebras from (S) as an ambient which contains the
exceptional one as a maximal subalgebra.

$$
\renewcommand{\arraystretch}{1.4}
\begin{tabular}{|l|}
\hline
$\fv\fle(4|3; r)=(\Pi(\Lambda(3))/\Cee\cdot 1, 
\fc\fvect(0|3))_{*}\subset\fvect(4|3), \quad r= 0, 1, K$\cr
\hline
$\fv\fas(4|4)=(\spin, \fas)_{*}\subset \fvect(4|4)$\cr
\hline
$\fk\fas^\xi(; r)\subset \fk(1|6; r),\quad r=0, 1, 
3\xi$\cr 
$\fk\fas^\xi(; 3\eta)=(\Vol_{0}(0|3), 
\fc(\fvect(0|3)))_{*}\subset \fsvect(4|3)$\cr 
\hline
$\fm\fb(4|5; r)=(\fab(4), \fc\fvect(0|3))_{*}^m\subset \fm(4|5), \quad 
r=0, 1, K$\cr
\hline
$\fk\fsle(9|6; r)=(\fhei(8|6), \fsvect(0|4)_{3, 4})_{*}^k\subset 
\fk(9|6), \quad r=0, 2$\cr 
$\fk\fsle(9|6; K)=(\id_{\fsl(5)}, 
\Lambda^2(\id_{\fsl(5)}^*), \fsl(5))_{*}^k\subset \fsvect(5|10; 2, 2, 2, 2, 2|1, 
\dots , 1)$\cr 
$\fk\fsle(9|6; CK)$ is described at the end of 1.2.2.\cr 
\hline
\end{tabular}\eqno{(E)}
$$

\ssec{1.2.1.  Nonstandard realizations} The following are all the
nonstandard gradings of the Lie superalgebras indicated.  In
particular, the gradings in the series $\fvect$ induce the gradings in
the series $\fsvect$, $\fsvect\degree$ and the exceptional algebras
$\fv\fle(4|3)$ and $\fv\fas(4|4)$; the gradings in $\fm$ induce the
gradings in $\fb_{\lambda}$, $\fle$, $\fsle$, $\fsle\degree$, $\fb$,
$\fs\fb$, $\fs\fb\degree$ and the exceptional algebra $\fm\fb$; the
gradings in $\fk$ induce the gradings in $\fpo$, $\fh$ and the
exceptional algebras $\fk\fas$ and $\fk\fsle$.

In what follows we consider $\fk (2n+1|m)$ as preserving the Pfaff 
equation $\tilde \alpha =0$, where 
$$
\tilde \alpha =dt+\mathop{\sum}\limits_{i\leq 
n}(p_idq_i-q_idp_i)+\mathop{\sum}\limits_{j\leq r} 
(\xi_jd\eta_j+\eta_jd\xi_j)+\mathop{\sum}\limits_{k\geq 
m-2r}\theta_kd\theta_k.
$$

The standard realizations correspond to $r=0$, they are marked by 
$(*)$.  Observe that the codimension of ${\cal L}_0$ attains its 
minimum in the standard realization.
  
\small
$$
\renewcommand{\arraystretch}{1.3}
\begin{tabular}{|c|c|}
\hline
Lie superalgebra & its $\Zee$-grading \\ 
\hline
$\fvect (n|m; r)$, & $\deg u_i=\deg \xi_j=1$  for any $i, j$
\hskip 5.5 cm
$(*)$\\ 
\cline{2-2}
$ 0\leq r\leq m$ & $\deg \xi_j=0$ for $1\leq j\leq r;$\\
&$\deg u_i=\deg \xi_{r+s}=1$ for any $i, s$ \\ 
\hline
 & $\deg \tau=2$, $\deg q_i=\deg \xi_i=1$  for any $i$ \hskip 4 cm $(*)$\\ 
\cline{2-2}
$\fm(n; r),$& $\deg \tau=\deg q_i=1$, $\deg \xi_i=0$ for any $i$\\ 
\cline{2-2}
$\; 0\leq r\leq n$& $\deg \tau=\deg q_i=2$, $\deg \xi_i=0$ for $1\leq i\leq r
<n$;\\ 
$r\neq n-1$& $\deg u_{r+j}=\deg \xi_{r+j}=1$ for any $j$\\ 
\hline
\cline{2-2}
$\fk (2n+1|m; r)$, & $\deg t=2$,\\
& $\deg p_i=\deg q_i=
\deg \xi_j=\deg \eta_j=\deg \theta_k=1$ for any $i, j, k$\qquad $(*)$ \\ 
\cline{2-2}
$0\leq r\leq [\frac{m}{2}]$& $\deg t=\deg \xi_i=2$, $\deg 
\eta_{i}=0$ for $1\leq i\leq r\leq [\frac{m}{2}]$; \\
$r\neq k-1$ for $m=2k$ and $n=0$&$\deg p_i=\deg q_i=\deg \theta_{j}=1$ for $j\geq 1$ and all $i$\\ 
\hline
$\fk(1|2m; m)$ & $\deg t =\deg \xi_i=1$, $\deg 
\eta_{i}=0$ for $1\leq i\leq m$ \\ \hline
\end{tabular}
$$
\vskip 0.2 cm

\normalsize 

\ssec{1.2.1.1.  The exceptional nonstandard regrading $E$} This is a 
regrading of $\fb_{a, b}(2; E)$ given by the formulas:
$$
\deg\tau=0; \; \; \deg \xi_1=\deg\xi_2= -1; \; \; \deg q_1=\deg 
q_2=1.\eqno{(r=E)}
$$
Then for the generic $a$ and $b$ we have:
$$
\fg_{-2}=\Pi(\Cee\xi_1\xi_2)\text{ and }\fg_{-1}=\Span\{\Le_{\xi_1}, 
\Le_{\xi_2}, \Le_{Q_1}, \Le_{Q_2}\}, \eqno{(*)}
$$
where $Q_1=A\xi_1\xi_2 q_1+B\tau\xi_2$, $Q_2=A\xi_1\xi_2 
q_2-B\tau\xi_1$ and where $A$ and $B$ are some coefficients determined 
by $a$ and $b$.  The bracket on $\fg_{-1}$ is determined by the odd 
form $\omega=c\sum dQ_i d\xi_i$, so $\fg_0$ must be contained in 
$\fm(2)_0$.  The direct calculations show that $\dim \fg_0=4|4$ and
$$
\fg_0=\fspe(2)\oplus \Cee X, \text{ where } X=\Le_{a\tau+b\sum 
q_i\xi_i}.
$$
Indeed, $\fspe(2)_0\cong\fsl(2)=\Span\{\Le_{q_1\xi_2}, \Le_{q_2\xi_1},
\Le_{q_1\xi_1-q_2\xi_2}\}$, $\fspe(2)_{-1}=\Cee\cdot \Le_{1}$ and
$\fspe(2)_1=\Cee \Le_{\alpha \xi_1\xi_2 P(q)+ \beta
\tau\Delta(\xi_1\xi_2P(q))}$, where $P(q)$ is a monomial of degree 2
and $\alpha, \beta$ are some coefficients.  The eigenvalues of $X$ on
$\fg_{-1}$ are $-a+b$ on the even part and $a+b$ an the odd part.  So
$\fb_{a, b}(2; E)\cong\fb_{-b, -a}(2)\cong \fb_{b, a}(2)$.

For $b=0$ and $a=b$ formula $(*)$ should be modified because then
$\fg_{-2}=0$.  In these cases $\fb_{a, 0}(2; E)\cong\fle(2)$ and
$\fb_{a, a}(2; E) \cong \fb_{\infty}^\circ(2)$, respectively.

For $n>2$, as well as for $\fm(n)$ for $n>1$, similar regradings are not 
Weisfeiler ones.

The exceptional grading $E$ of $\fb_{\lambda}(2)$ induces the
exceptional grading $E_{H}$ of the isomorphic algebra
$\fh_{\lambda}(2|2)$, see (OI).

\ssec{1.2.1.2.  The W-regradings of exceptional
algebras}

\begin{Theorem} 
{\em (\cite{Sh14}, \cite{CK2})} The W-regradings of the exceptional
simple vectorial Lie superalgebras are given by the following
regradings of their ``standard'' ambients listed in sec. $1.1$:
\end{Theorem}

\noindent 1) $\fv\fle(4|3; r)=(\Pi (\Lambda(3)/{\Cee}\cdot 1), \fcvect
(0|3))_*\subset \fvect(4|3), \quad r= 0, 1, K$;

\underline{$r= 0$}: $\deg y= \deg u_i= \deg\xi_i=1$

\underline{$r= 1$}: $\deg y =\deg\xi_1=0$, $ \deg u_2=\deg 
u_3=\deg\xi_2=\deg\xi_3=1$, $\deg u_1=2$

\underline{$r= K$}: $\deg y=0$, $\deg u_i=2$; $\deg\xi_i=1$

\noindent 2) $\fv\fas(4|4)=(\spin, \fas)_*\subset \fvect(4|4)$;

\noindent 3) $\fk\fas\subset \fk(1|6; r)$,\; $r=0, 1, 3\xi$; \quad 
$\fk\fas(;3\eta)\subset \fsvect(4|3)$;

\underline{$r=0$}: $\deg t=2$, \; $\deg \eta_i=1$; \; 
$\deg\xi_i=1$; 

\underline{$r=1$}: $\deg\xi_1=0$, \; $\deg\eta_1=\deg t =2$, \; 
$\deg\xi_2=\deg\xi_3= \deg\eta_2= \deg\eta_3=1$; 

\underline{$r=3\xi$}: $\deg\xi_i=0$, \; $\deg\eta_i=\deg t=1$; 

\underline{$r=3\eta$}: $\deg\eta_i=0$, $\deg\xi_i=\deg t=1$

\noindent 4) $\fm\fb(4|5; r)=(\fab(4), \fcvect (0|3))_*^{m}\subset \fm(4), \quad 
r=0, 1, K$;

\underline{$r= 0$}: $\deg \tau=2$, \; $\deg u_i= \deg\xi_i=1$ for   
$i=0$, 1, 2, 3;

\underline{$r= 1$}: $\deg \tau=\deg \xi_0=\deg u_1=2$,\; $\deg 
u_2=\deg u_3 =\deg\xi_2=\deg\xi_3=1$; $\deg\xi_1=\deg u_0=0$

\underline{$r= K$}: $\deg \tau=\deg \xi_0=3$, \; $\deg u_0=0$, \; 
$\deg u_i=2$; \; $\deg\xi_i=1$ for $i>0$

\noindent 5) $\fk\fsle(9|6; r)=(\fhei(8|6), \fsvect _{3, 4}(4))_*^{k} \subset 
\fk(9|6), \quad r=0$,  2,  K, CK

\underline{$r= 0$}: $\deg t=2$, \; $\deg p_i= \deg q_i= \deg\xi_i= 
\deg\eta_i=1$; 

\underline{$r= 2$}: $\deg t=\deg q_3=\deg q_4=\deg\eta_1= 2$,\; $\deg 
q_1=\deg q_2= \deg p_1=\deg p_2=\deg\eta_2=\deg\eta_3=\deg\zeta_2 
=\deg\zeta_3=1$; \; $\deg p_3=\deg p_4=\deg\zeta_1=0$; 

\underline{$r= K$}: $\deg t=\deg q_i=2$, \; $\deg p_i=0$;  \; 
$\deg\zeta_i=\deg\eta_i=1$; 

\underline{$r= CK$}: $\deg t=\deg q_1=3$, \; $\deg p_1=0$;  \; $\deg 
q_2=\deg q_3= \deg q_4=\deg\zeta_1=\deg\zeta_2 =
\deg\zeta_3=2$; \; $\deg p_2=\deg p_3= \deg p_4= 
\deg\eta_1=\deg\eta_2=\deg\eta_3=1$ 

Thus, from the point of view of classification 
of the W-filtered Lie superalgebras, there are five {\it families} of 
exceptional algebras consisting of 15 individual algebras.  

\ssec{1.2.2.  Several first terms that determine the Cartan prolongs}
To facilitate the comparison of various vectorial superalgebras, we
offer the following Table.  The most interesting phenomena occur for
extremal values of parameter $r$ and small values of superdimension
$m|n$.

The central element $z\in\fg_0$ is supposed to be chosen so that it 
acts on $\fg_k$ as $k\cdot\id$ and $\Lambda (r)=\Cee[\xi_1, \dots, 
\xi_r]$ is the Grassmann superalgebra generated by the $\xi_i$ of 
degree 0 for all $i$.  

We set $\Lambda (0)=\Cee$, the $\fvect(0|n)$-module $\Lambda (n)/\Cee 
\cdot 1$ is denoted by $T_0(\vec 0)$; set $\Vol _0(0|m)=\{v\in\Vol 
(0|m)\mid \int v=0\}$ whereas the $\fsvect$-module $T_0^0(\vec 0)$ is 
defined as $\Vol _0(0|m)/\Cee\cdot 1$ (for more elucidations see sec.  0.6).

Recall the range of the regrading parameter $r$: $0\leq r\leq m$,
where $m$ is the number of odd indeterminates, except for the series
$\fk$ and $\fh$ when $0\leq r\leq [\frac{m}{2}]$, and $\fm(n)$,
$\fb_{\lambda}(n)$, etc.  when $0\leq r\leq n$ with value $r=n-1$
excluded.  We exclude certain values of $r$, namely, $r=k-1$ for
$\fk(1|2k)$ and $r=n-1$ for $\fm(n)$ and their subalgebras because for
these values of $r$ the corresponding grading is not a W-grading: the
$\fg_{0}$ module $\fg_{-1}$ is reducible.
$$
\renewcommand{\arraystretch}{1.3}
\begin{tabular}{|c|c|c|c|}
\hline
$\fg$&$\fg_{-2}$&$\fg_{-1}$&$\fg_0$\cr
\hline
\hline
$\fvect(n|m; r)$&$-$&$\id\otimes\Lambda 
(r)$&$\fgl(n|m-r)\otimes\Lambda (r)\subplus\fvect(0|r)$\cr 
\hline
$\fvect(1|m; m)$&$-$&$\Lambda (m)$& $\Lambda 
(m)\subplus\fvect(0|m)$\cr 
\hline
\hline
$\fsvect(n|m; r)$&$-$&$\id\otimes\Lambda 
(r)$&$\fsl(n|m-r)\otimes\Lambda (r)\subplus\fvect(0|r)$\cr 
\hline
$\fsvect\degree(1|m; m)$&$-$&$\Vol_0(0|m)$&$\Lambda 
(m)\subplus\fsvect(0|m)$\cr 
\hline
$\fsvect\degree(1|2)$&$-$&$T_0(\vec 0)$&$\fvect(0|2)\cong\fsl(1|2)$\cr    
\hline
$\fsvect(2|1)$&$-$&$\Pi(T_0(\vec 0))$&$\fvect(0|2)\cong\fsl(2|1)$\cr  
\hline
\end{tabular} 
$$
$$
\renewcommand{\arraystretch}{1.3}
\begin{tabular}{|c|c|c|c|}
\hline
$\fh(2n|m)$&$-$&$\id$&$\fosp(m|2n)$\cr 
\hline
$\fh(2n|m; r)$&$T_0(\vec 0)$&$\id\otimes\Lambda
(r)$&$\fosp(m-2r|2n)\otimes\Lambda (r)\subplus\fvect(0|r)$\cr 
\hline
$\fh(2n|2r; r)$&$-$&$\id\otimes\Lambda(r)$&
$\fsp(2n)\otimes\Lambda (r)\subplus\fvect(0|r)$\cr 
\hline
\hline
$\fh_{\lambda}(2|2)$&$-$&$\Pi(\Vol^{\lambda}(0|2))$&
$\fosp(2|2)\cong\fvect(0|2)$\cr 
\hline
$\fh_{\lambda}(2|2; 1)$&$-$&
$\id_{\fsp(2)}\otimes\Vol^{\lambda}(0|1)$&$\fsp(2)\otimes \Lambda
(1)\subplus\fvect(0|1)$\cr \hline
\hline
$\fk(2n+1|m; r)$&$\Lambda(r)$&$\id\otimes\Lambda
(r)$&$\fc\fosp(m-2r|2n)\otimes\Lambda (r)\subplus\fvect(0|r)$\cr 
\hline
$\fk(1|2m; m)$&$-$&$\Lambda(m)$&$\Lambda (m) \subplus
\fvect(0|m)$\cr 
\hline
$\fk(1|2m+1; m)$&$\Lambda(m)$&$\Pi(\Lambda
(m))$&$\Lambda (m)\subplus\fvect(0|m)$\cr 
\hline
\end{tabular} 
$$
We set $p(\mu)\equiv n\pmod 2$, so $\mu$ can be odd indeterminate. 
The Lie superalgebras $\widetilde{\fsvect}_\mu (0|n)$ are isomorphic
for nonzero $\mu$'s; hence, so are the algebras
$\widetilde{\fs\fb}_{\mu}(2^{n-1}-1|2^{n-1})$.  So for $n$ even we can
set $\mu=1$, while if $\mu$ is odd we should consider it as an 
additional indeterminate on which the coefficients depend.
$$
\renewcommand{\arraystretch}{1.3}
\begin{tabular}{|c|c|c|c|}
\hline
$\fg$&$\fg_{-2}$&$\fg_{-1}$&$\fg_0$\cr
\hline
$\fm(n; r)$&$\Pi(\Lambda(r))$&$\id\otimes\Lambda(r)$&$\fc\fpe(n-r)
\otimes\Lambda(r)\subplus\fvect(0|r)$\cr 
\hline
$\fm(n; n)$&$-$&$\Pi(\Lambda(n))$&$ \Lambda(n)\subplus\fvect(0|n)$\cr  
\hline
\hline
$\widetilde{\fs\fb}_{\mu}(2^{n-1}-1|2^{n-1})$&$-$
&$\frac{\Pi(\Vol(0|n))}{\Cee(1+\mu\xi_1\dots\xi_n)\vvol(\xi)}$&
$\widetilde{\fsvect}_\mu (0|n)$\cr 
\hline
\end{tabular} 
$$
In what follows $\lambda=\frac{2a}{n(a-b)}\neq 0$, 1, $\infty$; the 
three exceptional cases (corresponding to the ``drop-outs'' $\fle(n)$, 
$\fb_{1}\degree(n)$ and $\fb\degree_{\infty}(n)$, respectively) are 
considered separately.  The irreducibility condition of the 
$\fg_0$-module $\fg_{-1}$ for $\fg=\fb\degree_{\infty}$ excludes $r=n-1$.  
The case $r=n-2$ is extra exceptional, so in the following tables
$$
0<r<n-2;\; \; \text{ additionally, unless specified, $a\neq b$ and $(a, b)\neq 
k(n, n-2)$.}
$$

To further clarify the following tables, denote the superspace of the 
identity $k|k$-dimensional representation of $\fspe(k)$ by $V$; let 
$d=\diag (1_{k}, -1_{k})\in \fpe(k)$.  Let $W=V\otimes \Lambda(r)$ and 
$D\in \fvect(0|r)$.  Let $\Xi=\xi_1\cdots\xi_n\in \Lambda(\xi_1, \dots 
,\xi_n)$.   Denote by $T^r$ the representation of $\fvect(0|r)$ in 
$\fspe(n-r)\otimes \Lambda(r)$ given by the formula
$$
T^r(D)=1\otimes D+d\otimes \frac{1}{n-r}\Div D. \eqno{(T^r)}
$$
\underline{In $\fsle\degree(n; r)_{0}$ for $r\neq n-2$}:
$$
\renewcommand{\arraystretch}{1.3}
\begin{array}{l}
\text{$\fvect(0|r)$ acts on the ideal $\fspe(n-r)\otimes \Lambda(r)$ via 
$T^r$};\\  
\text{ any $X\otimes f\in\fspe(n-r)\otimes \Lambda(r)$  acts in $\fg_{-1}$ as 
$\id\otimes f$ and in $\fg_{-2}$ as $0$};\\
\text{any $D\in\fvect(0|r)$ acts in $\fg_{-1}$ via $T^r$ and in 
$\fg_{-2}$ as $D$.}
\end{array}\eqno{(1.2.1.1)}
$$
\underline{In $\fsle\degree(n; n-2)_{0}$}, observe that 
$\fspe(2)\cong\Cee(\Le_{q_1\xi_1-q_2\xi_2})\subplus 
\Cee\Le_{\xi_1\xi_2}$, whereas $\fg_{-2}$ and $\fg_{-1}$ are as 
above, for $r< n-2$.  
Set $\fh=\Cee(\Le_{q_1\xi_1-q_2\xi_2})$.  In this case
$$
\fg_0\cong \underline{\left (\fh\otimes \Lambda(n-2)\subplus 
\Cee\Le_{\xi_1\xi_2}\otimes (\Lambda(n-2)\setminus 
\Cee\xi_3\cdots\xi_n)\right )}\subplus T^1(\fvect(0|n-2)).\eqno{(1.2.1.2)}
$$
The action of $\fvect(0|n-2)$, the quotient of $\fg_{0}$ modulo the 
underlined ideal is via (1.1.2.1).  In the subspace 
$\xi_1\xi_2\otimes\Lambda(n-2)\subset\fg_0$ this action is as in the 
space of volume forms.  So we can throw away $\Xi$.

In the following table the terms ``$\fg_{-1}$'' denote the superspace 
isomorphic to the listed one but with the action given by formulas 
(1.2.1.1) and (1.2.1.2), as indicated.
$$
\renewcommand{\arraystretch}{1.3}
\begin{tabular}{|c|c|c|c|}
\hline
$\fg$&$\fg_{-2}$&$\fg_{-1}$&$\fg_0$\cr
\hline
$\fle(n)$&$-$&$\id$&$\fpe(n)$\cr 
\hline
$\fle(n; r)$&$\Pi(T_0(\vec 0))$&$\id\otimes\Lambda
(r)$&$\fpe(n-r)\otimes\Lambda (r)\subplus\fvect(0|r)$\cr 
\hline
$\fle(n; n)$&$-$&$\Pi(T_{0}(\vec 0))$& $\fvect (0|n)$\cr 
\hline
\hline
$\fsle\degree(n)$&$-$&$\id$&$\fspe(n)$\cr 
\hline
$\fsle\degree(n; r)$&``$\Pi(T_0(\vec 0))$'' & ``$\id\otimes\Lambda
(r)$'' &$\fspe(n-r)\otimes\Lambda (r)\subplus\fvect(0|r)$\cr 
\hline
$\fsle\degree(n; n-2)$&``$\Pi(T_0(\vec 0))$'' & ``$\id\otimes\Lambda
(r)$''&see (1.2.1.2)\cr 
\hline
$\fsle\degree(n; n)$&$-$&$\Pi(T^0_{0}(\vec 0))$&$\fsvect(0|n)$\cr 
\hline
\end{tabular} 
$$

We consider $\fb_{a,b}(n; r)$ for $0<r<n-2$ and $ar-bn\neq 0$; in 
particular, this excludes $\fb\degree_{\infty}(n; n)=\fb\degree_{a, a}(n; 
n)$ and $\fb\degree_{1}(n; n-2)=\fb\degree_{n, n-2}(n; n-2)$.  Set  
$$
c=\frac{a}{ar-bn}.
$$
The case $ar=bn$, i.e., $\lambda=\displaystyle\frac{2}{n-r}$ is an 
exceptional one, the $\fvect(0|r)$-action on the ideal 
$\fc\fspe(n-r)\otimes\Lambda(r)\subplus \fvect(0|r)$ of $\fg_0$ and on 
$\fg_-$ is the same as for $\fsle\degree$, see (1.2.1.1).

If $z$ is the central element of 
$\fc\fspe(n-r)$ that acts on $\fg_{-1}$ as $\id$, then 
$$
z\otimes \psi\text{ acts on $\fg_{-1}$ as $\id\otimes \psi$, and on 
$\fg_{-2}$ as $2\id\otimes\psi$.  }
$$

{\protect
$$
\renewcommand{\arraystretch}{1.3}
\begin{tabular}{|c|c|c|c|}
\hline
$\fg$&$\fg_{-2}$&$\fg_{-1}$&$\fg_0$\cr
\hline
$\fb_{\lambda}(n)$&$\Pi(\Cee)$&$\id$&$\fspe(n)\subplus\Cee(az+bd)$\cr 
\hline
$\fb_{\lambda}(n; r)$&$\Pi((-c)\str)\otimes 
(\Vol(0|r))^{2c}$&$\left((-\frac{c}{2})\str)\otimes\id\right)\otimes
(\Vol(0|r))^{c}$&
$\left(\fpe(n-r)\otimes\Lambda (r)\right)
\subplus\fvect(0|r)$\cr 
\hline
$\fb_{\lambda}(n; n)$&$-$&$\Pi(\Vol^{\lambda}(0|n))$&$\fvect(0|n)$\cr 
\hline
\end{tabular} 
$$}

{\protect
$$
\renewcommand{\arraystretch}{1.3}
\begin{tabular}{|c|c|c|c|}
\hline
$\fg$&$\fg_{-2}$&$\fg_{-1}$&$\fg_0$\cr
\hline
$\fb_{2/(n-r)}(n; r)$&$\Pi(\Cee)\otimes \Lambda (r)$&$\id\otimes 
\Lambda (r)$& $\fc\fpe(n-r)\otimes\Lambda (r) \subplus\fvect(0|r)$\cr 
\hline
\hline
$\fb\degree_{\infty}(n)$&$\Pi(\Cee)$&$\id$&
$\fspe(n)_{a, a}$\cr 
\hline
$\fb\degree_{\infty}(n; r)$&$\Pi(\Cee)\otimes \Lambda (r)$&$\id\otimes\Lambda(r)$&
$\left((\fspe(n-r)_{a, a})\otimes\Lambda (r)\right)
\subplus\fvect(0|r)$\cr 
\hline
$\fb\degree_{\infty}(n; n), n>2$&$-$&$\Pi(\Lambda(n))$&$\left(\Lambda 
(n)\setminus\Cee\Xi\right )\subplus\fsvect(0|n)$\cr  
\hline
\hline
$\fb\degree_{1}(n)$&$\Pi(\Cee)$&$\id$&$(\fspe(n)_{n, n-2})$\cr 
\hline
$\fb\degree_{1}(n; r)$&``$\Pi(\Vol_{0}(0|r))$''& 
``$\id\otimes\Lambda(r)$'' &
$\left((\fspe(n-r)_{n, n-2})\otimes\Lambda (r)\right)
\subplus\fvect(0|r)$\cr 
\hline
$\fb\degree_{1}(n; n-2)$&``$\Pi(T_0(\vec 0))$'' & ``$\id\otimes\Lambda
(r)$'' &(1.1.2.1.2) with $\fc\fspe(2)$ instead of $\fspe(2)$\cr  
\hline
$\fb\degree_{1}(n; n)$&$-$&$\Pi(\Vol_{0}(0|n))$&$\fvect(0|n)$\cr  
\hline
\end{tabular} 
$$}

\ssec{1.2.3.  The exceptional vectorial Lie subsuperalgebras} Here are
the terms $\fg_{i}$ for $i\leq 0$ of 14 of the 15 exceptional
algebras, the last column gives $\dim \fg_{-}$: \footnotesize
$$
\renewcommand{\arraystretch}{1.3}
\begin{tabular}{|c|c|c|c|c|}
\hline
$\fg$&$\fg_{-2}$&$\fg_{-1}$&$\fg_0$&$\dim\fg_{-}$\cr
\hline
\hline
$\fv\fle(4|3)$&$-$&$\Pi(\Lambda(3)/\Cee 1)$&$\fc(\fvect(0|3))$&$4|3$\cr 
\hline
$\fv\fle(4|3; 1)$&$\Cee\cdot 1$&$\id\otimes\Lambda (2)$& 
$\fc(\fsl(2)\otimes\Lambda(2)\subplus 
T^{1/2}(\fvect(0|2))$&$5|4$\cr 
\hline
$\fv\fle(4|3; K)$&$\id_{\fsl(3)}$&$\id_{\fsl(3)}^*\otimes \id_{\fsl(2)}\otimes 
1$&$\fsl(3)\oplus\fsl(2)\oplus\Cee z$&$3|6$\cr 
\hline
\hline
$\fv\fas(4|4)$&$-$&$\spin$&$\fas$&$4|4$\cr 
\hline
\hline
$\fk\fas$&$\Cee\cdot 1$&$\Pi(\id)$& $\fc\fo(6)$&$1|6$\cr 
\hline
$\fk\fas(; 1)$ 
&$\Lambda(1)$&$\id_{\fsl(2)}\otimes(\id_{\fgl(2)}\otimes\Lambda(1))$& 
$(\fsl(2)\oplus\fgl(2)\otimes\Lambda(1))\subplus\fvect(0|1)$&$5|5$\cr 
\hline
$\fk\fas(; 3\xi)$&$-$& $\Lambda(3)$&$\Lambda(3)\oplus\fsl(1|3)$&$4|4$\cr 
\hline
$\fk\fas(; 3\eta)$&$-$&$\Vol_{0}(0|3)$& 
$\fc(\fvect(0|3))$&$4|3$\cr 
\hline
\hline
$\fm\fb(4|5)$&$\Pi(\Cee\cdot 1)$&$\Vol (0|3)$&$\fc(\fvect(0|3))$&$4|5$\cr  
\hline
$\fm\fb(4|5; 1)$&$\Lambda(2)/\Cee \cdot 1$ 
&$\id_{\fsl(2)}\otimes\Lambda(2)$ 
&$\fc(\fsl(2)\otimes\Lambda(2)\subplus T^{1/2}(\fvect(0|2))$&$5|6$\cr 
\hline
$\fm\fb(4|5; K)$&$\id_{\fsl(3)}$&$\Pi(\id_{\fsl(3)}^*\otimes 
\id_{\fsl(2)}\otimes 
\Cee)$&$\fsl(3)\oplus\fsl(2)\oplus\Cee z$&$3|8$\cr  
\hline 
\hline
$\fk\fsle(9|6)$&$\Cee\cdot 1$&$\Pi(T^0_{0}(\vec 0))$&$\fsvect(0|4)_{3, 
4}$&$9|6$\cr    
\hline
$\fk\fsle(9|6; 2)$&$\id_{\fsl(3|1)}$&$\id_{\fsl(2)}\otimes\Lambda(3)$&
$\left (\fsl(2)\otimes\Lambda(3)\right) \subplus \fsl(1|3)$&$11|9$\cr    
\hline
$\fk\fsle(9|6; K)$&$\id$&$\Pi(\Lambda^2(\id^*))$&$\fsl(5)$&$5|10$\cr    
\hline
\end{tabular} 
$$
\normalsize

\noindent Observe that none of the simple W-graded vectorial Lie
superalgebras is of depth $>3$ and only two algebras are of depth 3:
one of the above, $\fm\fb(4|5; K)$, for which we have $\fm\fb(4|5;
K)_{-3}\cong \Pi(\id_{\fsl(2)})$, and another one, $\fk\fsle(9|6;
CK)=\fc\fk(9|11)$.  

This $\fc\fk(9|11)$ is the 15-th exceptional simple vectorial Lie
superalgebra; its non-positive terms are as follows (we assume that 
the $\fsl(2)$- and $\fsl(3)$-modules are purely even):
$$
\renewcommand{\arraystretch}{1.4}
\begin{array}{l}
\fc\fk(9|11)_{0}\simeq \left 
(\fsl(2)\oplus\fsl(3)\otimes\Lambda(1)\right) \subplus \fvect(0|1);\\
\fc\fk(9|11)_{-1}\simeq \id_{\fsl(2)}\otimes\left 
(\id_{\fsl(3)}\otimes\Lambda(1)\right);\\
\fc\fk(9|11)_{-2}\simeq \id_{\fsl(3)}^*\otimes\Lambda(1);\\
\fc\fk(9|11)_{-3}\simeq \Pi(\id_{\fsl(2)}\otimes\Cee). \end{array}
$$

\ssec{1.2.3.1.  The exceptional Lie subsuperalgebra $\fk\fas$ of $\fk 
(1|6)$} This Lie superalgebra is not determined by its nonpositive 
part and requires a closer study.  The Lie superalgebra 
$\fg=\fk(1|2n)$ is generated by the functions from $\Cee[t, \xi_1, 
\dots, \xi_n, \eta_1, \dots, \eta_n]$.  The standard $\Zee$-grading of 
$\fg$ is induced by the $\Zee$-grading of $\Cee[t, \xi, \eta]$ given 
by $\deg t=2$, $\deg \xi_i=\deg\eta_i=1$; namely, $\deg K_f=\deg f-2$.  
Clearly, in this grading $\fg$ is of depth 2.  Let us consider the 
functions that generate several first homogeneous components of 
$\fg=\mathop{\oplus}\limits_{i\geq -2}\fg_i$:
$$
\renewcommand{\arraystretch}{1.3}
\begin{array}{|c|c|c|c|c|}
\hline
\text{component}&\fg_{-2}&\fg_{-1}&\fg_{0}&\fg_{1}\\
\hline
\text{its generators}&1&\Lambda^1(\xi, \eta)&\Lambda^2(\xi,
\eta)\oplus\Cee\cdot t&\Lambda^3(\xi,
\eta)\oplus t\Lambda^1(\xi, \eta)\\
\hline
\end{array}
$$
As one can prove directly, the component $\fg_1$ generates the whole 
subalgebra $\fg_+$ of elements of positive degree.  The component 
$\fg_1$ splits into two $\fg_0$-modules $\fg_{11}= \Lambda^3$ and 
$\fg_{12}=t\Lambda^1$.  It is obvious that $\fg_{12}$ is always 
irreducible and the component $\fg_{11}$ is trivial for $n=1$.

The partial Cartan prolongs of $\fg_{11}$ and 
$\fg_{12}$ are well-known:
$$
\renewcommand{\arraystretch}{1.4}
\begin{matrix}
(\fg_-\oplus\fg_0, \fg_{11})_*^{mk}\cong\fpo(0|2n)\oplus\Cee\cdot
K_t\cong\fd(\fpo(0|2n));\\
(\fg_-\oplus\fg_0,
\fg_{12})_*^{mk}=\fg_{-2}\oplus\fg_{-1}\oplus\fg_{0}
\oplus\fg_{12}\oplus\Cee\cdot
K_{t^{2}}\cong\fosp(2n|2).\end{matrix}
$$
Observe a remarkable property of $\fk(1|6)$: only for it does
the component $\fg_{11}$ split into 2 
irreducible modules that we will denote $\fg_{11}^\xi$ and 
$\fg_{11}^\eta$: one is generated by $\xi_{1}\xi_{2}\xi_{3}$, the other 
one by $\eta_{1}\eta_{2}\eta_{3}$.

Observe further, that $\fg_0=\fc\fo(6)\cong\fgl(4)$.  
As $\fgl(4)$-modules, $\fg_{11}^\xi$ and $\fg_{11}^\eta$ are the 
symmetric squares $S^2(\id)$ and $S^2(\id^*)$ of the identity 
4-dimensional representation and its dual, respectively.

\begin{Theorem} {\em (\cite{Sh5}, \cite{Sh14})} The Cartan prolong
$\fk\fas^\xi=(\fg_-\oplus\fg_0, \fg_{11}^\xi\oplus\fg_{12})_*^{mk}$ is
infinite dimensional and simple.  It is isomorphic to
$\fk\fas^\eta=(\fg_-\oplus\fg_0, \fg_{11}^\eta\oplus\fg_{12})_*^{mk}$.
\end{Theorem}

When it does not matter which of isomorphic algebras
$\fk\fas^\xi\simeq \fk\fas^\eta$ to take we will write $\fk\fas$.

For their explicit presentation by generating functions and for
related structures on the ``stringy'' Lie superalgebras see
\cite{GLS1}.  

\ssec{1.2.5.  Grozman's theorem and a description of $\fg$ as
$\fg_{\ev}$ and $\fg_{\od}$} In \cite{CK2} the exceptional algebras
are described as $\fg=\fg_{\ev}\oplus\fg_{\od}$.  For most of the
series such description is of little value because each homogeneous
component $\fg_{\ev}$ and $\fg_{\od}$ has a complicated structure. 
For the exceptions (and for twisted polyvector fields) the situation
is totally different!  

Observe that apart from being beautiful, such description is
useful for the construction of Volichenko algebras, cf.  \cite{LS}.

Recall in this relation a theorem of Grozman \cite{Gr}.  He completely
described bilinear differential operators acting in tensor fields and
invariant under all changes of coordinates.  It turned out that almost
all of the first order operators determine a Lie superalgebra on its
domain.  Some of these superalgebras are simple or close to simple. 
In the constructions below we use some of these invariant operators.

\noindent\underline{$\fg=\fe(5|10)$}:\; \; $\fg_\ev=\fsvect(5|0)\simeq 
d\Omega^{3}$, \; \; $\fg_\od=\Pi(d\Omega^{1})$ with the natural 
$\fg_\ev$-action on $\fg_\od$ and the bracketing of odd elements being 
their product; we identify for any permutation $(ijklm)$ 
of $(12345)$:
$$
dx_{i}\wedge dx_{j}\wedge dx_{k}\wedge dx_{l}\otimes
\vvol^{-1}=\sign(ijklm)\pder{x_{m}}.
$$

\noindent\underline{$\fg=\fv\fas(4|4)$}:\; \; $\fg_\ev=\fvect(4|0)$, 
and $\fg_\od=\Omega^{1}\otimes \Vol^{-1/2}$ with the natural 
$\fg_\ev$-action on $\fg_\od$ and the bracketing of odd elements being 
$$
[\omega_{1}\otimes \vvol^{-1/2}, \omega_{2}\otimes \vvol^{-1/2}]=
(d\omega_{1}\wedge\omega_{2}+ \omega_{1}\wedge d\omega_{2})\otimes
\vvol^{-1}],
$$
where we identify
$$
dx_{i}\wedge dx_{j}\wedge dx_{k}\otimes
\vvol^{-1}=\sign(ijkl)\pder{x_{l}}\text{ for any permutation $(ijkl)$ 
of $(1234)$}.
$$

\noindent\underline{$\fg=\fv\fle(3|6)$}:\; \;
$\fg_\ev=\fvect(3|0)\oplus \fsl(2)^{(1)}_{\geq 0}$, where
$\fg^{(1)}_{\geq 0}=\fg\otimes\Cee[t]$ , and 
$\fg_\od=\left(\Omega^{1}\otimes \Vol^{-1/2}\right )\otimes
\id_{\fsl(2)^{(1)}_{\geq 0}}$ with the natural 
$\fg_\ev$-action on $\fg_\od$.

Recall that $\id_{\fsl(2)}$ is the irreducible $\fsl(2)$-module $L^1$
with highset weight 1; its tensor square splits into $L^2\simeq
\fsl(2)$ and the trivial module $L^0$; accordingly, denote by
$v_{1}\wedge v_{2}$ and $v_{1}\bullet v_{2}$ the projections of
$v_{1}\otimes v_{2}\in L^1\otimes L^1$ onto the skew-symmetric and
symmetric components, respectively.  For $f_{1}, f_{2}\in\Omega ^0$,
$\omega_{1}, \omega_{2}\in\Omega ^1$ and $v_{1}, v_{2}\in L^1$ we set
$$
\renewcommand{\arraystretch}{1.4}
\begin{array}{l} 
{}[(\omega_{1}\otimes v_{1})\vvol^{-1/2}, (\omega_{2}\otimes
v_{2})\vvol^{-1/2}]=\\
\left(\omega_{1}\wedge \omega_{2})\otimes
(v_{1}\wedge v_{2})+d\omega_{1}\wedge \omega_{2}+\omega_{1}\wedge
d\omega_{2})\otimes (v_{1}\bullet v_{2})\right)\vvol^{-1},
\end{array}
$$
where we identify $\Omega ^0$ with $\Omega ^3\otimes_{\Omega ^0}
\Vol^{-1}$ and $\Omega ^2\otimes_{\Omega ^0}
\Vol^{-1}$ with $\fvect(3|0)$ by setting
$$
dx_{i}\wedge dx_{j}\otimes
\vvol^{-1}=\sign(ijk)\pder{x_{k}}\text{ for any permutation $(ijk)$ 
of $(123)$}.
$$

\noindent\underline{$\fg=\fm\fb(3|8)$}:\; \;
$\fg_\ev=\fvect(3|0)\oplus \fsl(2)^{(1)}_{\geq 0}$, and
$\fg_\od=\fg_{-1}\oplus \fg_1$, where 
$$
\fg_{-1}=
\left(\Pi\Vol^{-1/2}\right)\otimes \id_{\fsl(2)^{(1)}_{\geq 0}}\; \text{ and
}\; \fg_1= \left(\Omega^{1}\otimes \Vol^{-1/2}\right)\otimes
\id_{\fsl(2)^{(1)}_{\geq 0}};
$$
clearly, one can interchange $\fg_{\pm 1}$.

Multiplication is similar to that of $\fg=\fv\fle(3|6)$.  For $f_{1},
f_{2}\in\Omega ^0$, $\omega_{1}, \omega_{2}\in\Omega ^1$ and $v_{1},
v_{2}\in L^1$ we set
$$
\renewcommand{\arraystretch}{1.4}
\begin{array}{l} 
{}[(\omega_{1}\otimes v_{1})\vvol^{-1/2}, (\omega_{2}\otimes
v_{2})\vvol^{-1/2}]= 0, \\
{}[(f_{1}\otimes v_{1})\vvol^{-1/2}, (f_{2}\otimes v_{2})\vvol^{-1/2}]=
(df_{1}\wedge df_{2})\otimes (v_{1}\wedge v_{2})\vvol^{-1}, \\
{}[(f_{1}\otimes v_{1})\vvol^{-1/2}, (\omega_{1}\otimes 
v_{2})\vvol^{-1/2}]=\\
\left(f_{1}\omega_{1}\otimes (v_{1}\wedge v_{2})+(df_{1}\omega_{1}+ 
f_{1}d\omega_{1})\otimes (v_{1}\bullet v_{2})\right)\vvol^{-1}.
\end{array}
$$

\noindent\underline{$\fg=\fk\fas$}:\; \; $\fg_\ev=\fvect(1|0)\oplus
\fsl(4)^{(1)}_{\geq 0}$, and $\fg_\od=\fg_{-1}\oplus \fg_1$, where
$\fg_{-1}= \Pi\left(\Lambda ^2(\id_{\fsl(2)^{(1)}_{\geq 0}})\right)$ and
$\fg_1= \Pi\left(S^{2}(\id_{\fsl(2)^{(1)}_{\geq 0}})\right)$; clearly, one
can interchange $\fg_{\pm 1}$.

\noindent\underline{$\fg=\fb_{\lambda}(n; \bar n)$}:\; \; here $\bar
n$ denotes the grading given by the formulas $\deg q_{i}=0$, $\deg
\xi_{i}=1$ for $i=1, \dots , n$.  Then for $i=-1, 0, \dots , n-1$ we
have $\fg_i=\left(\Pi^i(\Lambda ^{i-1}(\fvect(n|0))\right)\otimes
\Vol^{-(i-1)\lambda}$.

Consider $n=2$ more attentively.  Clearly, one can interchange
$\fg_{\pm 1}$; this possibility explains a mysterious isomorphism
mentioned in (OI); if $\lambda=-1-\lambda$ we have additional
automorphisms, whereas for $\lambda=\frac12$ (and $\lambda=-\frac32$)
there is a nontrivial central extension missed in \cite{Ko1},
\cite{Ko2} and studied in \cite{LSh3}.

\section*{\S 2. Main result} 

Recall that we usually skip the wedge sign in the product of
supercommuting diferential forms; $\Pi$ is the change of parity sign. 
The relations will be devided as in introduction: into Serre relations (S); lowest (for
the positive part) or highest (for the negative part) weight
relations, (LW) and (HW), respectively; other, new type relations (N). 
Observe that (LW) and (HW) are of the same form as (S).

\underline{{\bf $\fk\fsle(5|10)$}} Set (positive generators):
$$
{X_i} = x_{i}\pder{x_{i+1}} \; \; \text{ for } i=1, 2, 3, 4,\; \text{ and }\; 
Z = \Pi x_{5}\, dx_{4} dx_{5}.
$$
{\bf Relations in $\fk\fsle_{+}$} are:  \footnotesize
$$
\renewcommand{\arraystretch}{1.4}
\begin{array}{l|l}
S&\left[{X_1},\,{X_3}\right] = 0,\quad \left[{X_1},\,{X_4}\right] = 0,\quad 
\ad_{X_4}^2{X_3}= 0, \quad 
\left[{X_2},\,{X_4}\right] = 0,\\
S&\ad_{X_1}^2{X_2}= 0,\quad 
 \ad_{X_2}^2{X_1}= 0,\\
S&\ad_{X_2}^2{X_3}= 0,\quad 
  \ad_{X_3}^2{X_2}= 0,\quad 
  \ad_{X_3}^2{X_4} = 0,\\
LW&\ad_{X_3}^2Z= 0,\quad 
 \left[{X_1},\,Z\right] = 0, \quad \left[{X_2},\,Z\right] = 0,\quad 
 \ad_{X_4}^2Z= 0,\\
N_{2}&\left[Z,\,Z\right] = 0,\quad 
  \left[\left[{X_3},\,Z\right],\,\left[{X_4},\,Z\right]
    \right] = 0,\quad 
\left[\left[{X_4},\,\left[{X_3},\,Z\right]
     \right],\,\left[\left[{X_3},\,{X_4}\right], \; Z\right]
    \right] = 0,\\
N_{3}&\left[\left[\left[{X_1},\,{X_2}\right],\,
     \left[{X_3},\,Z\right]\right],\,
    \left[\left[Z,\,\left[{X_2},\,{X_3}\right]\right],\,
     \left[Z,\,\left[{X_3},\,{X_4}\right]\right]\right]
     \right] = 0.
\end{array}
$$\normalsize

Set (negative generators):
$$
{Y_i} = x_{i+1}\pder{x_{i}} \; \text{ for } i=1, 2, 3, 4,\text{ and }
Z = \Pi\, dx_{1} dx_{2}.
$$
{\bf The relations in $\fk\fsle_{-}$} are: \footnotesize
$$
\renewcommand{\arraystretch}{1.4}
\begin{array}{l|l}
S&\left[{Y_1},\,{Y_3}\right] = 0,\quad 
  \left[{Y_1},\,{Y_4}\right] = 0,\quad 
  \left[{Y_2},\,{Y_4}\right] = 0,\\
S&  \ad_{Y_1}^2{Y_2}= 0,\quad 
  \ad_{Y_2}^2{Y_1}= 0,\quad 
  \ad_{Y_2}^2{Y_3}= 0,\\
S&\ad_{Y_3}^2{Y_2}= 0,\quad 
\ad_{Y_3}^2{Y_4} = 0,\quad 
\ad_{Y_4}^2{Y_3} = 0,\\
HW&\left[{Y_1},\,Z\right] = 0,\quad \left[{Y_3},\,Z\right] = 0,\quad 
  \left[{Y_4},\,Z\right] = 0,\quad 
 \ad_{Y_2}^2Z= 0,\quad \\
 N_{2}& \left[Z,\,Z\right] = 0;\quad 
  \left[\left[\left[{Y_2},\,Z\right],\,
     \left[{Y_3},\,{Y_4}\right]\right],\,
    \left[\left[Z,\,\left[{Y_1},\,{Y_2}\right]\right],\,
     \left[Z,\,\left[{Y_2},\,{Y_3}\right]\right]\right]
     \right] = 0.
\end{array}
$$ \normalsize
 
\underline{{\bf $\fm\fb(4|5)$}} Set (negative generators):
$$
 {Y_1} = \xi_0,\quad {Y_2} = q_2 \xi_1,\quad {Y_3} =q_3 \xi_2 ,\quad Z
 = -q_0 q_1 + \xi_2 \xi_3.
$$
{\bf The relations in $\fm\fb(4|5)_{-}$} are: \footnotesize
$$
\renewcommand{\arraystretch}{1.4}
\begin{array}{l|l}
S&\left[{Y_1},\,{Y_2}\right] = 0,\quad
  \left[{Y_1},\,{Y_3}\right] = 0,\quad
 \ad_{Y_3}^2{Y_2}= 0, 
 \quad
  \ad_{Y_2}^2Y_3= 0,\\
HW&\ad_{Y_1}^2Z= 0,\quad
  \ad_{Y_2}^2Z = 0,\quad
 \left[{Y_3},\,Z\right] = 0,\\
N_{2}&\left[Z,\,Z\right] = 0, \quad
  \left[\left[\left[{Y_1},\,Z\right],\,
     \left[{Y_2},\,Z\right]\right],\,
    \left[\left[{Y_1},\,Z\right],\,
     \left[Z,\,\left[{Y_2},\,{Y_3}\right]\right]\right]
     \right] = 0.
	 \end{array}
$$ \normalsize

Set (positive generators):
$$
\renewcommand{\arraystretch}{1.4}
\begin{array}{l}
	 {X_1} = {{q_0}^2} \xi_0 - q_0 q_1 \xi_1 - q_0 q_2 \xi_2 - q_0 q_3
	 \xi_3 + \tau q_0 + 2\xi_1 \xi_2 \xi_3 , \\
  {X_2} = q_1 \xi_2, {X_3} = q_2 \xi_3,\quad {Z_1} = {{q_3}^2},\quad
  {Z_2} = \xi_0 \xi_1.
\end{array}
$$ 
{\bf The relations in $\fm\fb(4|5)_{+}$} are: \footnotesize
$$
\renewcommand{\arraystretch}{1.4}
\begin{array}{l|l}
  S&\left[{X_1},\,{X_2}\right] = 0,\quad \left[{X_1},\,{X_3}\right] =
  0,\quad\ad_{X_2}^2{X_3}= 0,\quad
  \ad_{X_3}^2{X_2} = 0,\\
LW&  \quad
  \left[{X_2},\,{Z_1}\right] = 0,\quad
  \left[{X_3},\,{Z_2}\right] = 0,\quad
  \ad_{X_1}^2{Z_1}= 0,\\
LW&  \ad_{X_3}^3{Z_1}= 0,\quad\ad_{X_1}^{2}{Z_2}= 0,\quad
    \ad_{X_2}^2{Z_2}= 0,\\
N_{2}&\left[{Z_1},\,{Z_1}\right] = 0,\quad
  \left[{Z_1},\,{Z_2}\right] = 0,\quad
  \left[{Z_2},\,{Z_2}\right] = 0,\\
N_{2}& \left[{Z_1},\,\left[{X_3},\,\left[{Z_1}, \,{X_3}\right]
     \right]\right] = 0,\\
N_{2}&\left[\left[{X_1},\,{Z_1}\right],\,
	\left[{X_3},\,{Z_1}\right]\right] = 4\,
	\left[{Z_2},\,\left[{X_1},\,{Z_1}\right] \right] ,\quad
	\left[\left[{X_1},\,{Z_2}\right],\,
	\left[{X_2},\,{Z_2}\right]\right] = 0;\\
N_{3}&2\left[\left[{Z_2},\,\left[{X_1},\,{Z_1}\right]\right],\,
    \left[\left[{X_2},\,{X_3}\right],\,
     \left[{X_3},\,{Z_1}\right]\right]\right] = \\
& -   \left[\left[{Z_1},\,\left[{X_2},\,{X_3}\right]\right]
	   ,\,\left[\left[{X_1},\,{Z_2}\right],\,
	   \left[{X_3},\,{Z_1}\right]\right]\right] ;\\
N_{4}&\left[\left[\left[{X_3},\,{Z_1}\right],\,
     \left[{X_2},\,\left[{X_1},\,{Z_2}\right]\right]
     \right],\,\left[\left[{X_2},\,
      \left[{X_1},\,{Z_2}\right]\right],\,
     \left[{X_3},\,\left[{X_3},\,{Z_1}\right]\right]
     \right]\right] = \\
&-4\, \left[\left[\left[{X_1},\,{Z_2}\right],\,
\left[{X_2},\,{X_3}\right]\right],\,
\left[\left[{Z_2},\,\left[{X_1},\,{Z_1}\right]
\right],\,\left[{Z_2},\,
\left[{X_2},\,{X_3}\right]\right]\right]\right] \\
&- \left[\left[\left[{X_2},\,{Z_2}\right],\,
        \left[{X_3},\,\left[{X_1},\,{Z_1}\right]\right]
        \right],\,\left[
        \left[{X_2},\,\left[{X_1},\,{Z_2}\right]\right],\,
        \left[{X_3},\,\left[{X_3},\,{Z_1}\right]\right]
        \right]\right] . \end{array}
$$ \normalsize
 
\underline{{\bf $\fv\fle(4|3; K)$}} Set (negative generators):
$$
\renewcommand{\arraystretch}{1.4}
\begin{array}{l}
{Y_1} = -\, {{x_{4}}^2}\,\pder{x_{4}} -
x_{4} x_{5}\,\pder{x_{5}} - x_{4} x_{6}\,\pder{x_{6}} -
x_{4} x_{7}\,\pder{x_{7}} + x_{5} x_{6}\,\pder{x_{3}} -
x_{5} x_{7}\,\pder{x_{2}} + x_{6} x_{7}\,\pder{x_{1}},\\
{Y_2} =  -\,x_{2}\,\pder{x_{1}} + x_{5}\,\pder{x_{6}} 
,\quad {Y_3} = -\,x_{3}\,\pder{x_{2}} + x_{6}\,\pder{x_{7}}
,\quad Z = \pder{x_{5}}.\end{array}
$$ 

{\bf The relations in $\fv\fle(4|3; K)_{-}$} are: \footnotesize
$$
\renewcommand{\arraystretch}{1.4}
\begin{array}{l|l}
S&\left[{Y_1},\,{Y_2}\right] = 0,\quad \left[{Y_1},\,{Y_3}\right]
	= 0,\quad \ad_{Y_2}^2{Y_3}=
	0,\quad \ad_{Y_3}^2{Y_2}= 0,\\
HW&\ad_{Y_1}^2Z= 0,\quad
  \ad_{Y_2}^2Z= 0,\quad
  \left[{Y_3},\,Z\right] = 0,\\
N_{2}&\left[Z,\,Z\right] = 0,\quad
  \left[\left[Z,\,\left[{Y_2},\,{Y_3}\right]
     \right],\,\left[\left[{Y_1},\,Z\right]
     ,\,\left[{Y_2},\,Z\right]\right]\right]
    = 0.	\end{array}
$$ \normalsize
 
Set (positive generators):
$$
\renewcommand{\arraystretch}{1.4}
\begin{array}{l} 
  {X_1} = \pder{x_{4}},\quad {X_2} = -\,x_{1}\,\pder{x_{2}} +
  x_{6}\,\pder{x_{5}} ,\quad {X_3} = -\,x_{2}\,\pder{x_{3}} +
  x_{7}\,\pder{x_{6}},\\
	 {Z_1} = x_{3} x_{4}\, \pder{x_{7}} - x_{3} x_{5}\,\pder{x_{2}} +
	 x_{3} x_{6}\,\pder{x_{1}} + x_{5} x_{6}\,\pder{x_{7}},\\
{Z_2} = x_{2} x_{4}\,\pder{x_{7}} -
x_{2} x_{5}\,\pder{x_{2}} + x_{2} x_{6}\,\pder{x_{1}} +
x_{4} x_{5}\,\pder{x_{4}} + x_{5} x_{6}\,\pder{x_{6}} -
x_{3} x_{4}\,\pder{x_{6}} \\
	 - x_{3} x_{5}\,\pder{x_{3}} + x_{3} x_{7}\,\pder{x_{1}} + x_{4}
	 x_{5}\,\pder{x_{4}} + x_{5} x_{7}\,\pder{x_{7}}.
\end{array}
$$

{\bf The relations in $\fv\fle(4|3; K)_{+}$} are: \footnotesize
$$
\renewcommand{\arraystretch}{1.4}
\begin{array}{l|l} 
S&\left[{X_1},\,{X_2}\right] = 0,\quad \left[{X_1},\,{X_3}\right] =
  0,\quad \ad_{X_2}^2{X_3}= 0,\quad
\ad_{X_3}^2{X_2}= 0,\\
LW&  \left[{X_2},\,{Z_1}\right] = 0,\quad
  \left[{X_3},\,{Z_2}\right] = 0, \quad\ad_{X_1}^2{Z_1}= 0,\quad
\ad_{X_1}^2{Z_2}= 0,\\
LW&\ad_{X_3}^3{Z_1}= 0,\quad
  \ad_{X_2}^2{Z_2}= 0,\end{array}
$$ 
$$
\renewcommand{\arraystretch}{1.4}
\begin{array}{l|l} 
	N_{2}&\left[{Z_1},\,{Z_1}\right] = 0,\quad
  \left[{Z_1},\,{Z_2}\right] = 0,\quad
  \left[{Z_2},\,{Z_2}\right] = 0,\\
N_{2}&\left[{Z_1},\,\left[{X_3},\, \left[{X_3},\,{Z_1}\right]\right]\right]
= 0,\\
N_{2}&\left[\left[{X_1},\,{Z_1}\right],\,\left[{X_3},\,{Z_1}\right]\right]
	= \left[{Z_2},\,\left[{X_1},\,{Z_1}\right]\right], \\
N_{2}&\left[\left[{X_3},\,{Z_1}\right] , \left[{X_2},\,
\left[{X_1},\,{Z_2}\right]\right]\right ] = 2\,
\left[\left[{X_1},\,{Z_2}\right],\, \left[{X_2},\,{Z_2}\right]\right]
+ 3\, \left[\left[{X_1},\,{Z_2}\right] ,\,\left[{Z_1},\,
\left[{X_2},\,{X_3}\right]\right] \right] - \\
&2\, \left[\left[{X_2},\,{X_3}\right],\, \left[{Z_2},\,
\left[{X_1},\,{Z_1}\right]\right] \right] - 2\,
\left[\left[{X_2},\,{Z_2}\right],\, \left[{X_3},\,
\left[{X_1},\,{Z_1}\right]\right] \right] ; \\
N_{3}&\left[\left[{Z_1},\, \left[{X_2},\,{X_3}\right]\right],\,
\left[{Z_2},\,\left[{X_1},\,{Z_1}\right] \right]\right] = -3\, \left[
\left[{X_2},\,{Z_2}\right],\, \left[{Z_2},\,
\left[{X_1},\,{Z_1}\right]\right] \right] ;\\
N_{3}&\left[\left[{Z_2},\,
\left[{X_2},\,{X_3}\right]\right],\,
\left[\left[{X_1},\,{Z_2}\right],\,
\left[{X_2},\,{Z_2}\right]\right]\right ] = 0,\\
N_{4}&\left[\left[{Z_2},\, \left[{X_1},\,{Z_1}\right]\right],\,
\left[\left[{X_1},\,{Z_1}\right],\,
\left[{X_2}, \,{Z_2}\right]\right]\right ] = 0;\\
N_{4}&\left[\left[
\left[{X_3},\,{Z_1}\right],\, \left[{Z_2},\,
\left[{X_2},\,{X_3}\right]\right] \right],\,\left[ \left[{X_2},\,
\left[{X_1},\,{Z_2}\right]\right],\, \left[{X_3},\,
\left[{X_1},\,{Z_1}\right]\right] \right]\right] = \\
&6\, \left[
\left[\left[{X_1},\,{Z_2}\right],\, \left[{Z_1},\,
\left[{X_2},\,{X_3}\right]\right] \right],\,
\left[\left[{X_1},\,{Z_2}\right],\, \left[{Z_2},\,
\left[{X_2},\,{X_3}\right]\right] \right]\right] + \\
&4\, \left[\left[
\left[{X_1},\,{Z_2}\right],\, \left[{Z_2},\,
\left[{X_2},\,{X_3}\right]\right] \right],\,
\left[\left[{X_2},\,{X_3}\right],\, \left[{Z_2},\,
\left[{X_1},\,{Z_1}\right]\right] \right]\right] + \\
&2\, \left[\left[
\left[{X_1},\,{Z_2}\right],\, \left[{Z_2},\,
\left[{X_2},\,{X_3}\right]\right] \right],\,
\left[\left[{X_2},\,{Z_2}\right],\, \left[{X_3},\,
\left[{X_1},\,{Z_1}\right]\right] \right]\right] ;\\
N_{7}&6\left[\left[\left[{Z_2},\, \left[{X_1},\,{Z_1}\right]\right],\,
\left[\left[{X_1},\,{Z_2}\right],\, \left[{X_2},\,{Z_2}\right]\right]
\right],\,\left[ \left[{Z_2},\, \left[{X_2},\,{X_3}\right]\right],\,
\left[\left[{X_1},\,{Z_1}\right],\, \left[{X_2},\,{Z_2}\right]\right]
\right]\right] = \\
&\left[\left[\left[{X_1},\,{Z_2}\right]
,\,\left[{X_2},\,{Z_2}\right]\right], \,\left[\left[
\left[{X_1},\,{Z_2}\right],\, \left[{X_2},\,{Z_2}\right]\right],\,
\left[\left[{X_2},\,{Z_2}\right],\, \left[{Z_2},\,
\left[{X_1},\,{Z_1}\right]\right] \right]\right]\right] ;\\
N_{8}&\left[\left[\left[{Z_2},\, \left[{X_1},\,{Z_1}\right]\right],\,
\left[\left[{X_1},\,{Z_2}\right],\, \left[{X_2},\,{Z_2}\right]\right]
\right],\,\left[ \left[\left[{X_1},\,{Z_1}\right],\,
\left[{X_2},\,{Z_2}\right]\right],\,
\left[\left[{X_1},\,{Z_2}\right],\, \left[{X_2},\,{Z_2}\right]\right]
\right]\right] = 0.\end{array}
$$ \normalsize

\underline{$\fv\fa\fs(4|4)$} A preliminary step to 
description of presentation of $\fv\fa\fs(4|4)$ is the description of

{\bf Generators and relations for $\fa\fs$} We denote the positive
generators by $x$ and the negative by $y$; each even (odd) element has
even (odd) index.  Set
$$
\renewcommand{\arraystretch}{1.4}
\begin{array}{llll}
x_{1} = \eta _{3},& x_{2} = \zeta _{2}\eta _{1}& x_{4}  =\zeta _{3}\eta
_{2},& x_{3}  = \zeta _{1} \eta _{1} \eta _{3}- \zeta _{2}\eta _{2}
\eta _{3}\\
{y_1} = \zeta _{3},&
{y_2} = \zeta _{2}\eta _{3}&
{y_{4} } =  \zeta _{1}\eta _{2},\eta _{3},&
{y_{3} } = -\zeta _{1}\zeta _{3}\eta _{1}-\zeta _{2}\zeta _{3}\eta _{2};
\end{array}
$$ 
and let Cartan subalgebra be spanned by
$$
{h_0} = 1, \quad {h_i} = \zeta _{i} \eta _{i} .
$$

The relations for positive elements are: \footnotesize
$$
\renewcommand{\arraystretch}{1.4}
\begin{array}{l|l}
S&\ad_{x_{2}}^2x_{4} = 0,\quad
\ad_{x_{4}}^2x_{2}= 0, \quad
\ad_{x_{4}}^2x_{1} = 0, \\
LW&\left[x_{1},\,x_{2}\right] =0,\quad\ad_{x_{4}}^2x_{3} = 0,\quad 
\ad_{x_{2}}^2x_{3}= 0,\quad \\
N_{2}&\left[x_{1},\,x_{1}\right] = 0,\quad  \left[x_{1},\,x_{3}  \right] = 0,\quad
\left[x_{3}  ,\,x_{3}  \right] = 0,\\
N_{3}&2\left[x_{3}  ,\,
\left[x_{2},\,x_{4} \right] \right] = 
\left[x_{4} ,\, \left[x_{2},\,x_{3}  \right] \right] .
\end{array}
$$\normalsize

The relations for negative elements are \footnotesize
$$
\renewcommand{\arraystretch}{1.4}
\begin{array}{ll}
S&\left[{y_1},\,{y_1}\right] = 0,\quad
  \left[{y_1},\,{y_{3} }\right] = 0,\quad
  \left[{y_2},\,{y_{4} }\right] = 0,\\
S&\left[{y_{4} },\,{y_{4} }\right] = 0,\quad
  \left[{y_{4} },\,{y_{3} }\right] = 0,\quad
  \left[{y_{3} },\,{y_{3} }\right] = 0,\\
S&\ad_{y_2}^2{y_1}= 0,\quad
\ad_{y_2}^{2}y_{3} = 0,\\
N&\left[\left[{y_1},\,{y_{4} }\right],\,\left[{y_{4} },\,\left[{y_1},\,
{y_2}\right]\right]\right] = 0;\quad
\left[\left[{y_{3} },\,\left[{y_1},\,{y_2}\right]\right],\,
\left[\left[{y_1},\,{y_2}\right],\,
\left[{y_1},\,{y_{4} }\right]\right]\right] = 0 .
\end{array}
$$  \normalsize
The weights with respect to the $h_{i}$ are:\footnotesize
$$
\renewcommand{\arraystretch}{1.4}
\begin{array}{ccccc|ccccc}
	x_{1}&(0&0&0&1)&{y_1}&(0&0&0&-1)\\
  x_{2}&(0&1&-1&0)&y_2&(0&0&0&-1)\\
  x_{4} &(0&0&1&-1)&y_{4} &(0&-1&1&1)\\
  x_{3}  &(0&0&0&1)&y_{3} &(0&0&0&-1)
\end{array}
$$\normalsize

Other relations between positive and negative elements: \footnotesize
$$
\renewcommand{\arraystretch}{1.4}
\begin{array}{l}
\left[x_{1},\,{y_1}\right] = 
-\,{h_0} ,\quad
  \left[x_{1},\,{y_2}\right] = 0,\quad
  \left[x_{1},\,{y_{4} }\right] = 0,\\
  \left[x_{1},\,{y_{3} }\right] = 
 -\,{h_1} - {h_2} 
   ,\left[x_{2},\,{y_1}\right] = 0,\quad
  \left[x_{2},\,{y_2}\right] = 0,\quad
  \left[x_{2},\,{y_{4} }\right] = 
   x_{3}  ,\\
\left[x_{2},\,{y_{3} }
    \right] = 0,\quad
  \left[x_{4} ,\,{y_1}\right] = 0,\quad
  \left[x_{4} ,\,{y_2}\right] = 
 {h_2} - {h_3} ,\\
  \left[x_{4} ,\,{y_{4} }\right] = 0,\quad
  \left[x_{4} ,\,{y_{3} }\right] = 0,\quad
  \left[x_{3}  ,\,{y_1}\right] = 
-\,{h_1} + {h_2}, \\
   \left[x_{3}  ,\,{y_2}\right] = 0,\quad
  \left[x_{3}  ,\,{y_{4} }\right] = 0,\quad
  \left[x_{3}  ,\,{y_{3} }\right] = 0.
  \end{array}
$$\normalsize
  
\ssec{Generators and relations for $\fv\fa\fs(4|4)$} We denote
$\partial _{i}=\pder{x_i}$ and $\delta _{i}=\pder{\theta_i}$.  Set
(positive generators):
$$
\renewcommand{\arraystretch}{1.4}
\begin{array}{l}
{X_1} = {x_1}{{\delta }_4} - {x_4}{{\delta }_1} +
{{\theta}_3}{{\partial }_2} - {{\theta }_2}{{\partial }_3} ,\quad
{X_2} ={x_2}{{\partial }_1} - {{\theta }_1}{{\delta }_2} ,\\
{X_3} = {x_3}{{\delta }_3},\quad {X_4} ={x_1}{{\delta }_4} +
{x_4}{{\delta }_1},\\
Z = -\,{{{x_1}}^2}{{\partial }_4} + 2\, {x_1}{{\theta}_4}{{\delta }_1}
     \end{array}
$$ 

Relations in $\fv\fa\fs_+$ (since only $X_2$ is even and the
representations of $\fg_0$ in $\fg_{\pm 1}$ are not, strictly speaking
neither highest nor lowest weight one, all these relations are of (N)
type; the same applies for $\fv\fa\fs_-$ since only $Y_2$ and $Y_3$
are even): \footnotesize
$$
\renewcommand{\arraystretch}{1.4}
\begin{array}{l}
    \left[{X_1},\,{X_1}\right] = 0,\quad
  \left[{X_1},\,{X_4}\right] = 0,\quad
  \left[{X_2},\,{X_3}\right] = 0,\quad
  \left[{X_3},\,{X_3}\right] = 0,\\
  \left[{X_3},\,{X_4}\right] = 0,\quad
  \left[{X_3},\,Z\right] = 0,\quad
  \left[{X_4},\,{X_4}\right] = 0,\\
\left[{X_4},\,Z\right] = 3\, \left[{X_1},\,Z\right] ,\quad
 \ad_{X_2}^2{X_1}= 0,\\
  \ad_{X_2}^2{X_4} = 0,\quad
  \ad_Z^2X_1= 0,\\
2\left[Z,\,\left[{X_2},\,{X_4}\right]\right] =
-\left[{X_4},\,\left[{X_2},\,Z\right]\right];\\
\ad_Z^2{X_2}= 0,\quad
    \ad_{X_2}^3Z= 0,\\
    \left[\left[{X_1},\,{X_3}\right],\,
    \left[{X_3},\,\left[{X_1},\,{X_2}\right]\right]
    \right] = 0,\\
    \left[\left[{X_4},\,
     \left[{X_1},\,{X_2}\right]\right],\,
    \left[\left[{X_1},\,{X_2}\right],\,
     \left[{X_1},\,{X_3}\right]\right]\right] = 0,\\
  \left[\left[\left[{X_1},\,{X_2}\right],\,
     \left[{X_2},\,Z\right]\right],\,
    \left[\left[{X_1},\,{X_3}\right],\,
     \left[{X_2},\,Z\right]\right]\right] = 0 .
 \end{array}
$$ \normalsize
     
Generators in  $\fv\fa\fs(4|4)_-$. Set:
$$
\renewcommand{\arraystretch}{1.4}
\begin{array}{l}
{Y_1} = {x_2}{{\delta }_3} - {x_3}{{\delta }_2} +
{{\theta}_4}{{\partial }_1} - {{\theta }_1}{{\partial }_4},\quad 
{Y_2}
={x_2}{{\partial }_3} - {{\theta }_3}{{\delta }_2},\\
{Y_3} = {x_1}{{\partial }_2} - {{\theta }_2}{{\delta }_1},\quad {Y_4}
= {x_2}{{\delta }_3} + {x_3}{{\delta }_2},\quad Z = {{\delta }_4}.
\end{array}
$$ 

Relations in  $\fv\fa\fs(4|4)_-$: \footnotesize
$$
\renewcommand{\arraystretch}{1.4}
\begin{array}{l|l}
S& \ad_{Y_2}^2{Y_3}= 0,\quad
\ad_{Y_3}^2{Y_2}= 0,\\
N&\ad_{Y_3}^2{Y_1}= 0,\quad
\ad_{Y_2}^2{Y_4}= 0,\quad
  \ad_{Y_3}^2{Y_4}= 0,\\
N&\left[{Y_1},\,{Y_1}\right] = 0,\quad
  \left[{Y_1},\,{Y_2}\right] = 0,\quad
  \left[{Y_1},\,{Y_4}\right] = 0,\quad
  \left[{Y_2},\,Z\right] = 0,\\
N&\left[{Y_3},\,Z\right] = 0,\quad
  \left[{Y_4},\,{Y_4}\right] = 0,\quad
  \left[{Y_4},\,Z\right] = 0,\quad
  \left[Z,\,Z\right] = 0,\\
N&2\left[{Y_4},\,\left[{Y_2},\,{Y_3}\right]\right] = 
    \left[{Y_3},\,\left[{Y_2},\,{Y_4}\right]\right],\quad
     \left[\left[{Y_1},\,Z\right],\,
    \left[{Y_3},\,{Y_4}\right]\right] = 
   \left[\left[{Y_1},\,{Y_3}\right],\,
    \left[{Y_1},\,Z\right]\right] .
    \end{array}
$$\normalsize

\underline{{\bf $\fk\fas(4|3; K)$}} Set (negative generators):
$$
{Y_1} = {{\xi }_2} \eta _{1} ,\quad {Y_2} = {{\xi }_3} \eta _{2}
,\quad {Y_3} = \eta _{2} \eta _{3},\quad Z = {{\xi }_1}.
$$
{\bf The relations in $\fk\fas(4|3; K)_{-}$} are: \footnotesize
$$
\renewcommand{\arraystretch}{1.4}
\begin{array}{l|l}
S&\left[{Y_2},\,{Y_3}\right] = 0,\quad \ad_{Y_2}^2{Y_1}= 0,\quad 
\ad_{Y_3}^2Y_1= 0,\\
S&\ad_{Y_1}^2{Y_2}= 0,\quad
\ad_{Y_1}^2{Y_3}= 0,\\
HW&\left[{Y_2},\,Z\right] = 0,\quad
  \left[{Y_3},\,Z\right] = 0,\quad
\ad_{Y_1}^2Z= 0,\\
N_{2}&\left[Z,\,Z\right] = 0.
\end{array}
$$ \normalsize
    
Set (positive generators):
$$
{X_1} = {{\xi }_1} \eta _{2},\quad {X_2} = {{\xi }_2} \eta _{3} ,\quad
{X_3} = {{\xi}_2} {{\xi }_3},\quad {Z_1} = t \eta _{1} ,\quad {Z_2} =
{{\xi }_3} \eta _{1} \eta _{2} .
$$
{\bf The relations in $\fk\fas(4|3; K)_{+}$} (computed up to $\deg 
\leq 40$) consist of Serre-type relations: \footnotesize
$$
\renewcommand{\arraystretch}{1.4}
\begin{array}{l}
\left[{X_1},\,{Z_2}] = 0,\quad [{X_2},\,{X_3}] = 0,\quad [{X_2},\,{Z_1}] =
 0,\quad [{X_3},\,{Z_1}\right] = 0,\\
\left[{X_3},\,{Z_2}] = 0,\quad [{Z_1},\,{Z_1}] = 0,\quad [{Z_1},\,{Z_2}] =
 0,\quad [{Z_2},\,{Z_2}\right] = 0,\\
\ad_{X_1}^2{X_2}= 0,\quad \ad_{X_1}^{2}{X_3}=
0,\quad \ad_{X_1}^2{Z_1}= 0,\\
\ad_{X_2}^2{X_1}= 0,\quad \ad_{X_3}^2{X_1}
= 0,\quad \ad_{X_2}^3{Z_2}= 0,\end{array}
$$\normalsize 
and new (awful) type of relations: \noindent
\tiny
$$
\renewcommand{\arraystretch}{1.4}
\begin{array}{l}
\left[{Z_2},\, [{X_2},\, [{X_2},\,{Z_2}\right]]] = 0,\\
{}\\
\left[[ [{X_1},\,{X_3}],\, [{X_2},\,{Z_2}]],\, [[{X_1},\,{Z_1} ],\,
[{X_2},\,{Z_2}\right]]] = 
-\,[[{Z_1},\, [{X_1},\,{X_3}] ],\,[[{X_1},\,{Z_1} ],\, [{X_2},\,{Z_2}]]], \\
{}\\
4\left[[ [{Z_2},\,[{X_1},\,{X_2}] ],\, [{Z_1},\, [{X_3},\, [{X_1},\,{X_2}] ]]],\,
[[{Z_1},\, [{X_1},\,{X_3}] ],\,[[{X_2},\,{Z_2} ],\, [{Z_1},\,
[{X_1},\,{X_2}\right]]]]] = \\
{}[[ [{Z_1},\, [{X_1},\,{X_2}]],\, [{Z_1},\, [{X_1},\,{X_3}] ]],\,
[[{Z_1},\, [{X_1},\,{X_3}] ],\, [
[{X_2},\,{Z_2}] ,\,[{Z_1},\, [{X_1},\,{X_2}]]] ]] + \\
{}[[[{Z_1},\, [{X_1},\,{X_2} ]],\, [{Z_1},\,
[{X_1},\,{X_3} ]]],\, [ [{Z_1},\, [{X_3},\, [{X_1},\,{X_2} ]]],\ ,[
[{X_1},\,{Z_1} ],\, [{X_2},\,{Z_2}]]] ] , \\
{}\\
12\left[[[[{X_1},\,{Z_1} ],\, [{X_2},\,{Z_2}]],\, [[{Z_1},\, [{X_1},\,{X_2}] ],\,
[{Z_1},\, [{X_1},\,{X_3}] ]]],\, [[ [{X_1},\,{Z_1}],\, [{X_3},\,
[{X_1},\,{X_2}] ]],\, [[{X_2},\,{Z_2}],\, [{Z_1},\, [{X_1},\,{X_3}\right]]]]] = \\
{}[[[{Z_1},\, [{X_1},\,{X_2}] ],\, [{Z_1},\, [{X_1},\,{X_3}]]],\\
\left[[[{Z_1},\, [{X_1},\,{X_2}] ],\, [{Z_1},\, [{X_1},\,{X_3}] ]],\, [[{Z_1},\,
[{X_1},\,{X_3}] ],\, [[{X_1},\,{Z_1}] ,\,[{X_2},\, {Z_2}\right]]] ]] , \\
\left[[[[{Z_1},\, [{X_1},\,{X_2}] ],\, [{Z_1},\, [{X_1},\,{X_3}] ]],\,
[[{Z_2},\, [{X_1},\,{X_2}] ],\, [{Z_1},\, [{X_3},\, [{X_1},\, {X_2}\right]]]] ],\, \\
{}\\
24\left[ [[{Z_1},\, [{X_1},\,{X_2}] ],\, [{Z_1},\,[{X_1},\,{X_3}] ]],\, [[{Z_1},\,
[{X_1},\,{X_3}] ],\, [[{X_2},\,{Z_2}],\, [{Z_1},\, [{X_1},\,{X_2}\right]]]]]] = \\
{}[[ [{Z_1},\, [{X_1},\,{X_2}] ],\,
[{Z_1},\, [{X_1},\,{X_3}] ]],\, [[ [{Z_1},\, [{X_1},\,{X_2}]],\,
[{Z_1},\, [{X_1},\,{X_3} ]]],\, [[ [{Z_1},\, [{X_1},\,{X_2} ]],\,
[{Z_1},\, [{X_1},\,{X_3} ]]],\\
{}\\
24[[{Z_1},\, [{X_1},\,{X_3} ]],\, [[{X_2},\,{Z_2} ],\, 
[{Z_1},\, [{X_1},\,{X_2}]]] ]]] ] + \\
{}[[ [{Z_1},\, [{X_1},\,{X_2} ]],\,
[{Z_1},\, [{X_1},\,{X_3} ]]],\, [[ [{Z_1},\, [{X_1},\,{X_2} ]],\,
[{Z_1},\, [{X_1},\,{X_3} ]]],\, [[ [{Z_1},\, [{X_1},\,{X_2} ]],\, \\
\left[{Z_1},\, [{X_1},\,{X_3} ]]],\, [[{Z_1},\, [{X_3},\, [{X_1},\,{X_2}]]],\ ,
[[{X_1},\,{Z_1} ],\,
[{X_2},\,{Z_2} \right]]] ]]] , \\
{}[[ [[{Z_1},\, [{X_1},\,{X_2}]],\, [{Z_1},\, [{X_1},\,{X_3}]]],\,
[[ [{X_1},\,{Z_1}] ,\,[{X_2},\,{Z_2}]],\, [[{Z_1},\, [{X_1},\,{X_2}]],\, 
[{Z_1},\, [{X_1},\,{X_3}]]]] ],\\
{}\\
120[[[{Z_1},\, [{X_1},\,{X_2}] ],\, [{Z_1},\,
[{X_1},\,{X_3}] ]],\, [[[{X_1},\,{Z_1}] ,\,[{X_3},\,
[{X_1},\,{X_2}] ]],\, [[{X_2},\,{Z_2}],\, [{Z_1},\, [{X_1},\,{X_3}]]]] ]] = \\
{}[[ [{Z_1},\, [{X_1},\,{X_2}] ],\,
[{Z_1},\, [{X_1},\,{X_3}] ]],\, [[ [{Z_1},\, [{X_1},\,{X_2}]],\,
[{Z_1},\, [{X_1},\,{X_3}] ]],\, [[ [{Z_1},\, [{X_1},\,{X_2} ]],\,
[{Z_1},\, [{X_1},\,{X_3} ]]],\\ 
{}[[ [{Z_1},\, [{X_1},\,{X_2} ]],\,
[{Z_1},\, [{X_1},\,{X_3} ]]],\ ,[[{Z_1},\, [{X_1},\,{X_3} ]],\, 
[[{X_1},\,{Z_1} ],\, [{X_2},\,{Z_2}]]] ]]] ] ,\\
{}\\
120[[ [[{Z_1},\, [{X_1},\,{X_2}]],\, [{Z_1},\, [{X_1},\,{X_3}] ]],\,
[[ [{Z_1},\, [{X_1},\,{X_2}] ],\,[{Z_1},\, [{X_1},\,{X_3}] ]],\,
[[{Z_2},\, [{X_1},\,{X_2}] ],\,[{Z_1},\, [{X_3},\, 
[{X_1},\,{X_2}]]]] ]],\\ 
{}[[ [{Z_1},\,
[{X_1},\,{X_2}] ],\, [{Z_1},\, [{X_1},\,{X_3}] ]],\, 
[[[{Z_1},\, [{X_1},\,{X_2}] ],\, [{Z_1},\,
[{X_1},\,{X_3}] ]],\, [[{Z_1},\,[{X_1},\,{X_3}] ],\, 
[[{X_2},\,{Z_2}],\, [{Z_1},\, [{X_1},\,{X_2}]]]] ]]] = \\
{}[[ [{Z_1},\,[{X_1},\,{X_2}] ],\, [{Z_1},\,
[{X_1},\,{X_3}] ]],\, [[[{Z_1},\, [{X_1},\,{X_2} ]],\, [{Z_1},\,
[{X_1},\,{X_3} ]]],\, [ [[{Z_1},\, [{X_1},\,{X_2} ]],\, [{Z_1},\,
[{X_1},\,{X_3} ]]],\\
{}\\
120[ [[{Z_1},\, [{X_1},\,{X_2} ]],\, [{Z_1},\,
[{X_1},\,{X_3}]]],\,
\left[[[{Z_1},\,
[{X_1},\,{X_2} ]],\, [{Z_1},\, [{X_1},\,{X_3} ]]],\, [ [{Z_1},\,
[{X_1},\,{X_3} ]],\, [ [{X_2},\,{Z_2}],\, [{Z_1},\, [{X_1},\,{X_2} \right]]]]]] ]]] +\\
{}[[ [{Z_1},\, [{X_1},\,{X_2}]],\, [{Z_1},\, [{X_1},\,{X_3}]]],\, 
[ [ [{Z_1},\, [{X_1},\,{X_2} ]],\, [{Z_1},\,
[{X_1},\,{X_3} ]]],\ ,[[[{Z_1},\, [{X_1},\,{X_2} ]],\, [{Z_1},\,
[{X_1},\,{X_3} ]]];\\
{}[[[{Z_1},\, [{X_1},\,{X_2} ]],\, [{Z_1},\, [{X_1},\,{X_3} ]]],\, [[[{Z_1},\, [{X_1},\,{X_2} ]],\,
[{Z_1},\, [{X_1},\,{X_3} ]]],\ ,[[{Z_1},\, [{X_3},\, [{X_1},\,{X_2}
]]],\, [[{X_1},\,{Z_1} ],\, [{X_2},\,{Z_2}]]]]]]]].
\end{array}
$$ 

\normalsize

\ssec{Comment} Lie superalgebra $\fv\fas$ has only one W-grading, so
we consider it.  For the remaining exceptional simple Lie
superalgebras $\fg$ we take the consistent grading.  So $\fg_{0}$ is a
Lie algebra, moreover, a reductive one; $\fg_{1}$ generates the
positive part, and $\fg_{-1}$ generates the negative part, moreover,
$\fg_{-1}$ is an irreducible $\fg_{0}$-module.  Therefore, $\fg$ is
generated by the Chevalley generators $X_{i}^{\pm}$ and $H_{i}=
[X_{i}^+, X_{i}^-]$ of $\fg_{0}$, the lowest weight vectors
$v_{\lambda_{k}}$ of $\fg_{1}$ and the highest weight vector
$w_{\Lambda}$ of $\fg_{-1}$.  A number of relations between these
generators are already known: the Serre relations for $X_{i}^{\pm}$
and relations involving $H_{i}$; and also
$$
(\ad X_{i}^+)^{\lambda_{k}(X_{i}^+)+1}(v_{\lambda_{k}})=0\eqno{(LW)}
$$
$$
(\ad X_{i}^-)^{\Lambda(X_{i}^-)+1}(w_{\Lambda})=0\eqno{(HW)}
$$
and the weight relations
$H_{i}(v_{\lambda_{k}})=\lambda_{k}^{(i)}\cdot v_{\lambda_{k}}$, etc. 
Since the finite diminsional representations of the reductive algebras
are completely reducible, we have to compare $\fg_{1}\wedge \fg_{1}$
and $\fg_{2}$ (the lowest weight vectors of $\fg_{1}\wedge
\fg_{1}/\fg_{2}$ are relations) and similarly for the negative part. 
If this does not suffice ($\fg_{1}\wedge \fg_{1}\wedge \fg_{1}$ modulo
relations is greater than $\fg_{3}$) we add new relations, etc.

For positive parts of simple vectorial Lie algebras we can use Fuchs'
theorem with Kochetkov's correction (and direct calculations described
above for small dimensions).  For exceptional algebras we consider we
have no general theory, so having obtained the whole $\fg_{i}$ and no
new relations for several $i>0$ we conjecture that we got all.  For
$\fk\fas$ we keep getting new relations in every $i$.

\section*{Appendix}

\ssec{1.0.  Linear algebra in superspaces.  Generalities} A {\it 
superspace} is a $\Zee /2$-graded space; for any superspace 
$V=V_{\ev}\oplus V_{\od }$ denote by $\Pi (V)$ another copy of the 
same superspace: with the shifted parity, i.e., $(\Pi(V))_{\bar i}= 
V_{\bar i+\od }$.  The {\it superdimension} of $V$ is $\dim 
V=p+q\eps $, where $\eps ^2=1$ and $p=\dim V_{\ev}$, 
$q=\dim V_{\od }$.  (Usually, $\dim V$ is expressed as a pair $(p,q)$ 
or $p|q$; this obscures the fact that $\dim V\otimes W=\dim V\cdot 
\dim W$; this fact is clear with the use of $\eps $.)

A superspace structure in $V$ induces the superspace structure in the 
space $\End (V)$.  A {\it basis of a superspace} is always a basis 
consisting of {\it homogeneous} vectors; let $\Par=(p_1, \dots, 
p_{\dim V})$ be an ordered collection of their parities.  We call 
$\Par$ the {\it format} of (the basis of) $V$.  A square {\it 
supermatrix} of format (size) $\Par$ is a $\dim V\times \dim V$ matrix 
whose $i$th row and $i$th column are of the same parity $p_i$.  

\footnotesize One usually considers one of the simplest formats 
$\Par$, e.g., $\Par$ of the form $(\ev , \dots, \ev ; \od , \dots, \od 
)$ is denoted $(\dim V_{\ev}, \dim V_{\od})$ and called {\it standard}.  In this paper we can do without 
nonstandard formats.  But they are vital in the classification of 
systems of simple roots that the reader might be interested in in 
connection with applications to $q$-quantization or integrable 
systems.  Besides, systems of simple roots corresponding to distinct 
nonstandard formats are related by odd reflections --- analogs of our 
nonstandard regradings. (For an approach to superroots see \cite{S2}.)

\normalsize

The matrix unit $E_{ij}$ is supposed to be of parity $p_i+p_j$ and the 
bracket of supermatrices (of the same format) is defined via {\bf Sign 
Rule}:

{\it if something of parity $p$ moves past something of parity $q$ the 
sign $(-1)^{pq}$ accrues; the formulas defined on homogeneous elements 
are extended to arbitrary ones via linearity}.

Examples: by setting $[X, 
Y]=XY-(-1)^{p(X)p(Y)}YX$ we get the notion of the supercommutator and 
the ensuing notions of the supercommutative superalgebra and the Lie 
superalgebra (which in addition to superskew-commutativity satisfies 
the super Jacobi identity, i.e., the Jacobi identity amended with the 
Sign Rule).  The derivation (better say, superderivation) of a 
superalgebra $A$ is a linear map $D: A\tto A$ that satisfies the 
Leibniz rule (and Sign rule)
$$
D(ab)=D(a)b+(-1)^{p(D)p(a)}aD(b). 
$$
In particular, let $A=\Cee[x]$ be the free supercommutative polynomial 
superalgebra in $x=(x_{1}, \dots , x_{n})$, where the superstructure 
is determined by the parities of the indeterminates: $p(x_{i})=p_{i}$.  
Partial derivatives are defined (with the help of super Leibniz Rule) 
by the formulas $\pderf{x_{i}}{x_{j}}=\delta_{i,j}$.  Clearly, the 
collection $\fder A$ of all superdifferentiations of $A$ is a Lie 
superalgebra whose elements are of the form $\sum f_i(x)\pder{x_{i}}$.

Given the supercommutative superalgebra $\cF$ of ``functions'' in 
indeterminates $x$, define the supercommutative superalgebra $\Omega$ 
of differential forms as polynomial algebra over $\cF$ in $dx$, where 
$p(d)=\od$.  Since $dx$ is even for an odd $x$, we can consider not 
only polynomials in $dx$.  Smooth or analytic functions in the 
differentials of the $x$ are called {\it pseudodifferential forms} on 
the supermanifold with coordinates $x$, see \cite{BL1}.  We will need 
them to interpret $\fh_{\lambda}(2|2)$. The exterior differential is 
defined on (pseudo) differential forms by the formulas (mind Leibniz 
and Sign Rules):
$$
d(x_{i})=dx_{i}\text{ and } d^2=0.
$$
The Lie derivative is defined (minding same Rules) by the formula
$$
L_{D}(df)=(-1)^{p(D)}d(D(f)).
$$
In particular,
$$
L_{D}\left 
((df)^{\lambda}\right)=\lambda(-1)^{p(D)}d(D(f))(df)^{\lambda-1}\text{ 
for any }\lambda\in\Cee.
$$

The {\it general linear} Lie superalgebra of all supermatrices of size
$\Par$ is denoted by $\fgl(\Par)$ (usually, $\fgl(\dim V_{\bar 0}|\dim
V_{\bar 1})$).  Any matrix from $\fgl(\Par)$ can be expressed as the
sum of its even and odd parts; in the standard format this is the
following block expression:
$$
\begin{pmatrix}A&B\\ C&D\end{pmatrix}=\begin{pmatrix}A&0\\
0&D\end{pmatrix}+\begin{pmatrix}0&B\\ C&0\end{pmatrix},\quad 
p\left(\begin{pmatrix}A&0\\
0&D\end{pmatrix}\right)=\ev, \; p\left(\begin{pmatrix}0&B\\
C&0\end{pmatrix}\right)=\od.
$$

The {\it supertrace} is the map $\fgl (\Par)\tto \Cee$, 
$(A_{ij})\mapsto \sum (-1)^{p_{i}}A_{ii}$.  Since $\str [x, y]=0$, the 
subsuperspace of supertraceless matrices constitutes the {\it special 
linear} Lie subsuperalgebra $\fsl(\Par)$.

There are, however, two super versions of $\fgl(n)$, not one.  The 
other version is called the {\it queer} Lie superalgebra and is defined 
as the one that preserves the complex structure given by an {\it odd} 
operator $J$, i.e., is the centralizer $C(J)$ of $J$:
$$
\fq(n)=C(J)=\{X\in\fgl(n|n)\mid [X, J]=0 \}, \text{ where } J^2=-\id.
$$
It is clear that by a change of basis we can reduce $J$ to the form 
$J_{2n}=\begin{pmatrix}0&1_n\\ -1&0\end{pmatrix}$.  In the standard 
format we have
$$
\fq(n)=\left \{\begin{pmatrix}A&B\\ B&A\end{pmatrix}\right\}.
$$
On $\fq(n)$, the {\it queertrace} is defined: $\qtr: 
\begin{pmatrix}A&B\\
B&A\end{pmatrix}\mapsto
\tr B$. Denote by $\fsq(n)$ the Lie superalgebra of {\it queertraceless}
matrices.

Observe that the identity representations of $\fq$ and $\fsq$ in $V$, 
though irreducible in supersetting, are not irreducible in the 
nongraded sense: take homogeneous (with respect to parity) and 
linearly independent vectors $v_1$, \dots , $v_n$ from $V$; then 
$\Span (v_1+J(v_1), \dots , v_n+J(v_n))$ is an invariant subspace of 
$V$ which is not a subsuperspace.

A representation is {\it irreducible} \index{representation of Lie 
superalgebra irreducible} \index{$G$-type irreducible representation 
of Lie superalgebra} of {\it general type} or just of {\it $G$-type} 
if there is no invariant subspace, otherwise it is called {\it 
irreducible of $Q$-type} \index{$Q$-type irreducible representation of 
Lie superalgebra} ($Q$ is after the general queer Lie superalgebra --- 
a specifically ``superish'' analog of $\fgl$); an irreducible 
representation of $Q$-type has no invariant sub{\it super}space but 
{\it has} an invariant subspace.

{\bf Lie superalgebras that preserve bilinear forms: two types}.  To 
the linear map $F: V\tto W$ of superspaces there corresponds the dual 
map $F^*: W^*\tto V^*$ between the dual superspaces.  In a basis 
consisting of the vectors $v_{i}$ of format $\Par$, the formula 
$F(v_{j})=\mathop{\sum}\limits_{i}v_{i}A_{ij}$ assigns to $F$ the 
supermatrix $A$. In the dual bases, to $F^*$ the {\it supertransposed} matrix $A^{st}$ 
corresponds:
$$
(A^{st})_{ij}=(-1)^{(p_{i}+p_{j})(p_{i}+p(A))}A_{ji}.
$$

The supermatrices $X\in\fgl(\Par)$ such that 
$$
X^{st}B+(-1)^{p(X)p(B)}BX=0\quad \text{for an homogeneous matrix 
$B\in\fgl(\Par)$}
$$
constitute the Lie superalgebra $\faut (B)$ that preserves the 
bilinear form $B^f$ on $V$ whose matrix $B$ is given by the formula
$B_{ij}=(-1)^{p(B^{f})p(v_{i})}B^f(v_{i}, v_{j})$ for the basis vectors 
$v_{i}$.

Recall that the {\it supersymmetry} of the homogeneous form $B^f$
means that its matrix $B$ satisfies the condition $B^{u}=B$, where for
the matrix $B=\begin{pmatrix} R& S \\ T & U\end{pmatrix}$ we set
$B^{u}=
\begin{pmatrix} 
R^{t} & (-1)^{p(B)}T^{t} \\ (-1)^{p(B)}S^{t} & -U^{t}\end{pmatrix}$. 
Similarly, {\it skew-su\-per\-sym\-metry} of $B$ means that $B^{u}=-B$. 
Thus, we see that the {\it upsetting} of bilinear forms $u: \Bil (V, 
W)\tto\Bil(W, V)$, which for the {\it spaces} and when $V=W$ is expressed on 
matrices in terms of the transposition, is a new operation.

Most popular canonical forms of the nondegenerate supersymmetric form 
are the ones whose supermatrices in the standard format are the 
following canonical ones, $B_{ev}$ or $B'_{ev}$:
$$
B'_{ev}(m|2n)= \begin{pmatrix} 
1_m&0\\
0&J_{2n}
\end{pmatrix},\quad \text{where 
$J_{2n}=\begin{pmatrix}0&1_n\\-1_n&0\end{pmatrix}$},
$$
or
$$
B_{ev}(m|2n)= \begin{pmatrix} 
\antidiag (1, \dots , 1)&0\\
0&J_{2n}
\end{pmatrix}. 
$$
The usual notation for $\faut (B_{ev}(m|2n))$ is $\fosp(m|2n)$ or, 
more precisely, $\fosp^{sy}(m|2n)$.  Observe that the passage from $V$ 
to $\Pi (V)$ sends the supersymmetric forms to superskew-symmetric 
ones, preserved by the \lq\lq symplectico-orthogonal" Lie 
superalgebra, $\fsp'\fo (2n|m)$ or, better say, $\fosp^{sk}(m|2n)$, 
which is isomorphic to $\fosp^{sy}(m|2n)$ but has a different matrix 
realization.  We never use notation $\fsp'\fo (2n|m)$ in order not to 
confuse with the special Poisson superalgebra.

In the standard format the matrix realizations of these algebras
are: 
$$
\begin{matrix} 
\fosp (m|2n)=\left\{\left (\begin{matrix} E&Y&X^t\\
X&A&B\\
-Y^t&C&-A^t\end{matrix} \right)\right\};\quad \fosp^{sk}(m|2n)=
\left\{\left(\begin{matrix} A&B&X\\
C&-A^t&Y^t\\
Y&-X^t&E\end{matrix} \right)\right\}, \\
\text{where}\; 
\left(\begin{matrix} A&B\\
C&-A^t\end{matrix} \right)\in \fsp(2n),\quad E\in\fo(m).\end{matrix} 
$$

A nondegenerate supersymmetric odd bilinear form $B_{odd}(n|n)$ can be 
reduced to a canonical form whose matrix in the standard format is 
$J_{2n}$.  A canonical form of the superskew odd nondegenerate form in 
the standard format is $\Pi_{2n}=\begin{pmatrix} 
0&1_n\\1_n&0\end{pmatrix}$.  The usual notation for $\faut 
(B_{odd}(\Par))$ is $\fpe(\Par)$.  The passage from $V$ to $\Pi (V)$
establishes an isomorphism $\fpe^{sy}(\Par)\cong\fpe^{sk}(\Par)$.  
This Lie superalgebra is called, as A.~Weil suggested, {\it 
periplectic}.  The matrix realizations in the 
standard format of these superalgebras is shorthanded to:
$$
\begin{matrix}
\fpe ^{sy}\ (n)=\left\{\begin{pmatrix} A&B\\
C&-A^t\end{pmatrix}, \; \text{where}\; B=-B^t,\; 
C=C^t\right\};\\
\fpe^{sk}(n)=\left\{\begin{pmatrix}A&B\\ C&-A^t\end{pmatrix}, \;
\text{where}\; B=B^t,\;  C=-C^t\right\}.
\end{matrix}
$$
Observe that though the Lie superalgebras $\fosp^{sy} (m|2n)$ and 
$\fpe ^{sk} (2n|m)$, as well as $\fpe ^{sy} (n)$ and $\fpe ^{sk} (n)$, 
are isomorphic, the difference between them is sometimes crucial, see
\cite{Sh5}.

The {\it special periplectic} superalgebra is $\fspe(n)=\{X\in\fpe(n)\mid \str
X=0\}$. Of particular interest to us will be also $\fspe(n)_{a, 
b}=\fspe(n)\supplus\Cee(az+bd)$, where $z=1_{2n}$, $d=\diag(1_{n}, 
-1_{n})$.

\ssec{1.0.1.  What a Lie superalgebra is} Dealing with superalgebras 
it sometimes becomes useful to know their definition.  Lie superalgebras were
distinguished in topology in 1930's or earlier.  So when somebody
offers a ``better than usual'' definition of a notion which seemed to
have been established about 70 year ago this might look strange, to
say the least.  Nevertheless, the answer to the question ``what is a
Lie superalgebra?''  is still not a common knowledge.  Indeed, the
naive definition (``apply the Sign Rule to the definition of the Lie
algebra'') is manifestly inadequate for considering the (singular)
supervarieties of deformations and applying representation theory to
mathematical physics, for example, in the study of the coadjoint
representation of the Lie supergroup which can act on a supermanifold
but never on a superspace (an object from another category).  So, to
deform Lie superalgebras and apply group-theoretical methods in
``super'' setting, we must be able to recover a supermanifold from a
superspace, and vice versa.

A proper definition of Lie superalgebras is as follows, cf. 
\cite{L3}--\cite{L5}.  The {\it Lie superalgebra} in the category of
supermanifolds corresponding to the ``naive'' Lie superalgebra $L=
L_{\ev} \oplus L_{\od}$ is a linear supermanifold $\cL=(L_{\ev},
\cO)$, where the sheaf of functions $\cO$ consists of functions on
$L_{\ev}$ with values in the Grassmann superalgebra on $L_{\od}^*$;
this supermanifold should be such that for \lq\lq any" (say, finitely
generated, or from some other appropriate category) supercommutative
superalgebra $C$, the space $\cL(C)=\Hom (\Spec C, \cL)$, called {\it
the space of $C$-points of} $\cL$, is a Lie algebra and the
correspondence $C\tto \cL(C)$ is a functor in $C$.  (A.~Weil
introduced this approach in algebraic geometry in 1954; in super
setting it is called {\it the language of points} or {\it families},
see \cite{D}, \cite{L4}.)  This definition might look terribly complicated, but
fortunately one can show that the correspondence
$\cL\longleftrightarrow L$ is one-to-one and the Lie algebra $\cL(C)$,
also denoted $L(C)$, admits a very simple description: $L(C)=(L\otimes
C)_{\ev}$.

A {\it Lie superalgebra homomorphism} $\rho: L_1 \tto L_2$ in these 
terms is a functor morphism, i.e., a collection of Lie algebra 
homomorphisms $\rho_C: L_1 (C)\tto L_2(C)$ compatible with morphisms 
of supercommutative superalgebras $C\tto C'$.  In particular, a {\it 
representation} of a Lie superalgebra $L$ in a superspace $V$ is a 
homomorphism $\rho: L\tto \fgl (V)$, i.e., a collection of Lie algebra 
homomorphisms $\rho_C: L(C) \tto ( \fgl (V )\otimes C)_{\ev}$.

\begin{rem*}{Example} Consider a representation $\rho:\fg\tto\fgl(V)$. 
The tangent space of the moduli superspace of deformations of $\rho$
is isomorphic to $H^1(\fg; V\otimes V^*)$.  For example, if $\fg$ is
the $0|n$-dimensional (i.e., purely odd) Lie superalgebra (with the
only bracket possible: identically equal to zero), its only
irreducible representations are the trivial one, {\bf 1}, and
$\Pi({\bf 1})$.  Clearly, ${\bf 1}\otimes {\bf 1}^*\simeq \Pi({\bf
1})\otimes \Pi({\bf 1})^*\simeq {\bf 1}$, and because the superalgebra
is commutative, the differential in the cochain complex is trivial. 
Therefore, $H^1(\fg; {\bf 1})=E^1(\fg^*)\simeq\fg^*$, so there are
$\dim\,\fg$ odd parameters of deformations of the trivial
representation.  If we consider $\fg$ ``naively'' all of the odd
parameters will be lost.

Which of these infinitesimal deformations can be
extended to a global one is a separate much tougher question, usually
solved {\it ad hoc}. \end{rem*}

Examples that lucidly illustrate why one should always remember that a 
Lie superalgebra is not a mere linear superspace but a linear 
supermanifold are, e.g., deforms $\widetilde{\fsvect}(0|2n+1)$ and 
$\widetilde{\fs\fb}_{\mu}(2^{2n}-1|2^{2n})$ with odd parameters 
considered below and viewed as Lie algebras. In the category of 
supermanifolds these are simple Lie superalgebras.

\ssec{1.0.2.  Projectivization} If $\fs$ is a Lie algebra of scalar 
matrices, and $\fg\subset \fgl (n|n)$ is a Lie subsuperalgebra 
containing $\fs$, then the {\it projective} Lie superalgebra of type 
$\fg$ is $\fpg= \fg/\fs$.  Lie superalgebras $\fg_1\bigodot \fg_2$ 
described in sect. 3.1 are also projective.

Projectivization sometimes leads to new Lie superalgebras, for 
example: $\fpgl (n|n)$, $\fpsl (n|n)$, $\fpq (n)$, $\fpsq (n)$; 
whereas $\fpgl (p|q)\cong \fsl (p|q)$ if $p\neq q$.

\ssec{1.0.3.  What is a semisimple Lie superalgebra} These algebras
are needed in description of primitive Lie superalgebras of vector
fields --- a geometrically natural problem though wild for Lie
superalgebras, see \cite{ALSh}.  Recall that the Lie superalgebra $\fg$ without proper
ideals and of dimension $>1$ is called {\it simple}.  Examples:
$\fsl(m|n)$ for $m> n\geq 1$, $\fpsl(n|n)$ for $n>1$, $\fpsq(n)$ for
$n>2$, $\fosp(m|2n)$ for $mn\neq 0$ and $\fspe(n)$ for $n>2$.

\footnotesize

We will not need the remaining simple finite dimensional Lie
superalgebras of non-vectorial type.  These superalgebras, discovered
by I.~Kaplansky (a 1975-preprint, see \cite{K}) are
$\fosp_{\alpha}(4|2)$, the deforms of $\fosp(4|2)$, and the two
exceptions that we denote by $\fa\fg_{2}$ and $\fa\fb_{3}$.  For their
description we refer to \cite{K}, \cite{FK}, \cite{Sud}, see also \cite{GL2} for the
description of the system of simple roots see \cite{Ka4} completed in
\cite{vdL, S1, S2}.

\normalsize

We say that $\fh$ is {\it almost simple} if it can be sandwiched 
(non-strictly) between a simple Lie superalgebra $\fs$ and the Lie 
superalgebra $\fder~\fs$ of derivations of $\fs$: 
$\fs\subset\fh\subset\fder~\fs$.

By definition, $\fg$ is {\it semisimple} if its radical is zero. 
Literally following the description of semisimple Lie algebras over
the fields of prime characteristic, V.~Kac \cite{Ka4} gave the
following description of semisimple Lie superalgebras.  Let $\fs_1$,
\dots , $\fs_k$ be simple Lie superalgebras, let $n_1$, \dots , $n_k$
be {\it pairs} of non-negative integers $n_j=(n_j^\ev, n_j^\od)$, let
$\cF(n_j)$ be the supercommutative superalgebra of polynomials in
$n_j^\ev$ even and $n_j^\od$ odd indeterminates, and
$\fs=\mathop{\oplus}\limits_{j}\left (\fs_j\otimes\cF(n_j)\right)$. 
Then
$$
\fder~\fs=\mathop{\oplus}\limits_{j}\left((\fder~\fs_j)\otimes\cF(n_j)
\supplus \id_{\fs_j}\otimes
\fvect(n_j)\right). \eqno{(0.1.3)}
$$
{\sl Let $\fg$ be a subalgebra of $\fder\fs$ containing $\fs$.  If the 
projection of $\fg$ on $1\otimes\fvect(n_j)_{-1}$ is onto for each 
$j$, then $\fg$ is semisimple and all semisimple Lie superalgebras 
arise in the manner indicated}, i.e., as sums of {\it almost simple 
superalgebras} corresponding to the summands of $(0.1.3)$.

\ssec{1.0.4.  A.~Sergeev's central extension} In 70's A.~Sergeev
proved that there is just one nontrivial central extension of
$\fspe(n)$.  It exists only for $n=4$ and we denote it by $\fas$.  Let
us represent an arbitrary element $A\in\fas$ as a pair $A=x+d\cdot z$,
where $x\in\fspe(4)$, $d\in{\Cee}$ and $z$ is the central element. 
The bracket in $\fas$ is
$$
\left[\begin{pmatrix} a & b \cr 
c & -a^t \end{pmatrix}+d\cdot z, \begin{pmatrix}
a' & b' \cr 
c' & -a'{}^t \end{pmatrix} +d'\cdot z\right]=
\left[\begin{pmatrix} a & b \cr 
c & -a^t \end{pmatrix}, \begin{pmatrix}
a' & b' \cr 
c' & -a'{}^t \end{pmatrix}\right]+\tr~c\tilde c'\cdot z,\eqno{(0.1.4.1)}
$$
where $\, \tilde{}\, $ is extended via linearity from matrices
$c_{ij}=E_{ij}-E_{ji}$ on which $\tilde c_{ij}=c_{kl}$
for any even permutation $(1234)\mapsto(ijkl)$..

The Lie superalgebra $\fas$ can also be described with the help of the 
spinor representation.  For this we need several vectorial superalgebras 
defined in sect. 0.3.  Consider $\fpo(0|6)$, the Lie superalgebra 
whose superspace is the Grassmann superalgebra $\Lambda(\xi, \eta)$ 
generated by $\xi_1, \xi_2, \xi_3, \eta _1, \eta _2, \eta_3$ and the 
bracket is the Poisson bracket (0.3.6).  Recall that $\fh(0|6)=\Span (H_f\mid 
f\in\Lambda (\xi, \eta))$.

Now, observe that $\fspe(4)$ can be embedded into $\fh(0|6)$.  Indeed, 
setting $\deg \xi_i=\deg \eta _i=1$ for all $i$ we introduce a 
$\Zee$-grading on $\Lambda(\xi, \eta)$ which, in turn, induces a 
$\Zee$-grading on $\fh(0|6)$ of the form 
$\fh(0|6)=\mathop{\oplus}\limits_{i\geq -1}\fh(0|6)_i$.  Since 
$\fsl(4)\cong\fo(6)$, we can identify $\fspe(4)_0$ with $\fh(0|6)_0$.

It is not difficult to see that the elements of degree $-1$ in the 
standard gradings of $\fspe(4)$ and $\fh(0|6)$ constitute isomorphic 
$\fsl(4)\cong\fo(6)$-modules.  It is subject to a direct verification 
that it is possible to embed $\fspe(4)_1$ into $\fh(0|6)_1$.

Sergeev's extension $\fas$ is the result of the restriction to 
$\fspe(4)\subset\fh(0|6)$ of the cocycle that turns $\fh(0|6)$ into 
$\fpo(0|6)$.  The quantization deforms $\fpo(0|6)$ into 
$\fgl(\Lambda(\xi))$; the through maps $T_\lambda: 
\fas\tto\fpo(0|6)\tto\fgl(\Lambda (\xi))$ are representations of 
$\fas$ in the $4|4$-dimensional modules $\spin_\lambda$ isomorphic to 
each other for all $\lambda\neq 0$.  The explicit form of $T_\lambda$ 
is as follows:
$$
T_\lambda: \begin{pmatrix} a & b \cr 
c & -a^t \end{pmatrix}+d\cdot z\mapsto \begin{pmatrix}
a & b-\lambda \tilde c \cr 
c & -a^t \end{pmatrix}+\lambda d\cdot 1_{4|4}, \eqno{(0.1.4.2)}
$$
where $1_{4|4}$ is the unit matrix and $\tilde c$ is defined in 
formula (0.1.4.1).  Clearly, $T_\lambda$ is an irreducible 
representation for any $\lambda$.


\begin{thebibliography}{9999}
	
\bibitem[ALSh]{ALSh}
Alekseevsky D., Leites D., Shchepochkina 1., New examples of simple 
Lie superalgebras of vector fields.  C.r.  Acad.  Bulg.  Sci., v.  34, 
1980, no. 9, 1187--1190 (in Russian)

\bibitem[BL1]{BL1}
Bernstein J., Leites D., Invariant differential operators and
irreducible representations of Lie superalgebras of vector fields. 
Serdika, v. 7, 1981, 320--334 (in Russian); Sel.  Math.  Sov., v.  1,
no.  2, 1981, 143--160

\bibitem[CK1]{CK1}
Cheng Shun-Jen; Kac V., Generalized Spencer Cohomology and filtered
Deformations of $\Zee$-graded Lie Superalgebras, math-RT/9805039; Adv. 
Theor.  Math.  Phys.  2 (1998), no.  5, 1141--1182

\bibitem[CK2]{CK2}
Cheng S., Kac V., Structure of some $\Zee$-graded Lie superalgebras of
vector fields, Transformation groups, v.  4, 1999, 219--272

\bibitem[D]{D}
Deligne P. et al (eds.)  {\it Quantum fields and strings: a course for
mathematicians}.  Vol.  1, 2.  Material from the Special Year on
Quantum Field Theory held at the Institute for Advanced Study,
Princeton, NJ, 1996--1997.  AMS, Providence, RI; Institute for
Advanced Study (IAS), Princeton, NJ, 1999.  Vol.  1: xxii+723 pp.;
Vol.  2: pp.  i--xxiv and 727--1501

\bibitem[FF1]{FF1}
Feigin B. L., Fuks D. B., Stable cohomology of the algebra $W\sb{n}$
and relations in the algebra $L\sb{1}$.  (Russian) Funktsional.  Anal. 
i Prilozhen.  18 (1984), no.  3, 94--95

\bibitem[FF2]{FF2}
Feigin B. L., Fuks D. B., Cohomology of Lie groups
and Lie algebras.  (Russian) Current problems in mathematics. 
Fundamental directions, Vol.  21 (Russian), 121--209, 215, Itogi Nauki
i Tekhniki, Akad.  Nauk SSSR, Vsesoyuz.  Inst.  Nauchn.  i Tekhn. 
Inform., Moscow, 198


\bibitem[FLV]{FLV}
Floreanini R., Leites D.,  Vinet L., On defining relations of quantum
superalgebras. Lett. Math. Phys. 23, 1991, 127--131


\bibitem[FK]{FK}
Freund P., Kaplansky I., Simple supersymmetries. J. Math. Phys. 17
(1976), no. 2, 228--231

\bibitem[Fu]{Fu}
D. B. Fuks (Fuchs), {\it Cohomology of Infinite Dimensional Lie Algebras}.
Consultants Bureau, NY, 1987


\bibitem[GPS]{GPS}
Gomis J., Par\'is J., Samuel S., Antibracket, antifields and
gauge-theory quantization, Phys.  Rept.  259 (1995), no. 1--2, 1--191


\bibitem[Gr]{Gr}
Grozman P. , Classification of bilinear invariant operators on tensor 
fields.  (Russian) Funktsional.  Anal.  i Prilozhen.  14 (1980), no.  
2, 58--59



\bibitem[GL1]{GL1}
Grozman P., Leites D., {\it Mathematica}-aided study of Lie algebras 
and their cohomology.  From supergravity to ballbearings and magnetic 
hydrodynamics In: Ker\"anen V. (eds.)  {\em The second International 
Mathematica symposium}, Rovaniemi, 1997, 185--192

\bibitem[GL2]{GL2}
Grozman P., Leites D., Defining relations for classical Lie
superalgebras with Cartan matrix, hep-th 9702073; Czech.  J. Phys.,
Vol.  51, 2001, no.  1, 1--22

\bibitem[GL3]{GL3}
Grozman P., Leites D., Lie superalgebras of supermatrices of complex size.  Their
generalizations and related integrable systems. In:
by E. Ram'rez de Arellano, M. V. Shapiro, L. M. Tovar and N. L.
Vasilevski (eds.)  {\em Proc.  Internatnl.  Symp.  Complex Analysis
and related topics}, Mexico, 1996, Birkhauser Verlag, 1999, 73--105

\bibitem[GLP]{GLP}
Grozman P., Leites D., Poletaeva E., Defining relations for simple Lie
superalgebras.  Lie superalgebras without Cartan matrix.  In: Ivanov
E. et.  al.  (eds.)  {\em Supersymmetries and Quantum Symmetries}
(SQS'99, 27--31 July, 1999), Dubna, JINR, 2000, 387--396

\bibitem[GLS1]{GLS1}
Grozman P., Leites D., Shchepochkina I., Lie superalgebras of string 
theories, hep-th 9702120; Acta Mathematica Vietnamica, v. 
26, 2001, no. 1, 27--63

\bibitem[GSW]{GSW}
Green M., Schwarz J., Witten E. {\it Superstring theory}, vv.1, 2, 2nd
ed., Cambridge Monographs on Mathematical Physics., Cambridge Univ. 
Press, Cambridge, 1988.  x+470 pp., xii+596 pp.  

\bibitem[K1C]{K1C}
Kac V. G., Letter to editors, Funkcional.  Anal.  i prilozheniya 10, 
n.  2, 1976, 93

\bibitem[Ka1]{Ka1}
Kac V.G., {\it Infinite Dimensional Lie Algebras}.  3rd ed.  Cambridge 
Univ.  Press, Cambridge, 1990, xxii+400 pp.

\bibitem[Ka2]{Ka2}
Kac V. G., Classification of infinite-dimensional simple linearly 
compact Lie superalgebras. ESI-preprint 605 (www.esi.ac.at)

\bibitem[Ka3]{Ka3}
Kac V. G., Classification of infinite-dimensional simple groups of
supersymmetries and quantum field theory, math.QA/9912235

\bibitem[Ka4]{Ka4}
Kac V. G., Lie superalgebras. Adv. Math. v. 26, 1977, 8--96

\bibitem[Ka7]{Ka7}
Kac V., Classification of infinite-dimensional simple linearly compact
Lie superalgebras.  Adv.  Math.  139 (1998), no.  1, 1--55

\bibitem[KvdL]{KvdL}
Kac, V. G., van de Leur, J. W. On classification of superconformal
algebras.  In: S. J. Gates, Jr., C. R. Preitschopf and W. Siegel
(eds.)  {\it Strings '88} Proceedings of the workshop held at the
University of Maryland, College Park, Maryland, May 24--28, 1988. 
World Scientific Publishing Co., Inc., Teaneck, NJ, 1989, 77--106; Kac
V., Superconformal algebras and transitive group actions on quadrics. 
Comm.  Math.  Phys.  186 (1997), no.  1, 233--252; Erratum: 
Comm.  Math.  Phys.  217 (2001), no.  3, 697--698

\bibitem[K]{K}
Kaplansky I., Superalgebras.  Pacific J. Math.  86 (1980), no.  1,
93--98

\bibitem[Ki]{Ki}
Kirillov A. A., Ovsienko V. Yu., Udalova, O. D., 
Identities in the Lie
algebra of vector fields on the real line Selected translations. 
Selecta Math.  Soviet.  10 (1991), no.  1, 7--17

\bibitem[Ko1]{Ko1}
Kotchetkoff Yu.  D\'eformations de superalg\'ebres de Buttin et 
quantification.  C.R. Acad.  Sci.  Paris, ser.  I, 299: 14 (1984), 
643--645
\bibitem[Ko2]{Ko2}
Kochetkov Yu.  Deformations of Lie superalgebras.  VINITI 
Depositions, Moscow (in Russian) 1985, 384--85

\bibitem[L1]{L1}
Leites D., New Lie superalgebras and mechanics. Soviet Math. Dokl., v. 	
18, 1977, no. 5, 1277--1280

\bibitem[L2]{L2} 
Leites D., Lie superalgebras.  In: {\it Modern Problems of 
Mathematics.  Recent developments}, v.  25, VINITI, Moscow, 1984, 
3--49 (in Russian; English translation in: JOSMAR (J. Soviet Math.) v.  
30 (6), 1985, 2481--2512)

\bibitem[L3]{L3} 
Leites D., {\it Supermanifold theory}, Karelia Branch of the USSR Acad. 
Sci., Petrozavodsk, 1983, 200 pp.  (in Russian) (expanded in [L5])

\bibitem[L4]{L4} 
Leites D., Clifford algebra as a superalgebra and quantization. 
Theor.  Math.  Phys., v.  58, N2, 1984, 229--232; id., {\it
Quantization.  Supplement $3$}.  In: F. Berezin, M. Shubin.  {\it
Schr\" odinger equation}, Kluwer, Dordrecht, 1991, 483--522

\bibitem[L5]{L5} 
Leites D. (ed.)  {\it Seminar on Supermanifolds}, Reports of Stockholm 
University, 30/1988-13 and 31/1988-14.

\bibitem[LP]{LP}
Leites D., Poletaeva E., Defining relations for classical Lie algebras
of polynomial vector fields.  Math.  Scand.  81 (1997), no.  1, 5--19
(1998)


\bibitem[LSe]{LSe} 
Leites D., Serganova V., Defining relations for simple Lie
superalgebras.  I. Lie superalgebras with Cartan matrix, In:
J.~Mickelsson, O.~Peckkonen (eds.), {\em Proc.  of the conf. 
Topological methods in physics, 1991}, World Scientific, 1992,
194--201

\bibitem[LS]{LS}
Leites D., Serganova V., Symmetries wider than supersymmetries (with
V.~Serganova) In: S.~Duplij and J.~Wess (eds.)  {\em Noncommutative
structures in mathematics and physics}, Proc.  NATO Advanced Research
Workshop, Kiev, 2000.  Kluwer, 13--30

\bibitem[LSh0]{LSh0} 
Leites D., Shchepochkina I., Towards classification of simple
vectorial Lie superalgebras.  In: [L5], 31/1988-14; Leites D., Toward
classification of classical Lie superalgebras.  In: Nahm W., Chau L.
(eds.)  {\it Differential geometric methods in theoretical physics}
(Davis, CA, 1988), NATO Adv.  Sci.  Inst.  Ser.  B Phys., 245, Plenum,
New York, 1990, 633--651; Leites D., Shchepochkina I., Quivers and Lie
superalgebras, Czech.  J. Phys.  vol 47, no. 12, 1997, 1221--1229

\bibitem[LSh1]{LSh1} 
Leites D., Shchepochkina 1., Classification of simple Lie
superalgebras of vector fields, to appear; 

\bibitem[LSh3]{LSh3} 
Leites D., Shchepochkina I., How to quantize antibracket, preprint
ESI-875 (www.esi.ac.at); Theor.  and Math.  Physics, v.  126, no.  3,
339--369

\bibitem[LSh4]{LSh4} 
Leites D., Shchepochkina I., Deformations of simple vectorial Lie 
superalgebras, in preparation



\bibitem[OV]{OV} 
Onishchik A., Vinberg \'E., {\it Lie groups and algebraic groups}. 
Translated from the Russian and with a preface by D. A. Leites. 
Springer Series in Soviet Mathematics.  Springer-Verlag, Berlin, 1990. 
xx+328 pp.




\bibitem[Sh3]{Sh3} 
Shchepochkina 1., New exceptional simple Lie superalgebras.  C. R.
bulg.  Sci., 36, 3, 1983, 313--314

\bibitem[ShM]{ShM}
Shchepochkina 1., Maximal subalgebras of simple Lie superalgebras.
In: Leites D. (ed.) {\it Seminar on Supermanifolds} vv.1--34, 1987--1990,
v. 32/1988-15, Reports of Stockholm University, 1--43 (hep-th 9702120)

\bibitem[Sh5]{Sh5} 
Shchepochkina 1., The five exceptional simple Lie superalgebras of
vector fields, hep-th 9702120; id., The five exceptional simple Lie
superalgebras of vector fields.  Sov.  J. Funct.  Analisys, v.33,
1999, no.3, 14 pp.

\bibitem[Sh14]{Sh14}
Shchepochkina I., Five exceptional simple Lie superalgebras of vector 
fields and their fourteen regradings.  Represent.  Theory, v.  3, 
1999, 3 (1999), 373--415

\bibitem[ShP]{ShP} 
Shchepochkina 1., Post G., Explicit bracket in an exceptional simple 
Lie superalgebra, Explicit bracket in an exceptional simple Lie 
superalgebra.  Internat.  J. Algebra Comput.  8 (1998), no.  4, 
479--495; physics 9703022

\bibitem[SNR]{SNR} 
Scheunert M., Nahm W., Rittenberg V., Classification of all simple
graded Lie algebras whose Lie algebra is reductive.  I, II.
Construction of the exceptional algebras.  J. Mathematical Phys.  17
(1976), no.  9, 1626--1639, 1640--1644

\bibitem[Sch]{Sch}
Scheunert M., {\em The theory of 
Lie superalgebras.  An introduction}.  Lecture Notes in Mathematics, 
716.  Springer, Berlin, 1979.  x+271 pp

\bibitem[S1]{S1} 
Serganova V., Automorphisms of simple Lie superalgebras.  (Russian)
Izv.  Akad.  Nauk SSSR Ser.  Mat.  48 (1984), no.  3, 585--598; Feigin
B. L., Leites D. A.,Serganova,V. V., Kac-Moody superalgebras.  Group
theoretical methods in physics, Vol.  1--3 (Zvenigorod, 1982),
631--637, Harwood Academic Publ., Chur, 1985

\bibitem[S2]{S2} 
Serganova V., On generalizations of root systems.  Comm.  Algebra 24
(1996), no.  13, 4281--4299

\bibitem[Sm]{Sm}
Shmelev G. S., Differential operators that are invariant with respect
to the Lie superalgebra $H(2,\,2;\,\lambda )$ and its irreducible
representations.  (Russian) C. R. Acad.  Bulgare Sci.  35 (1982), no. 
3, 287--290

\bibitem[St]{St}
Sternberg S., {\em Lectures on differential geometry}, 
Chelsey, 2nd edition, 1985 


\bibitem[Sud]{Sud} 
Sudbery, A. Octonionic description of exceptional Lie superalgebras.  
J. Math.  Phys.  24 (1983), no.  8, 1986--1988

\bibitem[vdL]{vdL}
van de Leur J., A classification of contragredient Lie superalgebras
of finite growth.  Comm.  Algebra 17 (1989), no.  8, 1815--1841


\bibitem[WB]{WB}
Wess J., Bagger J., {\it Supersymmetry and supergravity}.  Princeton
Series in Physics.  Princeton University Press, Princeton, N.J., 1983. 
i+180 pp

\bibitem[WL]{WL} 
Woit P., String Theory: An Evaluation, physics/0102051; Larsson T.,
Symmetries of Everything, math-ph/0103013

\bibitem[WZ]{WZ}
Wess, J., Zumino, B., {\it Supergauge transformations in four
dimensions}.  Nuclear Phys.  {\bf B70} (1974), 39--50; Wess J., {\it
Supersymmetry-supergravity.  Topics in quantum field theory and gauge
theories} (Proc.  VIII Internat.  GIFT Sem.  Theoret.  Phys.,
Salamanca, 1977), pp.  81--125, Lecture Notes in Phys., 77, Springer,
Berlin-New York, 1978; Wess J., Zumino B., {\it Superspace formulation
of supergravity}.  Phys.  Lett.  {\bf B 66} (1977), no.  4, 361--364;
Wess J., {\it Supersymmetry/supergravity.  Concepts and trends in
particle physics} (Schladming, 1986), 29--58, Springer, Berlin, 1987;
Wess J., {\it Introduction to supersymmetric theories.  Frontiers in
particle physics '83} (Dubrovnik, 1983), 104--131, World Sci. 
Publishing, Singapore, 1984; Wess J., Bagger J., {\it Supersymmetry
and supergravity}.  Second edition.  Princeton Series in
Physics.  Princeton University Press, Princeton, NJ, 1992.  x+259 pp.

\bibitem[Y]{Y} 
Yamane H., On defining relations of affine Lie superalgebras and
affine quantized universal enveloping superalgebras.  q-alg 9603015;
Publ.  Res.  Inst.  Math.  Sci.  35 (1999), no.  3, 321--390


\end{thebibliography}
\end{document}